\documentclass[traditabstract]{aa} 

\usepackage{graphicx}
\usepackage{natbib}
\usepackage{lscape}
\usepackage{longtable}
\usepackage{txfonts}
\usepackage{caption}
\usepackage{subcaption}

\usepackage{float}

\def\mstar  {$M_{\star}$}
\def\macc   {$\dot{M}_{\rm acc}$}
\def\lacc   {$L_{\rm acc}$}

\def\msun {$M_{\odot}$}
\def\lsun {$L_{\odot}$}
\def\lstar {$L_\star$}
\def\rstar {$R_\star$}

\def\laccnoise   {$L_{\rm acc,noise}$}

\def\nodata {...}
\def\mdisk {$M_{\rm disk}$}

\defcitealias{Manara16}{MFH16}
\defcitealias{Herczeg14}{HH14}

\usepackage{color}
\usepackage{amstext}

\begin{document}

   \title{X-Shooter study of accretion in Chamaeleon I\thanks{This work is based on observations made with ESO Telescopes at the Paranal Observatory under programme ID 090.C-0253 and 095.C-0378. }}

\subtitle{II. A steeper increase of accretion with stellar mass for very low-mass stars?}

\titlerunning{X-Shooter study of accretion in Chamaeleon~I}
\authorrunning{Manara et al.}

   \author{C.F. Manara \inst{1}\fnmsep\thanks{ESA Research Fellow}, L. Testi\inst{2,3,4}, G.J. Herczeg\inst{5}, I. Pascucci\inst{6}, J.M. Alcal\'a\inst{7}, A. Natta\inst{3,8}, S. Antoniucci\inst{9}, D. Fedele\inst{3}, \\G.D. Mulders\inst{6,10}, T. Henning\inst{11}, S. Mohanty\inst{12}, T. Prusti\inst{1}, E. Rigliaco\inst{13}
          }

   \institute{Science Support Office, Directorate of Science, European Space Research and Technology Centre (ESA/ESTEC), Keplerlaan 1, 2201 AZ Noordwijk, The Netherlands \\
              \email{cmanara@cosmos.esa.int}
         \and
             European Southern Observatory, Karl-Schwarzschild-Strasse 2, D-85748 Garching bei M\"unchen, Germany
         \and
         INAF-Osservatorio Astrofisico di Arcetri, L.go E. Fermi 5, I-50125 Firenze, Italy
         \and
             Excellence Cluster Universe, Boltzmannstr. 2, D-85748 Garching, Germany
         \and
             Kavli Institute for Astronomy and Astrophysics, Peking University, Yi He Yuan Lu 5, Haidian Qu, Beijing 100871, China
\and
Lunar and Planetary Laboratory, The University of Arizona, Tucson, AZ 85721, USA
\and 
INAF-Osservatorio Astronomico di Capodimonte, via Moiariello 16, 80131 Napoli, Italy
\and
School of Cosmic Physics, Dublin Institute for Advanced Studies, 31 Fitzwilliams Place, Dublin 2, Ireland
\and
INAF-Osservatorio Astronomico di Roma, Via di Frascati 33, 00078 Monte Porzio Catone, Italy
\and
Earths in Other Solar Systems Team, NASA Nexus for Exoplanet System Science
\and
Max Planck Institute for Astronomy, K\"onigstuhl 17, D-69117 Heidelberg, Germany
\and 
Imperial College London, 1010 Blackett Lab, Prince Consort Rd., London SW7 2AZ, UK
\and
INAF Osservatorio Astornomico di Padova, Vicolo dell'Osservatorio 5, 35122 Padova, Italy
             }

   \date{Received November 28, 2016; accepted April 6, 2017}

 
\abstract{The dependence of the mass accretion rate on the stellar properties is a key constraint for star formation and disk evolution studies. Here we present a study of a sample of stars in the Chamaeleon~I star-forming region carried out using spectra taken with the ESO VLT/X-Shooter spectrograph. The sample is nearly complete down to stellar masses ($M_\star$) $\sim 0.1 M_\odot$ for the young stars still harboring a disk in this region. We derive the stellar and accretion parameters using a self-consistent method to fit the broadband flux-calibrated medium resolution spectrum. 
The correlation between accretion luminosity to stellar luminosity, and of mass accretion rate to stellar mass in the logarithmic plane yields slopes of 1.9$\pm$0.1 and 2.3$\pm$0.3, respectively. 
These slopes and the accretion rates are consistent with previous results in various star-forming regions and with different theoretical frameworks. However, we find that a broken power-law fit, with a steeper slope for stellar luminosity lower than $\sim 0.45$ \lsun \ and for stellar masses lower than $\sim$ 0.3 \msun\  is slightly preferred according to different statistical tests, but the single power-law model is not excluded. The steeper relation for lower mass stars can be interpreted as a faster evolution in the past for accretion in disks around these objects, or as different accretion regimes in different stellar mass ranges. 
Finally, we find two regions on the mass accretion versus stellar mass plane that are empty of objects: one region at high mass accretion rates and low stellar masses, which is related to the steeper dependence of the two parameters we derived. The second region is located just above the observational limits imposed by chromospheric emission, at $M_\star \sim 0.3-0.4 M_\odot$.
These are typical masses where photoevaporation is known to be effective. The mass accretion rates of this region are $\sim 10^{-10} M_\odot$/yr, which is compatible with the value expected for photoevaporation to rapidly dissipate the inner disk.
}



   \maketitle
%

\section{Introduction}\label{sect::introduction}

The circumstellar disk around young stars is the birthplace of planets, and the architecture of the forming planetary system depends on the evolution with time of the surface density of gas and dust in such disks \citep[e.g.,][]{Thommes08,Mordasini12}. 
Different processes are at play during the disk evolutionary phases, such as accretion of matter through the disk and on the central star \citep[e.g.,][]{Hartmann16}, removal of material through winds driven by the high-energy radiation from the central star, or the result of magnetic torques on disk material \citep[e.g.,][]{AlexanderPPVI,Gorti16,Armitage13,Bai16}, or even loss of material due to external disturbances, such as binarity and encounters \citep[e.g.,][]{Clarke93,Pfalzner05} or external photoevaporation \citep[e.g.,][]{Clarke07,Anderson13,Facchini16}. Each of these processes modifies the distribution of material in the disk and is thus relevant for the planet formation process. Models aiming at explaining the observed properties of exoplanets or our own solar system need constraints on the contribution of each of these disk evolution processes \citep[e.g.,][]{Adams10,Bitsch15,Pfalzner15}. These constraints are obtained by studying properties of young stars and their disks at different ages, evolutionary stages, and with a wide span of stellar properties. Here we focus on measuring the rate at which material is accreted onto the central star, which is crucial information for all the aforementioned processes, as a function of stellar mass and luminosity for a complete sample of young stars that are surrounded by disks in the nearby ($d$ = 160 pc\footnote{The recently released Gaia data \citep{DR1} include only eight objects that are confirmed members of Chamaeleon~I. Their measured parallaxes are compatible with the quoted distance when considering the statistical and systematic uncertainties of this first data release. Therefore, we adopt this commonly used value for the distance in our analysis.}) $\sim$2-3 Myr old Chamaeleon~I star-forming region \citep{LuhmanCha}.

Measurements of the mass accretion rate on the central star (\macc) are taken from the excess emission in spectra of young stars that is due to the shock of gas infalling from the disk onto the central star along the stellar magnetic field lines \citep[e.g.,][]{Calvet98}. This excess emission is especially strong in the ultraviolet (UV) spectral range, in particular in the Balmer continuum, as well as in the optical \citep[e.g.,][]{Fischer11}. Modern instruments mounted on 8 m class telescopes, such as the X-Shooter spectrograph \citep{Vernet11} at the ESO Very Large Telescope (VLT), allow us to access the Balmer continuum region at $\lambda\lesssim$ 346 nm with high sensitivity while simultaneously obtaining medium-resolution flux-calibrated spectra that cover up to $\lambda\sim$ 2.5 $\mu$m. This has allowed researchers to derive \macc \ for large samples of objects in different star-forming regions by modeling the Balmer continuum excess in the spectra of young stars (e.g., Taurus, \citealt{Herczeg08}; Lupus, \citealt{Alcala14}; $\sigma$-Orionis, \citealt{Rigliaco12}). None of these studies has been performed on a statistically complete sample of disk-bearing young stars. A complete sample is instead crucial to firmly constrain the effects of different mechanisms on the disk evolution.

In a previous work including $\sim$40\% of the young stars that
are surrounded by disks in the Chamaeleon~I region \citep[][hereafter \citetalias{Manara16}]{Manara16} we have derived stellar and accretion properties using VLT/X-Shooter spectra. In this sample \macc \ scales with stellar mass (\mstar) as a power law with exponent $\sim$2, in agreement with previous results in other star-forming regions \citep[e.g.,][]{Muzerolle03,Mohanty05,Natta06,Herczeg08}. Moreover, the young stars in the sample of \citetalias{Manara16}  whose disk present a dust-free region in their inner part, the so-called transition disks (TDs), have similar \macc \ as full disks with the same \mstar. This is consistent with previous results \citep[e.g.,][]{Manara14,Keane14}. The spread of values of \macc \ at any given \mstar \ was instead found to be smaller than those found in $\rho$-Ophiuchus \citep{Natta06} or in Taurus \citep{Herczeg08}, but larger than the sample comprising $\sim$50\% of the stars with disks in Lupus \citep{Alcala14}. However, complete samples, as analyzed here, are needed to constrain the real extent of the spread of \macc \ and the dependence of \macc \ on \mstar, to be then compared with expectations from different models of disk evolution.

In the following, we present our sample and the data acquisition and reduction in Sect.~\ref{sect::obs}, then we describe the stellar and accretion parameters derived for the sample in Sect.~\ref{sect::methodsect}. In Sect.~\ref{sect::results} we study the dependence between the accretion parameters and the stellar mass and luminosity, and we discuss the implications of our findings in Sect.~\ref{sect::discussion}. Finally, we present our conclusions in Sect.~\ref{sect::conclusions}.


\section{Data collection}\label{sect::obs}

\subsection{Sample selection}\label{sect::sample}

The sample of objects discussed here is based on the selection made for the accompanying ALMA survey of the disk population of the Chamaeleon~I region \citep{Pascucci16}. The ALMA sample includes all the objects displaying excess emission with respect to the photosphere in more than one infrared wavelength ranging from the near-infrared, that is, 2MASS data, to mid-infrared, which means \textit{Spitzer} or WISE data, and when available, far-infrared, that is, data obtained with \textit{Herschel} \citep{Luhman08a,Szucs10,Olofsson13}.
This selection excludes all the objects classified as Class~III, or disk-less, by \citet{Luhman08a}. Objects still surrounded by an optically thick envelope, the Class~0 and Class~I targets, were also excluded from the sample. These selection criteria result in a sample of 93 disk-bearing objects. According to previous spectral type classifications \citep{Luhman08a,Luhman08b,Luhman07}, this sample is complete down to M6, which roughly corresponds to \mstar$\sim$0.1\msun, and it includes three targets with later spectral type, as discussed by \citet{Pascucci16}. This sample includes some binary stars, as reported in \citet{Pascucci16}. 

The spectroscopic survey presented here similarly targets all the known young stellar objects in the region harboring a disk. Of the ALMA sample, only 5 targets were never observed with X-Shooter\footnote{\citet{Pascucci16} adopted values for the stellar parameters from \citet{Luhman07} for these five targets lacking X-Shooter spectra, and for three additional targets: the binary system 2MASS J11175211-7629392, ESO-H$\alpha$ 574, and ISO-ChaI 217.}: 4\footnote{These objects are 2MASS J11070925-7718471, 2MASS J11094260-7725578, 2MASS J11062942-7724586, and 2MASS J11082570-7716396.} because they are too faint to obtain a high enough signal-to-noise
ratio in the optical spectrum, and one (CHXR30B) as it was not in the slit when observing CHXR30A. Two of these targets have previously derived spectral types M6 and M8, which means that they are in the range where the sample is incomplete. 
All other targets were observed with X-Shooter. In particular, 43 ALMA targets observed in Pr.Id. 095.C-0378 (PI Testi) and 8 in Pr.Id. 090.C-0253 (PI Antoniucci) are presented here for the first time. Of these, 1 target is a newly discovered binary, as discussed in Sect.~\ref{sect::bin}. The remaining targets in the ALMA sample were observed with X-Shooter in the past. In particular, 35 stars were observed in Pr. Id. 084.C-1095 (PI Herczeg) and analyzed by \citet{Manara14,Manara16}. Three of the ALMA targets studied by \citetalias{Manara16} are in binary systems that were resolved and studied separately. The latter sample contained mainly solar-mass objects, while the more recent spectra analyzed here are mostly focused on the lower-mass objects and selected to complete the initial sample.
Two other targets were observed in Pr.Id. 085.C-0238 and 089.C-0143 (PI Alcal\'a). One of them is a well-known edge-on target, ESO-H$\alpha$ 574, which was analyzed by \citet{Bacciotti11} and is not discussed here, since the stellar and accretion parameters of edge-on targets are too uncertain to be included in the analysis. The other, ISO-ChaI 217, is a a well-known brown dwarf with a jet \citep{Whelan14} and is included in our sample.
With respect to the ALMA sample, our sample includes 2 additional targets. One is Sz18, which was not included in the ALMA sample because it was originally classified as a Class~III target by \citet{Luhman08a}, but was reclassified as a transition disk by \citet{Kim09} and thus studied by \citet{Manara14}. 
The other addition, Cha-H$\alpha$1, is a brown dwarf and in the spectral type range where the ALMA survey was incomplete, and it was studied by \citetalias{Manara16}. 

With respect to the ALMA sample, which is complete down to M6, we therefore lack only 3 targets with spectral type earlier than M6. We accordingly consider our total sample of 94 objects, which includes resolved components of binary systems, to comprise 97\% of the disk-bearing young stellar objects with spectral type earlier than M6 in the Chamaeleon~I region.

\subsection{New observations}

The new data presented in this work were obtained in two different observing runs using the VLT/X-Shooter spectrograph \citep{Vernet11}. The targets of Pr.Id. 095.C-0378 (PI Testi) were observed using both narrow (see Table~\ref{tab::log}) and large (5.0\arcsec x11\arcsec) slits. The narrow-slit observations were carried out by nodding the target along the slits on two slit positions A and B, with one exposure per position. The nodding cycle was ABBA, with a total of four exposures. The exposure times and slit widths were chosen to obtain enough signal based on the brightness of the targets, and are reported in Table~\ref{tab::log}. These exposures with narrow slits were obtained to achieve the highest spectral resolution, typically $R\sim$4000-10000 in the UVB arm ($\lambda\lambda\sim$300-550 nm), and $R\sim$6700-18000 in the VIS arm ($\lambda\lambda\sim$550-1000 nm). Simultaneous observations in the near-infrared (NIR) arm ($\lambda\lambda\sim$1000-2500 nm) will be described in a following paper. The large-slit observations were carried out in stare mode on the target immediately after the narrow-slit exposures, and they lasted for $\sim$10\% of the total exposure time of the observations taken with the narrow slit. 
These large-slit observations led to spectra with a much lower resolution, but without slit losses, and are thus crucial for a correct flux calibration of the spectra obtained with the narrow slit. 

The eight targets observed during Pr.Id. 090.C-0253 (PI Antoniucci) were similarly observed by nodding the telescope and using narrow slits, with widths reported in Table~\ref{tab::log}. All these spectra have a resolution of $R\sim$10000 in the UVB arm and $R\sim$18000 in the VIS arm. However, no exposures with the large slit were obtained in this observing run. We therefore used non-simultaneous photometric data to calibrate these spectra, as discussed in the next section.

\subsection{Data reduction}

Data reduction was performed with the ESO X-Shooter pipeline \citep{Modigliani10} version v.2.5.2 run through the Reflex workflow \citep{reflex}. The pipeline performs the usual reduction scheme, including bias subtraction, flat-fielding, wavelength calibration, flexure, and atmospheric dispersion correction, background removal (in stare mode) or combination of spectra obtained in a nodding cycle, and spectrum extraction. The latter is performed by the pipeline on the 1D background-subtracted spectra using a large extraction window. In order to maximize the signal in the UVB spectra, we manually extracted the spectrum using the apall task in IRAF\footnote{IRAF is distributed by National Optical Astronomy Observatories, which is operated by the Association of Universities for Research in Astronomy, Inc., under cooperative agreement with the National Science Foundation.}. The flux calibration of the spectra was also performed in the pipeline by deriving a spectral response function using a flux standard star observed during the same night. Typically, this procedure leads to a flux calibration accuracy of $\sim$2\% \citep[e.g.,][]{Alcala14,Manara16}. However, the spectra obtained using the narrow slit suffer slit losses. To correct for this effect, we scaled the flux-calibrated spectra obtained with the narrow slit to the flux-calibrated spectra obtained with the large slit using the median of the ratio between the two spectra as a correction factor in each arm when the narrow slit was 0.9\arcsec \ wide, or more, and a wavelength-dependent linear fit of the ratio in each arm when the narrow slit was 0.5\arcsec \ or 0.4\arcsec \ wide. The latter is needed to account for wavelength-dependent differential slit losses that are due to the high airmass of the targets at the time of the observations. Typically, the correction factors are $\sim 1.5 - 2$, and they are $\sim 3-5$ in a few cases when observations were performed with the narrowest slits and with large seeing. The agreement between the flux calibrated spectra in the overlapping region of the VIS and UVB arm is excellent. 
This same procedure to correct for slit losses was not possible for the eight targets observed in Pr.Id. 090.C-0253, as these were observed only with the narrow slits. We thus used available photometry to correct these spectra (see Table~\ref{tab::lit}). We first matched the spectra from the three arms and then scaled all of them to minimize the scatter to the photometric points. The correction factors found with this method are $\sim 3-5$, compatible with those found when the large slits were also adopted. 
Finally, telluric correction was performed in the VIS arm using telluric standard stars observed close in time and airmass to the target. The spectra of these standard stars were continuum normalized and the photospheric absorption features were removed before correcting the telluric lines using the telluric task of IRAF. 

The sample includes some unresolved close binaries, T5, CHXR30A, CHXR71, Cha-H$\alpha$2, T45, T46, CHXR 47, Hn13, CHXR79, and T43. Their spectra thus include both components.  
On the other hand, the two components of 2MASS J11175211$-$7629392 were separated, and we manually extracted the two spectra. Their analysis is discussed in Sect.~\ref{sect::bin}.

Three objects needed a particular procedure to reduce their spectra. CHXR71 was observed during a cloudy night, and the spectrum obtained with the large slit is fainter than the spectrum with the small slit. We thus corrected for slit losses using a correction factor of 3, which leads to a spectrum matching the available photometry. The ratio of the UVB spectra with the large and small slits for both T48 and T27 could not be fitted with a linear dependence to wavelength, and we thus used a second-order polynomial to fit this ratio and correct the small-slit spectra.


\section{Data analysis}\label{sect::methodsect}

The spectra analyzed here contain multiple features typical of young stellar objects. We checked in particular whether they present the H$\alpha$ line in emission and the Li absorption line at $\lambda$670.8 nm. These features are usually considered a confirmation of the nature of young stars for the target. All objects except for ISO-ChaI 79 present a clear H$\alpha$ line in emission. This line is visible in the spectrum of this target only with some smoothing, as the spectrum is very noisy, probably because the disk is viewed edge-on, obscuring the star (see Sect.~\ref{sect::hrd}) The lithium absorption line is detected in 36 targets in the sample. These are the objects whose spectrum has signal-to-noise S/N$\gtrsim$10 in the continuum at $\lambda\sim$ 700 nm (see Table~\ref{tab::log}). We thus confirm that all the targets are young stellar objects, since the spectra of 36 targets present the lithium line in absorption and the H$\alpha$ line in emission, while the other targets show the H$\alpha$ line in emission and the lithium line is not detected due to the low signal-to-noise ratio of the spectra. 

In the following, we describe the method used to derive stellar and accretion parameters for the targets. 

\begin{table*}  
\begin{center}  
\footnotesize  
\caption{\label{tab::results} Names, coordinates, and properties for the Chamaeleon~I targets included in this work }  
\begin{tabular}{l|l|cc|ccc| cc | cc | l }    
\hline \hline  
 2MASS & Object &  RA(2000)  & DEC(2000) &  SpT & T$_{\rm eff}$ & A$_V$ & \lstar & log\lacc & \mstar & log\macc & Notes \\    
   &   &  h \, :m \, :s & $^\circ$ \, ' \, ''   &  \hbox{} & [K] & [mag] & [\lsun] & [\lsun] & [\msun] & [\msun/yr]  &  \\     
\hline  

\multicolumn{12}{c}{Sample from Pr.Id. 095.C-0378 (PI Testi)}\\   
\hline  
J10533978-7712338  &   \nodata   &  10:53:39.78  &  $-$77:12:33.8 & M2 & 3560 & 1.8 & 0.02 & -4.56 & 0.33 & -11.95 & UL,$^m$,$^*$ \\ 
J10561638-7630530   &   ESO H$\alpha$ 553  &  10:56:16.38  &  $-$76:30:53.0 & M6.5 & 2935 & 0.3 & 0.08 & -4.55 & 0.11 & -10.95 & $^\dag$ \\ 
J10574219-7659356   &   T5        &  10:57:42.19  &  $-$76:59:35.6 & M3 & 3415 & 1.4 & 0.53 & -1.98 & 0.28 & -8.51 & \nodata \\ 
J10580597-7711501         &   \nodata   &  10:58:05.97  &  $-$77:11:50.1 & M5.5 & 3060 & 1.2 & 0.01 & -5.07 & 0.11 & -11.87 & $^\dag$,$^m$ \\ 
J11004022-7619280         &   T10    &  11:00:40.22  &  $-$76:19:28.0 & M4 & 3270 & 1.1 & 0.10 & -2.45 & 0.23 & -9.22 & \nodata \\ 
J11023265-7729129         &   CHXR71   &  11:02:32.65  &  $-$77:29:12.9 & M3 & 3415 & 1.4 & 0.26 & -3.81 & 0.29 & -10.52 & $^\dag$ \\ 
J11040425-7639328          &   CHSM1715   &  11:04:04.25  &  $-$76:39:32.8 & M4.5 & 3200 & 1.5 & 0.03 & -3.91 & 0.18 & -10.84 & $^m$ \\ 
J11045701-7715569          &   T16    &  11:04:57.01  &  $-$77:15:56.9 & M3 & 3415 & 4.9 & 0.28 & -1.11 & 0.29 & -7.80 & \nodata \\ 
J11062554-7633418         &   ESO H$\alpha$ 559  &  11:06:25.54  &  $-$76:33:41.8 & M5.5 & 3060 & 2.5 & 0.03 & -4.09 & 0.12 & -10.79 & $^m$ \\ 
J11063276-7625210          &   CHSM 7869   &  11:06:32.76  &  $-$76:25:21.0 & M6.5 & 2935 & 0.7 & 0.02 & -4.49 & 0.07 & -11.06 & $^m$ \\ 
J11063945-7736052   &   ISO-ChaI 79   &  11:06:39.45  &  $-$77:36:05.2 & M5 & 3125 & 2.7 & 0.002 & $<$-5.57 & \nodata & \nodata & UL,$^m$,$^*$ \\ 
J11064510-7727023          &   CHXR20   &  11:06:45.10  &  $-$77:27:02.3 & K6 & 4205 & 3.4 & 0.53 & -1.49 & 0.90 & -8.71 & \nodata \\ 
J11065939-7530559          &   \nodata   &  11:06:59.39  &  $-$75:30:55.9 & M5.5 & 3060 & 0.4 & 0.01 & -4.26 & 0.10 & -11.12 & \nodata \\ 
J11071181-7625501         &   CHSM 9484   &  11:07:11.81  &  $-$76:25:50.1 & M5.5 & 3060 & 0.8 & 0.01 & -4.98 & 0.10 & -11.84 & $^\dag$,$^m$ \\ 
J11072825-7652118   &   T27    &  11:07:28.25  &  $-$76:52:11.8 & M3 & 3415 & 1.2 & 0.34 & -1.72 & 0.29 & -8.36 & \nodata \\ 
J11074245-7733593          &   Cha-H$\alpha$-2   &  11:07:42.45  &  $-$77:33:59.3 & M5.5 & 3060 & 2.4 & 0.04 & -3.39 & 0.13 & -10.05 & \nodata \\ 
J11074366-7739411         &   T28    &  11:07:43.66  &  $-$77:39:41.1 & M1 & 3705 & 2.8 & 0.30 & -0.97 & 0.48 & -7.92 & \nodata \\ 
J11074656-7615174          &   CHSM 10862   &  11:07:46.56  &  $-$76:15:17.4 & M6.5 & 2935 & 0.5 & 0.01 & -5.30 & 0.07 & -12.03 & $^\dag$,$^m$ \\ 
J11075809-7742413          &   T30   &  11:07:58.09  &  $-$77:42:41.3 & M3 & 3415 & 3.8 & 0.16 & -1.49 & 0.30 & -8.31 & \nodata \\ 
J11080002-7717304          &   CHXR30A   &  11:08:00.02  &  $-$77:17:30.4 & K7 & 4060 & 8.0 & 0.97 & -3.23 & 0.69 & -10.17 & $^\dag$,$^m$ \\ 
J11081850-7730408   &   ISO-ChaI 138   &  11:08:18.50  &  $-$77:30:40.8 & M6.5 & 2935 & 0.0 & 0.01 & -5.17 & 0.07 & -11.81 & $^\dag$ \\ 
J11082650-7715550          &   ISO-ChaI 147   &  11:08:26.50  &  $-$77:15:55.0 & M5.5 & 3060 & 2.5 & 0.01 & -4.35 & 0.11 & -11.15 & $^m$ \\ 
J11085090-7625135   &   T37   &  11:08:50.90  &  $-$76:25:13.5 & M5.5 & 3060 & 0.8 & 0.03 & -4.05 & 0.12 & -10.74 & \nodata \\ 
J11085367-7521359          &   \nodata   &  11:08:53.67  &  $-$75:21:35.9 & M1 & 3705 & 1.5 & 0.19 & -1.06 & 0.51 & -8.15 & \nodata \\ 
J11085497-7632410         &   ISO-ChaI 165    &  11:08:54.97  &  $-$76:32:41.0 & M5.5 & 3060 & 1.8 & 0.03 & -3.95 & 0.12 & -10.65 & $^m$ \\ 
J11092266-7634320         &   C 1-6    &  11:09:22.66  &  $-$76:34:32.0 & M1 & 3705 & 8.0 & 0.08 & -2.20 & 0.58 & -9.54 & $^m$ \\ 
J11095336-7728365          &   ISO-ChaI 220   &  11:09:53.36  &  $-$77:28:36.5 & M5.5 & 3060 & 5.2 & 0.01 & -3.65 & 0.11 & -10.45 & $^m$ \\ 
J11095873-7737088          &   T45    &  11:09:58.73  &  $-$77:37:08.8 & M0.5 & 3780 & 3.0 & 0.61 & -0.12 & 0.49 & -6.95 & \nodata \\ 
J11100369-7633291         &   Hn11  &  11:10:03.69  &  $-$76:33:29.1 & M0 & 3850 & 5.0 & 0.23 & -2.45 & 0.63 & -9.62 & $^m$ \\ 
J11100704-7629376          &   T46   &  11:10:07.04  &  $-$76:29:37.6 & K7 & 4060 & 1.2 & 0.53 & -1.59 & 0.75 & -8.70 & \nodata \\ 
J11100785-7727480          &   ISO-ChaI 235   &  11:10:07.85  &  $-$77:27:48.0 & M5.5 & 3060 & 6.0 & 0.04 & -4.28 & 0.13 & -10.96 & $^m$ \\ 
J11103801-7732399   &   CHXR 47   &  11:10:38.01  &  $-$77:32:39.9 & K4 & 4590 & 3.9 & 1.90 & -0.94 & 1.32 & -8.12 & \nodata \\ 
J11104141-7720480          &   ISO-ChaI 252   &  11:10:41.41  &  $-$77:20:48.0 & M5.5 & 3060 & 3.6 & 0.01 & -3.12 & 0.11 & -9.91 & $^m$ \\ 
J11105333-7634319         &   T48   &  11:10:53.33  &  $-$76:34:31.9 & M3 & 3415 & 1.2 & 0.16 & -1.14 & 0.30 & -7.96 & \nodata \\ 
J11105359-7725004   &   ISO-ChaI 256   &  11:10:53.59  &  $-$77:25:00.4 & M5 & 3125 & 5.5 & 0.04 & -3.55 & 0.15 & -10.32 & $^m$ \\ 
J11105597-7645325         &   Hn13  &  11:10:55.97  &  $-$76:45:32.5 & M6.5 & 2935 & 1.3 & 0.13 & -3.24 & 0.12 & -9.57 & \nodata,$^*$ \\ 
J11111083-7641574   &   ESO H$\alpha$ 569  &  11:11:10.83  &  $-$76:41:57.4 & M1 & 3705 & 2.2 & 0.003 & -2.90 & \nodata & \nodata & UL,$^*$ \\ 
J11120351-7726009   &   ISO-ChaI 282   &  11:12:03.51  &  $-$77:26:00.9 & M5.5 & 3060 & 2.8 & 0.07 & -3.32 & 0.14 & -9.89 & \nodata \\ 
J11120984-7634366   &   T50   &  11:12:09.84  &  $-$76:34:36.6 & M5 & 3125 & 0.1 & 0.14 & -2.82 & 0.17 & -9.34 & \nodata \\ 
J11175211-7629392\_one    &   \nodata   &  11:17:52.11  &  $-$76:29:39.e & M4.5 & 3200 & 0.8 & 0.07 & -4.39 & 0.20 & -11.16 & $^\dag$ \\ 
J11175211-7629392\_two    &   \nodata   &  11:17:52.11  &  $-$76:29:39.o & M4.5 & 3200 & 0.3 & 0.06 & -4.66 & 0.19 & -11.44 & $^\dag$ \\ 
J11183572-7935548    &  \nodata    &  11:18:36.72  &  $-$79:35:55.8 & M5 & 3125 & 0.0 & 0.26 & -2.50 & 0.19 & -8.95 & TD \\ 
J11241186-7630425          &   \nodata   &  11:24:11.86  &  $-$76:30:42.5 & M5.5 & 3060 & 1.0 & 0.03 & -3.90 & 0.12 & -10.59 & TD \\ 
J11432669-7804454         &   \nodata   &  11:43:26.69  &  $-$78:04:45.4 & M5.5 & 3060 & 0.4 & 0.09 & -2.19 & 0.14 & -8.71 & \nodata \\ 
\hline  
\multicolumn{12}{c}{Sample from Pr.Id. 090.C-0253 (PI Antoniucci)}\\   
\hline  
J11072074-7738073          &   Sz19   &  11:07:20.74  &  $-$77:38:07.3 & K0 & 5110 & 1.5 & 5.10 & -0.37 & 2.08 & -7.63 & \nodata,$^*$ \\ 
J11091812-7630292          &   CHXR79   &  11:09:18.12  &  $-$76:30:29.2 & M0 & 3850 & 5.0 & 0.25 & -1.91 & 0.62 & -9.05 & $^m$ \\ 
J11094621-7634463   &   Hn 10e   &  11:09:46.21  &  $-$76:34:46.3 & M3 & 3415 & 2.1 & 0.06 & -2.43 & 0.34 & -9.51 & \nodata \\ 
J11094742-7726290          &   ISO-ChaI 207   &  11:09:47.42  &  $-$77:26:29.0 & M1 & 3705 & 5.0 & 0.10 & -1.93 & 0.58 & -9.21 & $^m$ \\ 
J11095340-7634255          &   Sz32   &  11:09:53.40  &  $-$76:34:25.5 & K7 & 4060 & 4.3 & 0.48 & 0.06 & 0.78 & -7.08 & \nodata \\ 
J11095407-7629253         &   Sz33   &  11:09:54.07  &  $-$76:29:25.3 & M1 & 3705 & 1.8 & 0.11 & -2.10 & 0.56 & -9.35 & \nodata \\ 
J11104959-7717517         &   Sz37   &  11:10:49.59  &  $-$77:17:51.7 & M2 & 3560 & 2.7 & 0.15 & -0.81 & 0.41 & -7.82 & \nodata \\ 
J11123092-7644241         &   CW Cha   &  11:12:30.92  &  $-$76:44:24.1 & M0.5 & 3780 & 2.1 & 0.18 & -0.85 & 0.59 & -8.03 & \nodata \\ 
\hline 
\end{tabular} 
\tablefoot{UL = objects located well below the 30 Myr isochrone. $^\dag$ Objects with low accretion, compatible with chromospheric noise. $^m$ Stellar and accretion parameters not derived from UV excess. TD = transition disks. All stellar parameters have been derived using the \citet{Baraffe15} evolutionary models except for objects with an asterisk, for which the models of \citet{Siess00} were used.       } 
\end{center} 
\end{table*}  

\subsection{Determining the stellar and accretion parameters}\label{sect::method}

The spectral type of the target, its stellar luminosity (\lstar), extinction ($A_V$), and accretion luminosity (\lacc) are obtained by modeling the spectrum of the targets from $\lambda \sim$ 330 nm to $\lambda \sim$ 715 nm following the automatic procedure described by \citet{Manara13b}. This same procedure was used by \citetalias{Manara16}, \citet{Manara14}, and \citet{Alcala14,Alcala17}, among others. Briefly, the observed spectrum is modeled as a sum of a photospheric template and of a slab model to reproduce the excess emission due to accretion \citep[as in, e.g., ][]{Valenti93,Herczeg08}, which are reddened assuming a typical extinction law with $R_V$=3.1 \citep{Cardelli}. The best match is found by minimizing a $\chi^2_{\rm like}$ distribution over the spectral type of the template, the parameters of the slab model, and the values of $A_V$. With respect to \citetalias{Manara16}, here we include additional photospheric templates of non-accreting young stars in addition to those by \citet{Manara13a,Manara14}. They have spectral types between G5 and K6, and between M6.5 and M8 \citep{Manara16d}. The spectral type of the best-matching photospheric template is adopted for the target, and the flux ratio between the template and the target spectrum corrected for extinction using the best-fit $A_V$ is used to derive \lstar \ \citep{Manara13b}. These values are reported in Table~\ref{tab::results}. Typical uncertainties on $A_V$ are $<$ 0.5 mag, on the spectral type the uncertainties
are $\pm$ 0.5 subclass for M-type young stars and $\pm$1 subclass for earlier-type stars, and on log\lstar \ the uncertainties
are 0.2 dex. The final $T_{\rm eff}$ for the targets is obtained from the spectral type using the relation by \citet{Luhman03} for M-type stars and the relation by \citet{KH95} for earlier-type stars. Finally, \lacc \ is derived by directly integrating the flux of the best-fit slab model, with typical uncertainties on \lacc \ of $\sim$0.25 dex.  

These results from the automatic procedure are also validated against independent methods. The spectral type is compared with the type derived using different spectral indices by \citet{Riddick07}, \citet{Jeffries07}, and \citet{Herczeg14}. Of the indices by \citet{Riddick07}, we selected the same as have been used by \citet{Manara13a}, while we used the indices $G$-band, R5150, TiO-7140, TiO-7700, and TiO-8465 from \citet{Herczeg14} depending on the spectral type of the target. The spectral types derived with these indices are reported in Table~\ref{tab::spt_ind}, where we list the mean value when multiple indices from the same author are adopted. In general, these values agree within $\pm$1 subclass with our estimates, which is within their validity ranges. The estimated \lstar \ agree within $\sim$0.2 dex with those obtained using the bolometric correction by \citet{Herczeg14}. Finally, \lacc \ from the method just described, which is directly obtained from the continuum UV-excess emission, is always in agreement with the one derived by converting the luminosity of different emission lines into \lacc \ using the relation by \citet{Alcala14} within the uncertainties. 

This method is only applicable when the spectrum has signal down to $\lambda\sim$ 330 nm, and it cannot be applied to 19 late-M brown dwarfs with non-negligible extinction. 
For these young stars, marked in Table~\ref{tab::results} with ``$m$'', the stellar parameters are derived using the spectral indices by \citet{Riddick07}, \citet{Jeffries07}, and \citet{Herczeg14} and by comparing the observed spectrum with the spectra of photospheric templates reddened by increasing values of $A_V$ in steps of 0.1 mag, until a best match is found visually. The emission line fluxes are then used to derive \lacc \ using the relations between the line luminosity and \lacc \ by \citet{Alcala14}. 

In the following sections, we discuss the results for the whole Chamaeleon~I sample, including the targets studied by \citetalias{Manara16}, \citet{Manara14}, and \citet{Whelan14}. 
For these objects we also report the spectral type, $T_{\rm eff}$, $A_V$, \lstar, and \lacc \ in Table~\ref{tab::resultsold}. These values have been derived with the automatic method described in this section, or with the other method we described in the case of ISO-ChaI 217 \citep{Whelan14}.

\begin{table*}  
\begin{center}  
\footnotesize  
\caption{\label{tab::resultsold} Names, coordinates, and properties for the additional Chamaeleon~I targets studied with X-Shooter }  
\begin{tabular}{l|l|cc|ccc| cc | cc | l }    
\hline \hline  
 2MASS & Object &  RA(2000)  & DEC(2000) &  SpT & T$_{\rm eff}$ & A$_V$ & \lstar & log\lacc & \mstar & log\macc & Notes \\    
   &   &  h \, :m \, :s & $^\circ$ \, ' \, ''   &  \hbox{} & [K] & [mag] & [\lsun] & [\lsun] & [\msun] & [\msun/yr]  &  \\     
\hline  

\hline  
\multicolumn{12}{c}{Data from \citet{Manara16} }\\   
\hline  
J10555973-7724399 & T3 & 10:55:59.73 & -77:24:39.9 & K7 & 4060 & 2.6 & 0.18 & -1.25 & 0.77 & -8.61 & \nodata \\ 
... & T3 B & ..:.:. & -..:.:. & M3 & 3415 & 1.3 & 0.19 & -1.66 & 0.29 & -8.43 & \nodata \\ 
J10563044-7711393 & T4 & 10:56:30.44 & -77:11:39.3 & K7 & 4060 & 0.5 & 0.43 & -2.24 & 0.78 & -9.41 & \nodata \\ 
J10590108-7722407 & TW Cha & 10:59:01.08 & -77:22:40.7 & K7 & 4060 & 0.8 & 0.38 & -1.66 & 0.79 & -8.86 & \nodata \\ 
J10590699-7701404 & CR Cha & 10:59:06.99 & -77:01:40.4 & K0 & 5110 & 1.3 & 3.26 & -1.42 & 1.77 & -8.71 & \nodata,$^*$ \\ 
J11025504-7721508 & T12 & 11:02:55.04 & -77:21:50.8 & M4.5 & 3200 & 0.8 & 0.15 & -2.12 & 0.19 & -8.70 & \nodata \\ 
J11040909-7627193 & CT Cha A & 11:04:09.09 & -76:27:19.3 & K5 & 4350 & 2.4 & 1.50 & 0.37 & 0.98 & -6.69 & \nodata \\ 
J11044258-7741571 & ISO-ChaI 52 & 11:04:42.58 & -77:41:57.1 & M4 & 3270 & 1.2 & 0.09 & -3.79 & 0.23 & -10.59 & $^\dag$ \\ 
J11064180-7635489 & Hn 5 & 11:06:41.80 & -76:35:48.9 & M5 & 3125 & 0.0 & 0.05 & -2.56 & 0.16 & -9.28 & \nodata \\ 
J11065906-7718535 & T23 & 11:06:59.06 & -77:18:53.5 & M4.5 & 3200 & 1.7 & 0.32 & -1.65 & 0.21 & -8.11 & \nodata \\ 
J11071206-7632232 & T24 & 11:07:12.06 & -76:32:23.2 & M0 & 3850 & 1.5 & 0.40 & -1.48 & 0.58 & -8.49 & \nodata \\ 
J11071668-7735532 & Cha H$\alpha$1 & 11:07:16.68 & -77:35:53.2 & M7.5 & 2795 & 0.0 & 0.00 & -5.11 & 0.04 & -11.68 & \nodata \\ 
J11071860-7732516 & Cha H$\alpha$ 9 & 11:07:18.60 & -77:32:51.6 & M5.5 & 3060 & 4.8 & 0.03 & -4.19 & 0.12 & -10.91 & \nodata \\ 
J11075792-7738449 & Sz 22 & 11:07:57.92 & -77:38:44.9 & K5 & 4350 & 3.2 & 0.51 & -1.03 & 1.01 & -8.34 & \nodata \\ 
J11080148-7742288 & VW Cha & 11:08:01.48 & -77:42:28.8 & K7 & 4060 & 1.9 & 1.64 & -0.78 & 0.67 & -7.60 & \nodata \\ 
J11080297-7738425 & ESO H$\alpha$ 562 & 11:08:02.97 & -77:38:42.5 & M1 & 3705 & 3.4 & 0.12 & -2.01 & 0.56 & -9.24 & \nodata \\ 
J11081509-7733531 & T33 A & 11:08:15.09 & -77:33:53.1 & K0 & 5110 & 2.5 & 1.26 & -1.62 & 1.26 & -8.97 & \nodata \\ 
... & T33 B & ..:.:. & -..:.:. & K0 & 5110 & 2.7 & 0.69 & -1.32 & 1.00 & -8.69 & \nodata \\ 
J11082238-7730277 & ISO-ChaI 143 & 11:08:22.38 & -77:30:27.7 & M5.5 & 3060 & 1.3 & 0.03 & -3.38 & 0.12 & -10.07 & \nodata \\ 
J11083952-7734166 & Cha H$\alpha$6 & 11:08:39.52 & -77:34:16.6 & M6.5 & 2935 & 0.1 & 0.07 & -3.86 & 0.10 & -10.25 & \nodata \\ 
J11085464-7702129 & T38 & 11:08:54.64 & -77:02:12.9 & M0.5 & 3780 & 1.9 & 0.13 & -2.02 & 0.63 & -9.30 & \nodata \\ 
J11092379-7623207 & T40 & 11:09:23.79 & -76:23:20.7 & M0.5 & 3780 & 1.2 & 0.55 & -0.48 & 0.49 & -7.33 & \nodata \\ 
J11100010-7634578 & T44 & 11:10:00.10 & -76:34:57.8 & K0 & 5110 & 4.1 & 2.68 & 0.62 & 1.65 & -6.68 & \nodata,$^*$ \\ 
J11100469-7635452 & T45a & 11:10:04.69 & -76:35:45.2 & K7 & 4060 & 1.1 & 0.34 & -2.59 & 0.80 & -9.83 & \nodata \\ 
J11101141-7635292 & ISO-ChaI 237 & 11:10:11.41 & -76:35:29.2 & K5 & 4350 & 4.1 & 0.61 & -2.47 & 1.03 & -9.74 & $^\dag$ \\ 
J11113965-7620152 & T49 & 11:11:39.65 & -76:20:15.2 & M3.5 & 3340 & 1.0 & 0.29 & -0.81 & 0.25 & -7.41 & \nodata \\ 
J11114632-7620092 & CHX18N & 11:11:46.32 & -76:20:09.2 & K2 & 4900 & 0.8 & 1.03 & -0.74 & 1.25 & -8.09 & \nodata \\ 
J11122441-7637064 & T51 & 11:12:24.41 & -76:37:06.4 & K2 & 4900 & 0.1 & 0.64 & -0.79 & 1.04 & -8.16 & \nodata \\ 
... & T51 B & ..:.:. & -..:.:. & M2 & 3560 & 0.5 & 0.09 & -1.92 & 0.44 & -9.07 & \nodata \\ 
J11122772-7644223 & T52 & 11:12:27.72 & -76:44:22.3 & K0 & 5110 & 1.0 & 2.55 & -0.19 & 1.62 & -7.48 & \nodata,$^*$ \\ 
J11124268-7722230 & T54 A & 11:12:42.68 & -77:22:23.0 & K0 & 5110 & 1.2 & 2.51 & -2.29 & 1.62 & -9.60 & $^\dag$,TD,$^*$ \\ 
J11124861-7647066 & Hn17 & 11:12:48.61 & -76:47:06.6 & M4.5 & 3200 & 0.4 & 0.11 & -3.05 & 0.20 & -9.71 & \nodata \\ 
J11132446-7629227 & Hn18 & 11:13:24.46 & -76:29:22.7 & M4 & 3270 & 0.8 & 0.11 & -3.05 & 0.24 & -9.81 & \nodata \\ 
J11142454-7733062 & Hn21W & 11:14:24.54 & -77:33:06.2 & M4.5 & 3200 & 2.2 & 0.12 & -2.40 & 0.20 & -9.04 & \nodata \\ 
\hline  
\multicolumn{12}{c}{Data from \citet{Manara14} }\\   
\hline  
J10581677-7717170 & Sz Cha & 10:58:16.77 & -77:17:17.0 & K2 & 4900 & 1.3 & 1.17 & -0.48 & 1.31 & -7.82 & TD \\ 
J11022491-7733357 & CS Cha & 11:02:24.91 & -77:33:35.7 & K2 & 4900 & 0.8 & 1.45 & -0.97 & 1.40 & -8.29 & TD \\ 
J11071330-7743498 & CHXR22E & 11:07:13.30 & -77:43:49.8 & M4 & 3270 & 2.6 & 0.07 & -4.06 & 0.23 & -10.90 & $^\dag$,TD \\ 
J11071915-7603048 & Sz18 & 11:07:19.15 & -76:03:04.8 & M2 & 3560 & 1.3 & 0.26 & -1.85 & 0.38 & -8.70 & TD \\ 
J11083905-7716042 & Sz27 & 11:08:39.05 & -77:16:04.2 & K7 & 4060 & 2.9 & 0.33 & -1.63 & 0.80 & -8.86 & TD \\ 
J11173700-7704381 & Sz45 & 11:17:37.00 & -77:04:38.1 & M0.5 & 3780 & 0.7 & 0.42 & -1.16 & 0.51 & -8.09 & TD \\ 
\hline  
\multicolumn{12}{c}{Data from \citet{Whelan14} }\\   
\hline  
J11095215-7639128 & ISO-ChaI217 & 11:17:37.00 & -77:04:38.1 & M6.5 & 2940 & 2.5 & 0.03 & -4.20 & 0.08 & -10.70 & \nodata,$^*$ \\ 
\hline 
\end{tabular} 
\tablefoot{UL = objects located well below the 30 Myr isochrone. $^\dag$ Objects with low accretion, compatible with chromospheric noise. $^m$ Stellar and accretion parameters not derived from UV-excess. TD = transition disks. All stellar parameters have been derived using the \citet{Baraffe15} evolutionary models apart from objects with an asterisk, for which the \citet{Siess00} models were used.} 
\end{center} 
\end{table*}   

\subsection{Determining stellar mass and mass accretion rate}\label{sect::hrd}

\begin{figure}[!t]
\centering
\includegraphics[width=0.5\textwidth]{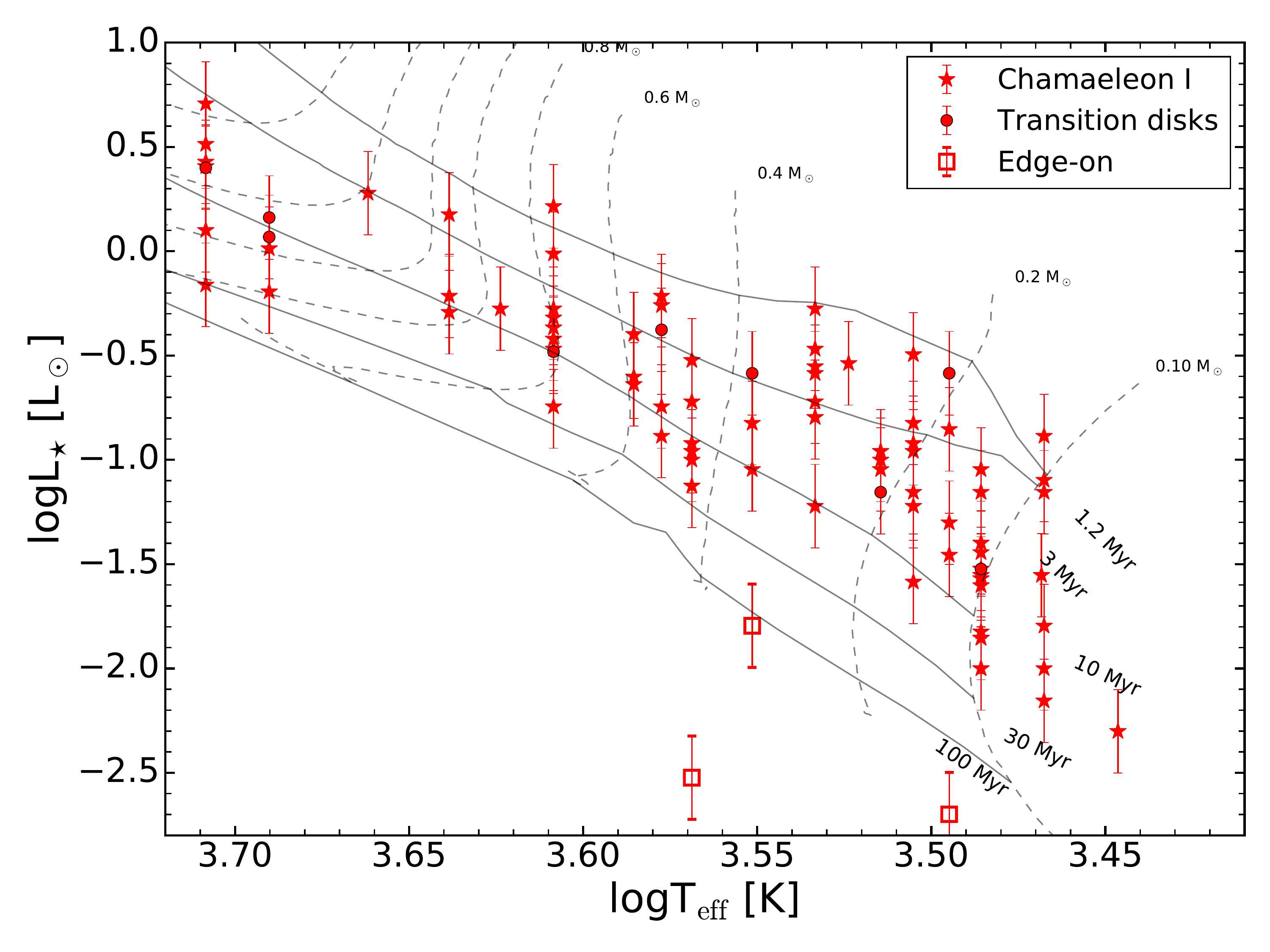}
\includegraphics[width=0.5\textwidth]{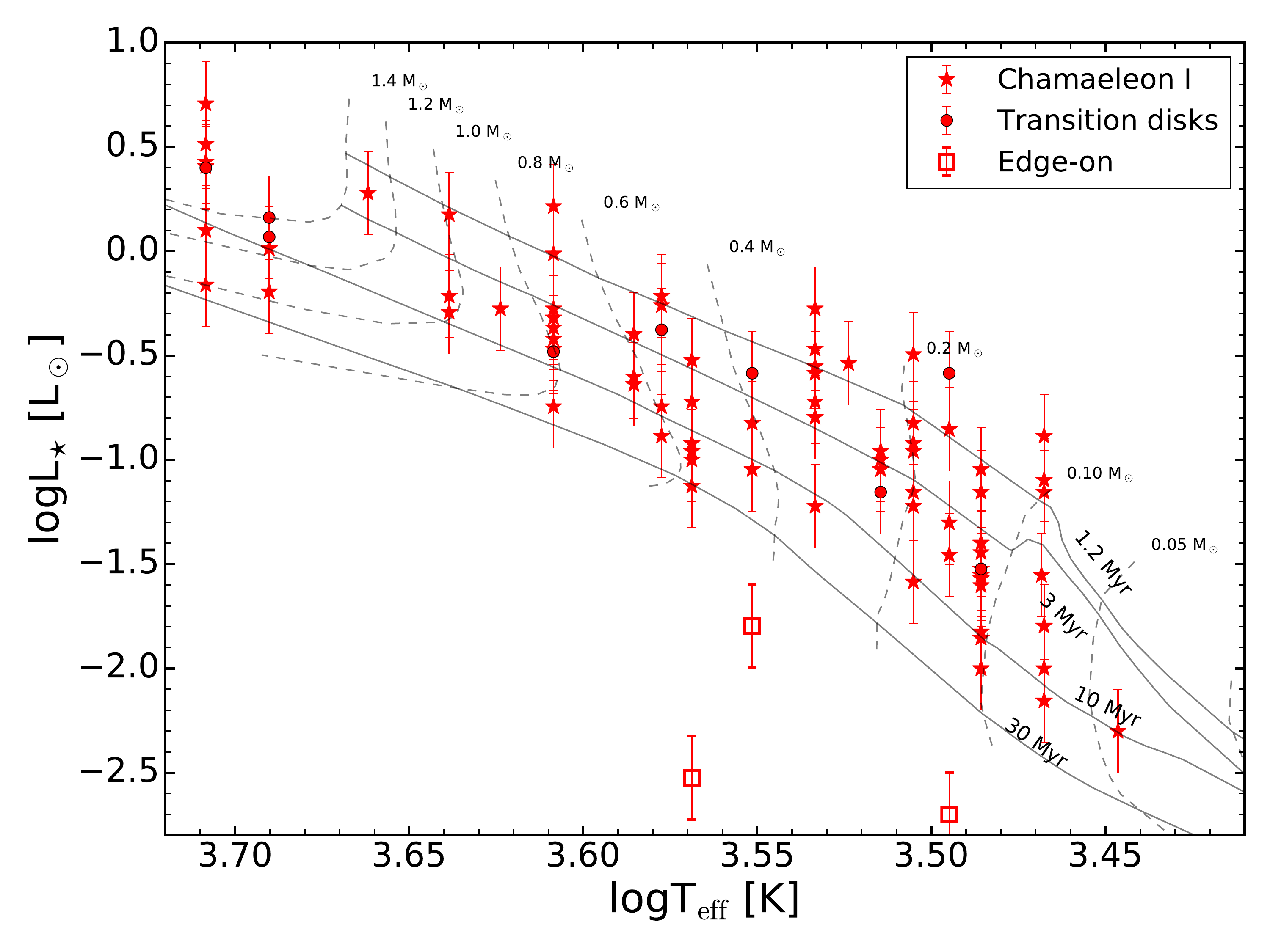}
\caption{HR diagram of the Chamaeleon~I objects discussed here overplotted on evolutionary models from \citet{Siess00} (top panel) and \citet{Baraffe15} (bottom panel). Symbols are reported in the legend. 
     \label{fig::hrd}}
\end{figure}

In order to derive \mstar \ for the observed targets, pre-main sequence evolutionary models are necessary. 
The choice of model has an effect on the estimate of \mstar \ \citep[see, e.g.,][]{Stassun14,Rizzuto16}. We show in Fig.~\ref{fig::hrd} the Hertzsprung-Russel diagram (HRD) for the whole Chamaeleon~I sample and two different sets of isochrones from the models by \citet{Siess00} and \citet{Baraffe15}. These models lead to similar values of \mstar, while the isochronal ages are different. 
We decide to use the more recent models by \citet{Baraffe15} to derive \mstar \ by interpolating the models to the position of the targets on the HRD. 
These models are an improvement from past models in their ability to reproduce the observed distribution of low-mass targets between 3200-5000 K on the HRD for different young associations.
However, these models are limited to \mstar$<$1.4 \msun\  and can therefore not be used for five targets whose position on the HRD indicate higher \mstar. For these targets we use \mstar \ derived using the evolutionary models by \citet{Siess00}, since they lead to similar \mstar \ as in \citet{Baraffe15} in the overlapping mass range, and they extend to higher \mstar. We do not consider the individual isochronal ages of the targets in this work, therefore the large differences in the age estimates using one or the other model do not affect our analysis. The values of \mstar \ derived with these models are reported in Tables~\ref{tab::results} and~\ref{tab::resultsold} and have typical uncertainties of $\sim$ 0.1 dex. 
The method used to derive \mstar \ as well as the evolutionary model adopted for \mstar$>$1.4 \msun \ differ from the choice of the companion paper that analyzes the ALMA data for Chamaeleon~I objects \citep{Pascucci16}. There we used a Bayesian approach to derive the stellar mass and its uncertainty for each object using the \citet{Baraffe15} models complemented with those from \citet{Feiden16}. 
However, the differences in \mstar \ are very small, typically $\Delta(\log$\mstar)$<$0.05 dex, with only seven objects with differences up to 0.07 dex.

The targets in the Chamaeleon~I region studied here are distributed in the HRD along the 3 Myr isochrone of the \citet{Baraffe15} models, but with a wide spread of up to $\sim$ 1 dex in log\lstar \ at all $T_{\rm eff}$. This spread is wider than the typical uncertainties on \lstar \ and would in turn imply a wide spread in ages. However, ages derived from evolutionary models are uncertain, therefore it is not straightforward to derive an age spread from the HRD \citep[e.g.,][]{SoderblomPPVI}. 

The positions on the HRD of three objects are clearly below the main locus of the Chamaeleon~I sample and below the 30 Myr isochrone of the \citet{Baraffe15} model. They are marked in the plots with open squares and in Table~\ref{tab::results} as ``UL'', which stands for underluminous. Two of these targets are known to present an edge-on disk (ESO H$\alpha$ 569, \citet{Robberto12}; 2MASS J10533978$-$7712338, \citet{Luhman07}). We also consider the third object, ISO-ChaI 79, to be seen behind an edge-on disk, although future studies are needed to confirm this finding. These objects will not be considered in the following analysis, as their stellar and accretion parameters are more uncertain because
of the high gray extinction on the line of sight caused by the edge-on disk. 
One additional object, ESO-H$\alpha$ 574, also has an edge-on disk, but is not included in our sample or discussed further.
These four objects, together with the target 2MASS J11082570-7716396, which was not observed with X-Shooter, are the same five underluminous objects in the ALMA sample as were reported by \citet{Pascucci16}. The fact that stars seen through an edge-on disk have \lstar \ corresponding to a position on the HRD below the 30 Myr isochrone is consistent with what is found in the Lupus star-forming region by \citet{Alcala14,Alcala17}. 

One target, Hn13, has \lstar \ significantly higher than the luminosity corresponding to the youngest isochrone for its temperature according to the models by \citet{Baraffe15}, and we adopt the value of \mstar \ obtained using the models by \citet{Siess00} for this target. This choice is possible since \mstar \ derived with the two models is very similar in this stellar mass range.

The values of \lacc \ measured by fitting the observed spectrum are finally combined with \mstar \ and the stellar radius (\rstar), derived from \lstar \ and $T_{\rm eff}$, to derive \macc \ with the usual relation \macc = \lacc \rstar / (0.8$\cdot$G\mstar) \citep[e.g.,][]{Hartmann98}. These values of \macc \ are reported both for the sample analyzed here and for the remaining targets in Chamaeleon~I in Tables~\ref{tab::results} and \ref{tab::resultsold}, and their typical uncertainties are $\sim$0.35 dex.

\subsection{Non-accreting targets}\label{sect::noacc}

\begin{figure}[!t]
\centering
\includegraphics[width=0.5\textwidth]{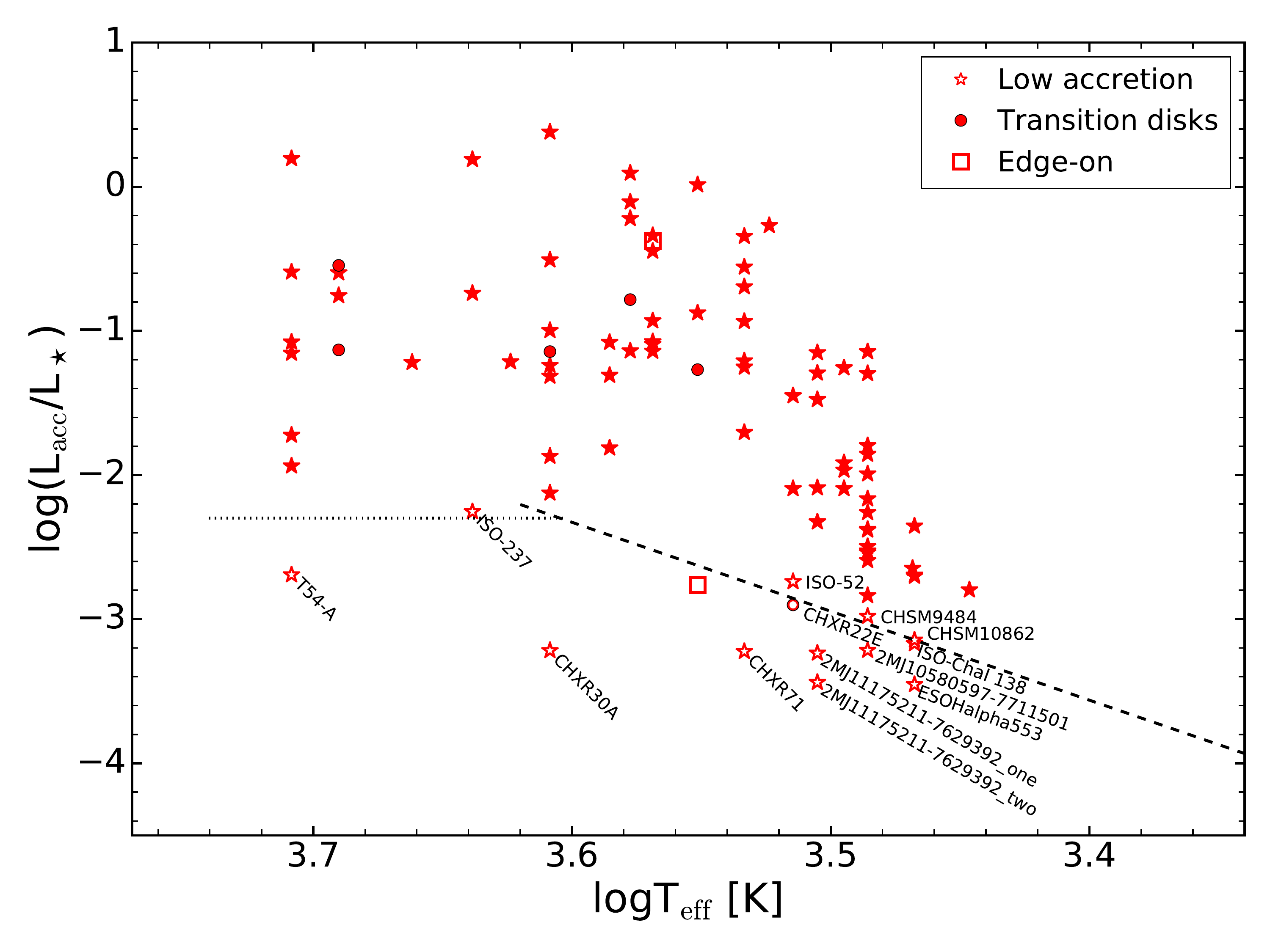}
\caption{Accretion luminosity divided by stellar luminosity as a function of the target temperature. The dashed and dotted lines represent the typical noise on the measurements of accretion luminosity that is due to chromospheric emission. While most of the targets have accretion rates much higher than this threshold, 13 objects with measured excess emission are compatible with being partially or totally dominated by chromospheric emission (open symbols). 
     \label{fig::lacc_noise}}
\end{figure}

Young stars are known to have a very active chromosphere. The emission from this chromosphere contributes in the spectra of young stars to both the continuum emission \citep[e.g.,][]{Ingleby11} and the line emission \citep[e.g.,][]{Manara13a}. The choice of using spectra of non-accreting young stars as photospheric templates for the analysis is also driven by the fact that this allows us to consider the approximate chromospheric contribution to the observed continuum emission of the target star. When objects are strongly accreting, the chromospheric emission represents a negligible contribution to the total excess emission. On the other hand, there are objects for which the excess emission is small and where most, if not all, of this emission may come from a pure chromosphere. \citet{Manara13a} have shown that the chromospheric emission measured from the luminosity of emission lines in the spectra of young stars can be converted into a typical bias, or noise, on the accretion luminosity (\laccnoise). 
The typical intensity of the chromospheric emission, when converted into \laccnoise \ and measured as the ratio with \lstar, decreases for stars with later spectral type \citep{Manara13a}, while it is constant for late-G and K-type young stars \citep{Manara16d}. 

In order to constrain the importance of chromospheric emission with respect to the total measured excess emission, we compare the values of log(\lacc/\lstar) measured for our targets with the expected chrosmopheric emission at the same $T_{\rm eff}$, log(\laccnoise/\lstar) in Fig.~\ref{fig::lacc_noise}. 
The values of log(\laccnoise/\lstar) expected at each $T_{\rm eff}$ are shown with a dashed line for M-type stars \citep{Manara13a} and a dotted line for earlier-type stars \citep{Manara16d}. The vast majority of the targets have \lacc/\lstar \ much higher than the typical \laccnoise/\lstar. However, 13 targets have \lacc \ comparable or below the expected emission of an active chromosphere. This number includes the two components of the binary 2MASS J11175211-7629392, which are then excluded in the following analysis. It is possible that the excess emission, or the emission line luminosity, measured for these targets is mainly due to chromospheric emission. However, it is also possible that the chromosphere of these targets is less active than typical young stars, and thus this emission really originates from the accretion of matter onto the central star. Since it is not possible to distinguish between these possibilities, we refer to these targets as ``dubious accretors'' or low-accretors in the text, and we mark them in Tables~\ref{tab::results} and \ref{tab::resultsold}, and with open symbols in the following plots. The measured values of \lacc \ for the dubious accretors are not upper limits on their accretion rate, but a measurement that is most likely strongly contaminated by another process. We also show in Appendix~\ref{sect::lowacc_stat} that considering their measured \lacc \ as detection or upper limit does not affect our results. We stress that because of our selection criteria, all these objects show evidence of infrared excess due to the presence of warm dust around them, and that millimeter continuum emission from a disk has been detected for five of them\footnote{These non-accreting objects with emission from a disk detected at millimeter wavelengths are ESO H$\alpha$553, 2MASS J10580597-7711501, CHSM 10862, ISO-ChaI 52, and ISO-ChaI 237.} by \citet{Pascucci16}. Interestingly, objects with an ALMA counterpart are mostly located on the \lacc/\lstar \ vs $T_{\rm eff}$ plane very close to the threshold for being considered accretors, while most of the objects not detected with ALMA are found even at much lower values of \lacc/\lstar \ than this threshold. We note that objects of this type, that is, those showing infrared excess and/or millimeter continuum emission from a disk while not showing a signature of accretion, are well known in the literature \citep[e.g.,][]{Mohanty05,McCabe06,Fedele10,Wahhaj10,Furlan11,Hernandez14}. It is as yet unclear, however, whether they represent a typical evolutionary stage of disk evolution or are peculiar. However, this discussion is beyond the scope of this paper.

\citetalias{Manara16} used a different criterion to asses the status of accretor for the targets, namely the width at 10\% of the peak and the equivalent width of the H$\alpha$ line, which is a criterion usually adopted in the literature \citep[e.g.,][]{WB03,Muzerolle03,Mohanty05}. This leads to some discrepancies with the results just discussed. In particular, only T54-A and ISO-ChaI 52 are considered dubious accretors here and also by \citetalias{Manara16}. Three other objects that seem to be non-accreting according to the width of their H$\alpha$ line, that is, T45a, T4, and T33-A are located very close to the \laccnoise \ value for their$T_{\rm eff}$. Finally, Hn17 and Hn18, whose H$\alpha$ line width indicates
that their are not accreting, are instead found to be well above the typical \laccnoise \ for their $T_{\rm eff}$ , and they are thus accreting. The method of comparing the measured \lacc \ with \laccnoise \ is less affected by uncertainties on the real peak value of the H$\alpha$ line, which is needed to derive the width at 10\% of the peak, and by the effects of a strong photospheric absorption line, which would modify the measured equivalent width of the emission line. Moreover, the width of the H$\alpha$ emission line is subject to the rotation of the star itself, and non-accreting objects could present very wide H$\alpha$ emission lines if they rotate very fast. This is the case, for example, of Sz121 \citep{Manara13a}. In the following we consider as dubious accretors only those classified comparing \lacc \ with the typical \laccnoise.

\begin{table*}  
\begin{center}  
\footnotesize  
\caption{\label{tab::spt_ind} Spectral type from spectral indices }  
\begin{tabular}{l|l|c|ccc }    
\hline \hline  
 2MASS & Object &  SpT & SpT HH14 & SpT TiO & SpT Rid  \\    
\hline  

J10533978-7712338 &  \nodata  & M2 & M2.7 & M2.3 & M3.4 \\ 
J10561638-7630530  &  ESO H$\alpha$ 553  & M6.5 & M6.2 & M5.9 & M6.1 \\ 
J10574219-7659356  &  T5         & M3 & M3.7 & M3.4 & M3.8 \\ 
J10580597-7711501        &  \nodata  & M5.5 & M5.7 & M5.5 & M5.7 \\ 
J11004022-7619280        &  T10   & M4 & M3.8 & M3.5 & M4.0 \\ 
J11023265-7729129        &  CHXR71  & M3 & M3.4 & M2.7 & M3.8 \\ 
J11040425-7639328         &  CHSM1715  & M4.5 & M4.2 & M4.2 & M4.6 \\ 
J11045701-7715569         &  T16   & M3 & M1.7 & M1.3 & M3.9 \\ 
J11062554-7633418        &  ESO H$\alpha$ 559  & M5.5 & M5.3 & M5.3 & M5.7 \\ 
J11063276-7625210         &  CHSM 7869  & M6.5 & M7.1 & M6.3 & M6.8 \\ 
J11063945-7736052  &  ISO-ChaI 79  & M5 & M5.2 & M5.2 & M5.7 \\ 
J11064510-7727023         &  CHXR20  & K6 & K9.1 & K5.4 & M4.6 \\ 
J11065939-7530559         &  \nodata  & M5.5 & M5.7 & M5.4 & M5.7 \\ 
J11071181-7625501        &  CHSM 9484  & M5.5 & M5.4 & M5.5 & M5.6 \\ 
J11072825-7652118  &  T27   & M3 & M3.7 & M3.2 & M3.7 \\ 
J11074245-7733593         &  Cha-H$\alpha$-2  & M5.5 & M5.3 & M5.2 & M5.7 \\ 
J11074366-7739411        &  T28   & M1 & \nodata & K7.3 & M3.7 \\ 
J11074656-7615174         &  CHSM 10862  & M6.5 & M6.5 & M6.4 & M6.2 \\ 
J11075809-7742413         &  T30  & M3 & M1.7 & M1.3 & M4.0 \\ 
J11080002-7717304         &  CHXR30A  & K7 & \nodata & K5.1 & M2.9 \\ 
J11081850-7730408  &  ISO-ChaI 138  & M6.5 & M7.5 & M7.1 & M7.0 \\ 
J11082650-7715550         &  ISO-ChaI 147  & M5.5 & M6.4 & M6.1 & M6.3 \\ 
J11085090-7625135  &  T37  & M5.5 & M5.5 & M5.4 & M5.5 \\ 
J11085367-7521359         &  \nodata  & M1 & M0.4 & K8.9 & M3.2 \\ 
J11085497-7632410        &  ISO-ChaI 165   & M5.5 & M5.9 & M5.5 & M5.9 \\ 
J11092266-7634320        &  C 1-6   & M1 & M0.6 & K8.3 & M4.0 \\ 
J11095336-7728365         &  ISO-ChaI 220  & M5.5 & M6.0 & M5.2 & M6.2 \\ 
J11095873-7737088         &  T45   & M0.5 & \nodata & K6.9 & M3.7 \\ 
J11100369-7633291        &  Hn11 & M0 & \nodata & K7.2 & M3.9 \\ 
J11100704-7629376         &  T46  & K7 & \nodata & K7.5 & M3.2 \\ 
J11100785-7727480         &  ISO-ChaI 235  & M5.5 & M5.2 & M4.8 & M5.5 \\ 
J11103801-7732399  &  CHXR 47  & K4 & K4.8 & K4.4 & M5.8 \\ 
J11104141-7720480         &  ISO-ChaI 252  & M5.5 & M6.4 & M5.3 & M6.2 \\ 
J11105333-7634319        &  T48  & M3 & M3.0 & M0.9 & M3.7 \\ 
J11105359-7725004  &  ISO-ChaI 256  & M5 & M4.0 & M3.8 & M5.1 \\ 
J11105597-7645325        &  Hn13 & M6.5 & M6.3 & M5.8 & M6.1 \\ 
J11111083-7641574  &  ESO H$\alpha$ 569  & M1 & M0.7 & K9.0 & M3.8 \\ 
J11120351-7726009  &  ISO-ChaI 282  & M5.5 & M5.3 & M5.2 & M5.5 \\ 
J11120984-7634366  &  T50  & M5 & M5.2 & M5.0 & M5.1 \\ 
J11175211-7629392\_one   &  \nodata  & M4.5 & M4.5 & M4.8 & M4.5 \\ 
J11175211-7629392\_two   &  \nodata  & M4.5 & M4.6 & M4.8 & M4.6 \\ 
J11183572-7935548   & \nodata   & M5 & M5.1 & M4.8 & M5.0 \\ 
J11241186-7630425         &  \nodata  & M5.5 & M5.3 & M5.5 & M5.3 \\ 
J11432669-7804454        &  \nodata  & M5.5 & M4.8 & M4.6 & M5.2 \\ 
\hline  
J11072074-7738073         &  Sz19  & K0 & \nodata & K4.8 & \nodata \\ 
J11091812-7630292         &  CHXR79  & M0 & \nodata & K7.4 & M4.0 \\ 
J11094621-7634463  &  Hn 10e  & M3 & M2.8 & M2.8 & M3.9 \\ 
J11094742-7726290         &  ISO-ChaI 207  & M1 & M1.0 & M0.5 & M4.6 \\ 
J11095340-7634255         &  Sz32  & K7 & K1.4 & K4.3 & M3.9 \\ 
J11095407-7629253        &  Sz33  & M1 & M1.5 & M0.9 & M3.7 \\ 
J11104959-7717517        &  Sz37  & M2 & \nodata & K7.2 & M5.6 \\ 
J11123092-7644241        &  CW Cha  & M0.5 & \nodata & K7.9 & M3.5 \\ 
\hline 
\end{tabular} 
\tablefoot{Values of the spectral type derived in this work are reported in Col. 3. Columns 4, 5, and 6 report the mean value obtained using different sets of spectral indices \citep[][respectively]{Herczeg14,Jeffries07,Riddick07}.} 
\end{center} 
\end{table*}  

\section{Results}\label{sect::results}

This section is focused on the relationships between the accretion and stellar properties for the whole sample of Chamaeleon~I young stars. Different statistical tests are run to determine the shape of the \lacc-\lstar \ and \macc-\mstar \ relations. Additional tests are reported in Appendix~\ref{sect::add_stats}. In the following subsections, objects whose position on the HRD is well below the 30 Myr isochrone (see Sect.~\ref{sect::hrd}) are excluded from the analysis and are not shown on the plots.

\subsection{Accretion luminosity and stellar luminosity dependence}\label{sect::lacc_lstar}

\begin{figure}[!t]
\centering
\includegraphics[width=0.5\textwidth]{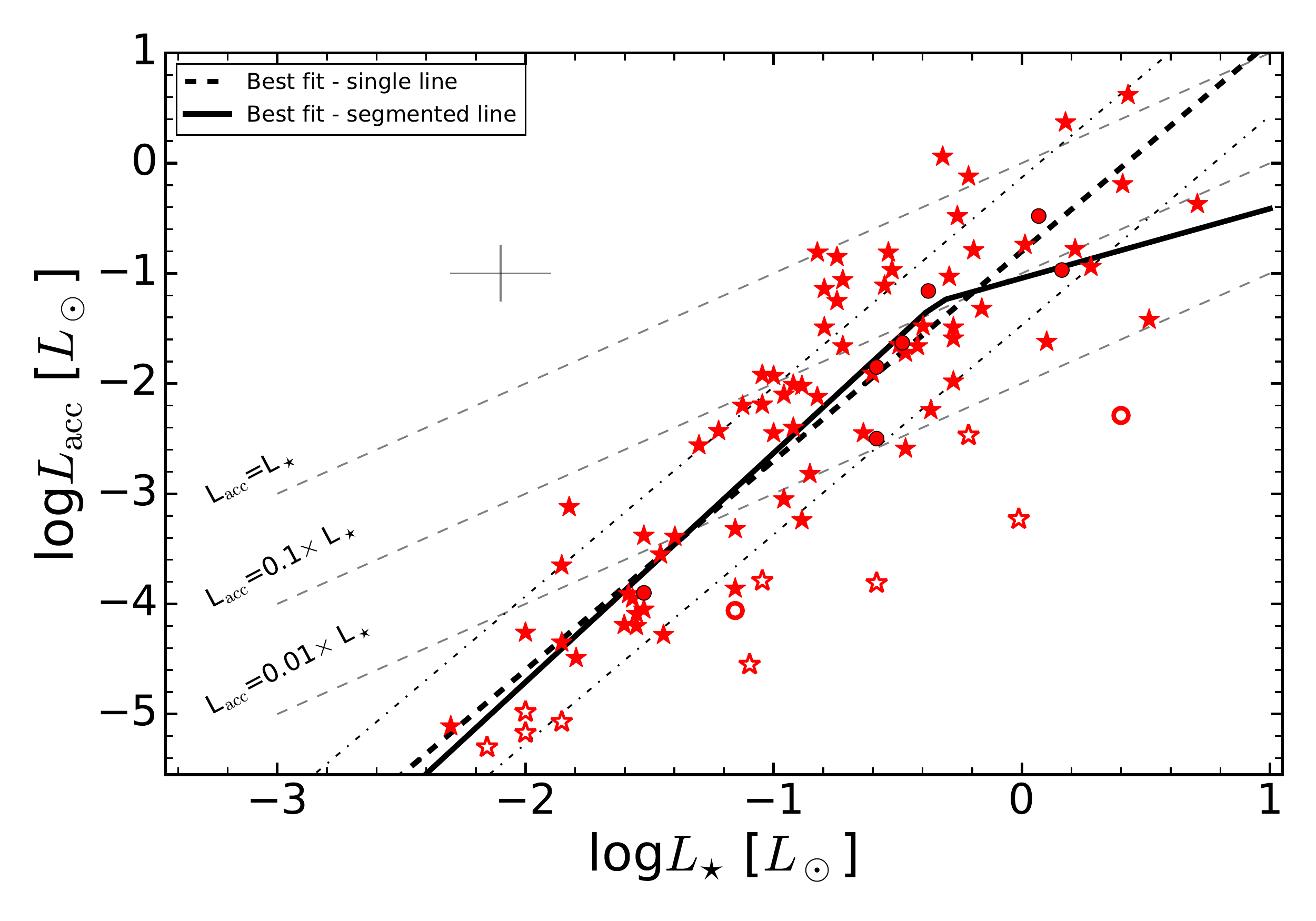}
\caption{Accretion vs stellar luminosity for the objects with a disk in the Chamaeleon~I star-forming region. Red stars are used for objects surrounded by a full disk, and circles for transition disks. Empty symbols are used for objects with low accretion, as described in Sect.~\ref{sect::noacc}. Lines of equal \lacc/\lstar \ from 1, to 0.1, to 0.01 are labeled. The best fit with a single linear fit or a segmented line is shown. For the former, the 1$\sigma$ deviation around the best fit is also reported.
     \label{fig::lacc_lstar_all}}
\end{figure}

The logarithmic dependence of \lacc \ on \lstar \ is shown in Fig.~\ref{fig::lacc_lstar_all}. In general, \lacc \ increases with \lstar. However, objects with \lstar$\gtrsim$ 0.1 \lsun \ reach higher ratios of the accretion to stellar luminosity (\lacc/\lstar), up to $\sim$1, than lower luminosity stars. The lines of constant \lacc/\lstar \ ratio are overplotted in Fig.~\ref{fig::lacc_lstar_all} for increasing values from 0.01 to 1. No targets with \lstar$\lesssim$ 0.1 \lsun \ have ratios \lacc/\lstar \ $\gtrsim$ 0.1. In contrast, $\sim$50\% of the objects with \lstar$\gtrsim$ 0.1 \lsun \ have an accretion luminosity higher than 10\% of the stellar luminosity.
This suggests a change in the accretion properties for objects with a stellar luminosity above and below \lstar$\sim$0.1 \lsun.

We first explore our data using non-parametric tools, which have the advantage that they do not rely on prior assumptions on the real model describing the data. Both a Nadaraya-Watson (spatial averaging) and a local-polynomial fit carried out using the PyQt-fit module on the \lacc \ vs \lstar \ data show two different regimes in the \lacc-\lstar \ relation: a steep increase for log(\lstar/\lsun)$\lesssim -0.6$, and a flatter relation at higher \lstar, with a slope $\sim$1. A similar result is derived when computing the median values of log(\lacc/\lsun) as a function of log(\lstar/\lsun) by dividing the sample into bins containing the same number of objects. The results from these tests, discussed in Appendix~\ref{sect::add_stats}, suggest that the dependence of \lacc \ to \lstar \ is possibly more complex than a single power-law distribution, and that two power laws, a segmented line on the logarithmic plane, are a possible representation of the observations. Therefore we consider in the following both hypotheses, and we try to quantify which model is a better fit using different statistical tests.

The fit is performed using two different Python modules, \textit{linmix} by \citet{Kelly07} and \textit{scipy.optimize.curve\_fit}. The former allows one to consider the uncertainties on the measurements on both axes, fits the data with a fully Bayesian analysis, but can only be used to fit linear relations. The latter does not include a treatment of the measurement errors, but is able to fit more complex relations between the quantities. It uses the nonlinear least-squares Levenberg-Marquardt algorithm to fit a user-defined function to the data. We show in Appendix~\ref{sect::add_stats} that the two tools lead to results that are compatible with each other since the scatter in the values dominates the outcome of the fit more than the measurement errors. Moreover, we also show in Appendix~\ref{sect::add_stats} that treating the accretion rate measured for the dubious accretors as detections is equivalent to considering these values as upper limits in the fit.  

The two models considered here have a different number of free parameters. The single power law is described by two parameters, a slope and intercept in the logarithmic plane. The double power law is a segmented line in the logarithmic plane, which is described by the following function:
\begin{equation}\label{eqfit}
\begin{split}
y = \theta_0 + \theta_1 \cdot x~~~~~~~~{\rm if}~x\le x_c \\
y = \theta_0 + \theta_1 \cdot x  + \theta_2 \cdot [x - x_c]~~~~~~~~{\rm if}~x> x_c,
\end{split}
\end{equation}
where $\theta_i ~(i = 0,1,2)$ and $x_c$ are free parameters of the fit, $y$ is $\log (L_{\rm acc}/L_\odot)$ and $x$ is $\log (L_\star/L_\odot)$. Hence, this model has four free parameters. To properly quantify whether the single or double power-law is to be preferred, we use three different information criteria to estimate the goodness of fit, namely $R^2$, the Akaike information criterion (AIC), and the Bayesian information criterion (BIC), as discussed by  \citet{FeigelsonBOOK}, for instance. While the former does not consider the number of free parameters in the model, both AIC and BIC take this aspect into account. In general, the model that better reproduces the data should maximize $R^2$, and it should minimize both AIC and BIC\footnote{Here we define AIC and BIC with a minus sign with respect to the definition by \citet{FeigelsonBOOK}, therefore the best model minimizes these quantities.}. As discussed for example by \citet{BIC}  and \citet{Riviere-Marichalar16}, a difference in the values of BIC between 2 and 6 shows that the model with lower BIC is more plausible, while a difference of 10 or more excludes the model with the higher value of BIC with a high probability. Similarly, \citet{AIC} discussed that differences in AIC values of 14 or more generally firmly exclude the model with the higher AIC value, while differences of $\sim$4-8 correspond to a $p$-value of 0.05, thus to a distinction between the models at a lower significance. 

We first fit the data with \textit{scipy.optimize.curve\_fit}. The slope of the best fit with a single line is 1.75, and the intercept is $-$0.96. The fit with a segmented line leads to a lower AIC (208 vs 212), a higher $R^2$ (0.78 vs 0.70), and the same value of BIC (217). According to the AIC criterion, the segmented line is preferred, and $R^2$ also points toward a better fit with the segmented line. However, the latter does not consider the number of degrees of freedom in the model, and the BIC criterion does not prefer any of the two models. Therefore, the statistical tests on this relation are not conclusive.
The best fit with the segmented line is shown in Fig.~\ref{fig::lacc_lstar_all}, and more discussion of the results with this method are reported in Appendix~\ref{sect::add_stats}. The best-fit parameters for the segmented line model are $\theta_0 = -0.55$, $\theta_1 =  2.08$, $\theta_2 = -1.45$, and $x_c = -0.34$, which imply

\begin{equation}
\begin{split}
\log (L_{\rm acc}/L_\odot) = -0.55 + 2.08 \cdot \log (L_\star/L_\odot) ~{\rm if}~\log (L_\star/L_\odot)\le -0.34 \\
\log (L_{\rm acc}/L_\odot) = -1.04 + 0.63 \cdot \log (L_\star/L_\odot) ~{\rm if}~\log (L_\star/L_\odot)> -0.34.
\end{split}
\end{equation}

The \textit{linmix} tool is also used to test the two models. The single-line fit is shown in Fig.~\ref{fig::lacc_lstar_all} and is
\begin{equation}
\log (L_{\rm acc}/L_\odot) = (-0.8\pm0.2) + (1.9\pm 0.1) \cdot \log (L_\star/L_\odot),
\end{equation}
with $R^2$ = 0.69, BIC = 219, and AIC = 214, and a 1$\sigma$ dispersion of 0.67$\pm$0.08 around the best fit.
We then use the value of log(\lstar/\lsun) at which the \textit{scipy.optimize.curve\_fit} module finds a change in the slope of the segmented line fit as initial parameter to divide the sample into two subsamples. These two subsamples are then fitted separately with the \textit{linmix} module (see Appendix~\ref{sect::add_stats}). 
The two slopes for high and low \lstar \ are compatible with those obtained with \textit{scipy.optimize.curve\_fit}, namely 2.5$\pm$0.2 below log(\lstar/\lsun) = $-0.34$, and 1.1$\pm$1.3 above. However, the value of the slope for higher \lstar \ is not constrained by the fit, since the uncertainty is too large. This is probably due to the large scatter of $\sim$0.9 dex in this region of the plot, which then results in a low correlation (r=0.3$\pm$0.3), in contrast to the very high correlation found below the break value (r=0.91$\pm$0.04). This fit with a segmented line leads to a slightly higher $R^2$ = 0.71, to a higher value of BIC = 222, and to a lower AIC = 212. Therefore, in this case the statistical tests are not conclusive either in the choice of best model to describe the data, since the better difference in AIC values is not significant, and the BIC statistics would prefer the linear model with a low confidence.

We can then conclude that the fit with a segmented line, with an exact value of the slope in the range of value log(\lacc/\lstar)$\ge -0.34,$ which is very uncertain, is slightly preferred by some statistical tests, but not by others, and never with results that clearly exclude the single power-law model. Therefore, both models are plausible.

\subsection{Mass accretion rate and stellar mass dependence}\label{sect::macc_mstar}

\begin{figure}[!t]
\centering
\includegraphics[width=0.5\textwidth]{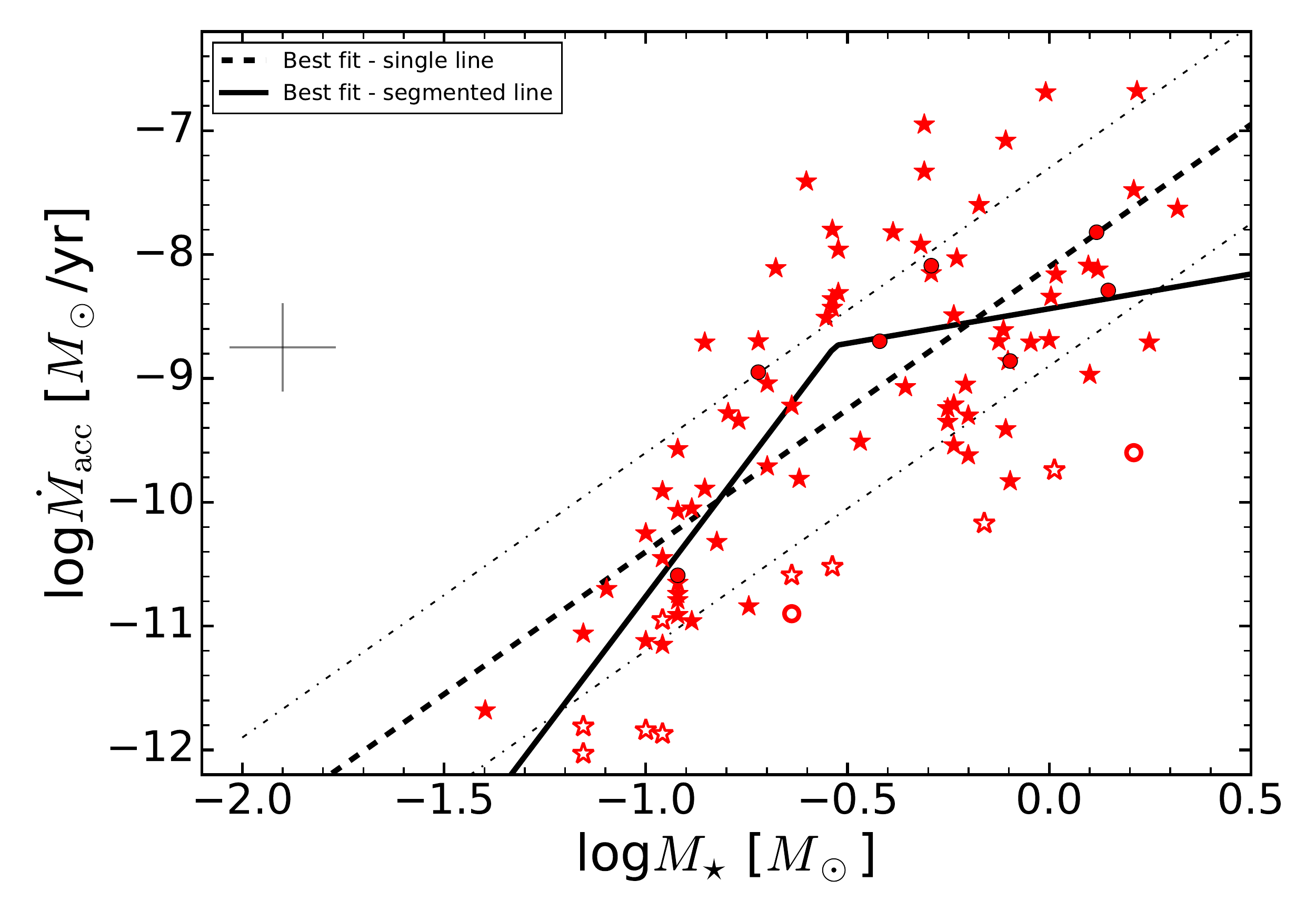}
\caption{Accretion rate vs stellar mass for the objects with a disk in the Chamaeleon~I star-forming region. Symbols are the same as Fig.~\ref{fig::lacc_lstar_all}.
     \label{fig::macc_mstar_chaI}}
\end{figure}

While the relation between \lacc \ and \lstar \ is derived independently of evolutionary models, the physical quantity that characterize the accretion process is instead the amount of material accreted onto the central star per unit time, \macc. It is thus important to study how this varies with \mstar \ to constrain models of disk evolution. We show the logarithmic relation between these two quantities in Fig.~\ref{fig::macc_mstar_chaI}. Stars with higher \mstar \ have generally higher \macc. Similarly to the findings by \citet{Manara14,Manara16}, as well as others \citep[e.g.,][]{Sicilia-Aguilar09,Fang13,Keane14,Alcala17}, transition disks are well mixed with objects with a full disk. 
Non-parametric analyses (see Appendix~\ref{sect::add_stats}) suggest that a break in the distribution of \macc \ for increasing \mstar \ is present, with a steeper slope for log(\mstar/\msun)$\lesssim-0.5$ than above this value, where \macc \ is almost constant. Here we proceed similarly as in Sect.~\ref{sect::lacc_lstar} to quantify whether a single power-law or a double power-law model better describes the data.

The fit with \textit{scipy.optimize.curve\_fit} is performed first, using both a single linear relation in the logarithmic plane and a segmented line equivalent to the line used for the \lacc-\lstar \ relation (Eq.~\ref{eqfit}).
We find that the segmented line leads to a better fit to the data with respect to a single line, with a higher value of $R^2$ (0.6 vs 0.5), and both a lower AIC (223 vs 236) and BIC (233 vs 241). The single-line fit has a slope of 2.2 and an intercept of $-$8.2. The best-fit values for the segmented line are $\theta_0 = -6.45$, $\theta_1 = 4.31$, $\theta_2 = -3.75$, and $x_c = -0.53$, and this best fit is shown in Fig.~\ref{fig::macc_mstar_chaI}. This implies two lines with equation:
\begin{equation}
\begin{split}
\log (\dot{M}_{\rm acc}/M_\odot) = -6.45 + 4.31 \cdot \log (M_\star/M_\odot) \\~{\rm if}~\log (M_\star/M_\odot)\le -0.53 \\
\log (\dot{M}_{\rm acc}/M_\odot) = -8.44 + 0.56 \cdot \log (M_\star/M_\odot) \\~{\rm if}~\log (M_\star/M_\odot)> -0.53.
\end{split}
\end{equation}

We then fit the data using the \textit{linmix} tool, starting with a single power-law fit. The best fit in this case is shown in Fig.~\ref{fig::macc_mstar_chaI} and is
\begin{equation}
\log (\dot{M}_{\rm acc}/M_\odot) = (-8.1\pm0.2) + (2.3\pm 0.3) \cdot \log (M_\star/M_\odot) 
,\end{equation}
and the 1$\sigma$ dispersion around the best fit is 0.8$\pm$0.1. This dispersion is consistent with the one reported by \citetalias{Manara16} for a subsample of Chamaeleon~I targets. Moreover, the slope of this relation is slightly steeper than those reported in the past \citep[e.g.,][]{Mohanty05,Natta06,Herczeg08,Alcala14,Manara16}, but still compatible within the uncertainty with the typical values of 1.8-2 reported in the past. This small difference in the slope is only partially due to the evolutionary model chosen here. For comparison, using the models by \citet{Baraffe98} on this same dataset would result in a best-fit slope of $2.2\pm0.2$. This fit results in values of $R^2$ = 0.5, BIC = 241, and AIC = 236. 
Finally, we test the segmented line fit also using the \textit{linmix} tool on the two subsamples, considering either the objects with log(\macc/\mstar)$\ge -0.53$, or lower than this value, where this is selected based on the results of the fit with \textit{scipy.optimize.curve\_fit}. The result is as follows. The correlation between the two quantities is strong in the sample of objects with \mstar \ lower than the value of the break (r=0.9$\pm$0.1), with a slope of 5$\pm$1. Conversely, the spread is wide at higher values of \mstar, and the fit results in a slope of 0.7$\pm$0.8, with r=0.2$\pm$0.2. This fit with a segmented line is a better representation of the data according to all the three information criteria we use, as it leads to a higher $R^2$ = 0.6, and both a lower BIC = 238 and AIC = 228 than the single-line fit. However, the differences in the values of these information criteria do not rule out any of the two models.

Even if the statistical improvement of the segmented line is small, the statistical information criteria seem to prefer this model, which leads to a steeper slope at lower \mstar\  and a lack of correlation at higher \mstar. However, none of the two description of the data can be ruled out.

There is a discrepancy in the results of the statistical tests between the best model to describe the \macc-\mstar \ and the \lacc-\lstar \ relations. The model with a double power-law is slightly preferred by all the information criteria in the case of the \macc-\mstar \ relation, while this is not necessarily the case for the \lacc-\lstar \ relation. This discrepancy could be attributed to the fact that both \mstar \ and \macc \ are derived from the measured quantities \lstar \ and \lacc \ using evolutionary models, in this case, from \citet{Baraffe15} and \citet{Siess00}. In particular, it is possible that the conversion from \lstar \ and $T_{\rm eff}$ into \mstar \ enhances the break in the \macc-\mstar \ relation. Several studies have shown the limits of the various evolutionary models in estimating \mstar \ \citep[e.g.,][]{Stassun14,Herczeg15,Rizzuto16}, and this issue may add a word of caution on the double power-law model, but also on the values of the single power-law fit. Furthermore, we checked that there is a small difference in the typical values of $A_V$ in the lower \mstar \ and higher \mstar \ subsamples, with the latter having a higher median $A_V$ by $\sim$0.9 mag. This could explain the wider spread in value of \macc \ at higher \mstar. However, the distributions of $A_V$ at low and higher \mstar \ are similar, and thus this difference in median values is probably not the origin of the possible bimodal distribution in the \macc-\mstar \ plane. Finally, we do not expect strong effects on our results because of the incompleteness in the sample. We discussed in Sect.~\ref{sect::sample} that our sample comprises nearly all the stars with disks in the Chamaeleon~I region down to \mstar$\sim$0.1 \msun. Unless we systematically missed the strongly accreting brown-dwarfs, for example, when they are still embedded in the parental cloud and are thus classified as Class~I or are
undetected at infrared wavelengths, the sample incompleteness would not affect the result. 
This said, at the present time we cannot rule out either of the two models, that is, whether a single or a double power-law are the best description of the data, and the implications of both possibilities are discussed in the next section.


\section{Discussion}\label{sect::discussion}

We have shown in Sect.~\ref{sect::results} that the sample of stars with a disk in Chamaeleon~I shows an increase in \lacc \ with \lstar, and likewise in \macc \ with \mstar. We have discussed that these relations can be modeled either with a single or double power-law. 
The latter leads to a steeper slope at lower \lstar \ or \mstar \ than for higher \lstar \ or \mstar \ objects. Our statistical tests cannot firmly rule out any of the two hypotheses, although the double power-law description is preferred in particular for the \macc-\mstar \ relation. 
In the following, we thus discuss the implications of our results in light of both descriptions of the data. We then discuss the overall distribution of data on the \macc-\mstar \ plane, and how this affects our understanding of disk evolution.

\subsection{Single power-law describing the relation of accretion to stellar parameters}

When fitting the logarithmic relations between accretion and stellar parameters with a single linear fit, we derive \lacc$\propto$\lstar$^{1.9\pm0.1}$ and \macc$\propto$\mstar$^{2.3\pm0.3}$, which is consistent with previous studies \citep[e.g.,][]{Muzerolle03,Natta06,Mohanty05,Herczeg08,Rigliaco11a,Natta14,Alcala14,Alcala17,Venuti14,Kalari15,Manara12,Manara16}. These values are in broad agreement with expectations from different theoretical models. 

The \lacc-\lstar \ relation was studied by \citet{Tilling08}. The authors showed that in the context of viscously evolving disks the slope of this relation can be related to the index of the exponential decay of \macc \ with time in the self-similar late phase of disk evolution. This index depends on the scaling of the viscosity ($\nu$) with the disk radius ($R$). A scaling in the form $\nu\propto R$ \citep{Hartmann98} implies \lacc$\propto$\lstar$^{1.7}$. A steeper slope of the \lacc-\lstar \ relation would be reproduced by a steeper dependence of the viscosity on the disk radius. However, the authors argue that different assumptions on the scaling of the viscosity with radius similarly fill the whole distribution of the observed \lacc-\lstar, thus it is not possible to derive constraints on the viscosity law from this distribution, unless there are regions of the allowed parameter space that
are empty of observed points. We return to this point below. 

The dependence of \macc \ to \mstar \ with a power law with exponent $\sim$2 has been discussed by several authors. \citet{Alexander06} and \citet{Dullemond06} argued that this dependence is a result of the imprint of initial conditions on the subsequent viscous evolution of disks. The former authors were able to reproduce the observed \macc$\propto$\mstar$^{2}$ relation under the assumption that the viscous timescale ($t_\nu$) scales with the inverse of \mstar. This implies a longer viscous timescale for lower mass objects, thus, in a viscously evolving system, a longer disk lifetime. However, this could be in contrast with our results for a fit with a double power-law, as we discuss in the following. \citet{Dullemond06} instead assumed that cores with significantly different masses rotate similarly to breakup rate ratios, and this resulted in a power-law dependence with exponent 1.8$\pm$0.2 for the \macc-\mstar \ relation, which is compatible with our result. As a consequence of viscous evolution, the authors suggest that \macc \ and the disk mass (\mdisk) must also be correlated with their assumptions on initial conditions. This correlation was recently observed in both the Lupus star-forming region \citep{Manara16b} and in the Chamaeleon~I region (Mulders et al., in prep.). 

Another theoretical argument proposed by \citet{Padoan05} to explain the \macc $\propto$ \mstar$^2$ relation is that accretion onto the star is a consequence of Bondi-Hoyle accretion of the gas in the surrounding star-forming region onto the disk-star system. However, the actual rate of accretion onto the central star may be lower than the Bondi-Hoyle rate \citep[e.g.,][]{Mohanty05}, and a dependence of \macc \ on the properties of the gas surrounding the young stars has never been observed \citep[e.g.,][]{Mohanty05, Hartmann06}. 

Photoevaporation by high-energy radiation from the central star can also explain the dependence of \macc \ on \mstar. In this context, our results are more in agreement with expectations from X-ray-driven photoevaporation \citep[][\macc$\propto$\mstar$^{1.6-1.9}$]{Ercolano14} than UV-driven photoevaporation \citep[][\macc$\propto$\mstar$^{1.35}$]{Clarke06}. 

Finally, the dependence of \macc \ on \mstar \ can also be explained if the ionization of the disk, and thus the magnetorotational instability that generates the viscosity driving the accretion in the disk, is strongly dependent on \mstar \ \citep{Mohanty05}. As discussed by \citet{Hartmann06}, in a disk model where only the surface layer is ionized, and hence accretion is driven in this layer, \macc \ is independent on \mstar. By including additional heating by irradiation from the central star in a disk model with ongoing layered accretion, the authors were able to predict a dependence \macc$\propto$\mstar. They suggest that this could explain the properties of disks around solar-mass stars. 
They then suggest that disks around very low-mass stars may be magnetically active either because of a very small initial disk radius that is due to a strong dependence of the disk radius on stellar mass ($R_{\rm d} \propto \Omega_0^2 M_\star^3$), or because the viscosity parameter $\alpha$ is high. This scenario leads to a steep dependence of \macc \ on \mstar, which is compatible with observations. In this scenario, very low-mass stars and brown dwarfs will have a small and magnetically active disk that quickly evolves viscously and settles at a lower rate of \macc. This predicts a faster evolution for low-mass than for solar mass objects. We note that recent observations of disks around brown dwarfs with ALMA have found that these disks are small in the young $\rho$-Ophiuchus region \citep[e.g.,][]{Testi16}.
As we discuss in the following, this could be in line with our results.

\subsection{Implications of a bimodal distribution in the relation
of accretion to stellar parameters}

The possibility that the \lacc-\lstar \ and \macc-\mstar \ relations are described by a double power-law with a steeper slope at \mstar $\lesssim$ 0.3 \msun \ is considered here. This possibility has been little explored in the past and could have strong implications on our theoretical understanding of disk accretion and evolution. We have presented some possible caveats to this finding in Sect.~\ref{sect::macc_mstar}, but based on our current knowledge, we cannot exclude that this bimodal distribution is indeed the correct representation of the data. We thus qualitatively discuss some possible explanation for this bimodality. 
We also note that a similar behavior is observed in the complete sample of young stars with disks in the Lupus star-forming region that has been studied with the same method by \citet{Alcala17}, and is possibly present in the survey of NGC2264 by \citet{Venuti14}. Both regions have a very similar age as Chamaeleon~I \citep[][and references therein]{Alcala14,Venuti14}. Moreover, \citet{Fang13} discuss that the same behavior of a steeper dependence at lower \mstar \ is compatible with their data in the similarly young L1641 region, although they did not perform a fit of this relation.
\citet{Vorobyov09} also suggested that the observed \macc-\mstar \ relation, when very many objects are considered, is better reproduced by a segmented line with a break around \mstar$\sim$ 0.2 \msun. While this qualitatively matches our observed values, they derived a shallower slope than we obtained, although the two values are compatible. Finally, we note that the linear slope derived for the higher \lstar \ and \mstar \ subsample of objects is consistent with the almost linear relation between \lacc \ and \lstar \ observed in samples of Herbig Ae-Be stars \citep[e.g.,][]{Mendigutia15}, which have stellar masses higher than the objects studied here, but are also in general older. 

A possible interpretation for this bimodal distribution on the \macc-\mstar \ plane is a different evolutionary timescale for disk accretion around stars with different masses. In this view, disks around stars with \mstar$\lesssim$ 0.3 \msun \ will faster
evolve to lower values of \macc . The complete $U$-band photometric surveys in $\sigma$-Orionis \citep{Rigliaco11a} and in the Orion Nebula Cluster \citep{Manara12} and the spectroscopic survey of L1641 by \citet{Fang13} came to the similar conclusion of a faster evolutionary timescale of accretion for lower-mass stars than for solar-mass stars.
This is a possible explanation for the difference with the results obtained by \citet{Manara15} with a similar procedure on a sample of very low-mass stars and brown dwarfs in the younger $\rho$-Ophiuchus star-forming region. \citet{Manara15} found that the observed \macc-\mstar \ relation for the objects in $\rho$-Ophiuchus follows the same \macc$\propto$\mstar$^{\sim 1.8}$ relation as was found for objects of higher stellar masses derived with an incomplete sample in the Lupus star-forming region by \citet{Alcala14}. Thus, \citet{Manara15} argued that the \macc-\mstar \ relation is the same from brown dwarfs up to solar-mass stars. However, these objects in $\rho$-Ophiuchus have higher \macc \ than those derived here for objects in the Chamaeleon~I region with similar \mstar, and a simple explanation is then that targets in $\rho$-Ophiuchus are younger than those in the Chamaeleon~I or in the Lupus region. However, the sample studied by \citet{Manara15} is highly incomplete and biased toward stronger accretors, therefore additional data are needed to confirm this hypothesis. 

This hypothesis slightly contradicts the well-established result that the \textit{\textup{dusty}} inner disks evolve on longer timescales for very low-mass stars than for solar- and higher-mass stars \citep[e.g.,][]{Carpenter06,Bayo12,Ribas15}. It should be considered that here we included only objects with \mstar$\lesssim$ 2 \msun, which means that they are in a different mass range than was considered by \citet{Ribas15}. In this context, the result of Ribas and collaborators of a shorter evolutionary timescale for disks around stars with \mstar$>$ 2 \msun \ cannot be compared with our findings. However, different studies in various star-forming regions have shown that several objects still show evidence of the presence of a \textit{\textup{dusty}} inner disk from infrared excess, but show no signatures of ongoing accretion \citep[e.g.,][]{Mohanty05,McCabe06,Fedele10,Wahhaj10,Furlan11,Hernandez14}. Similarly, the dubious accretors discussed here show no evidence of ongoing accretion, but they all have non-negligible infrared excess, and their disks are detected with ALMA in five cases (see Sect.~\ref{sect::noacc}). Therefore, this discrepancy can be ascribed to a different evolutionary timescale for the \textit{\textup{dusty}} inner disk with respect to the timescale of the process of accretion of material onto the central star.  

A faster evolution for lower-mass stars is opposite to predictions by \citet{Alexander06}, who postulated that the viscous timescale increases with stellar mass to explain the \macc$\propto$\mstar$^2$ relation.
As discussed, a linear relation between \macc \ and \mstar \ similar to the relation we observe for \mstar$\ge$0.3 \msun \ is predicted in the context of centrally irradiated accretion disks around solar-mass stars with an active accretion layer \citep[e.g.,][]{Mohanty05, Hartmann06}. Intriguingly, the models by \citet{Hartmann06} imply a faster evolution of disks around low-mass stars and brown dwarfs, which is in line with our results. 

An additional possibility is that the bimodal distribution is a result of two different accretion regimes at different stellar masses, as initially suggested by \citet{Vorobyov09} to explain the bimodality of \macc \ vs \mstar. The authors modeled the evolution of disks considering the self-consistently generated gravitational torques, which efficiently drive accretion onto the central star in solar-mass stars, as well as an effective turbulence implemented in the models, needed to model accretion in the very low-mass stars where the effects of disk self-gravity are weak. Their models result in slightly higher \macc \ than those observed here, but they are compatible within the spread, in particular in the solar-mass range. The measured disk masses for this same sample of targets in the Chamaeleon~I region shows that these disks are \textit{\textup{currently}} not gravitationally unstable \citep{Pascucci16}. It is nevertheless possible that these disks were gravitationally unstable at earlier ages. 

Further investigations of the effect of different magnetic field topologies in different ranges of \mstar, or other differences between solar-mass and very low-mass stars, should be pursued to explain our findings.

\begin{figure}[!t]
\centering
\includegraphics[width=0.5\textwidth]{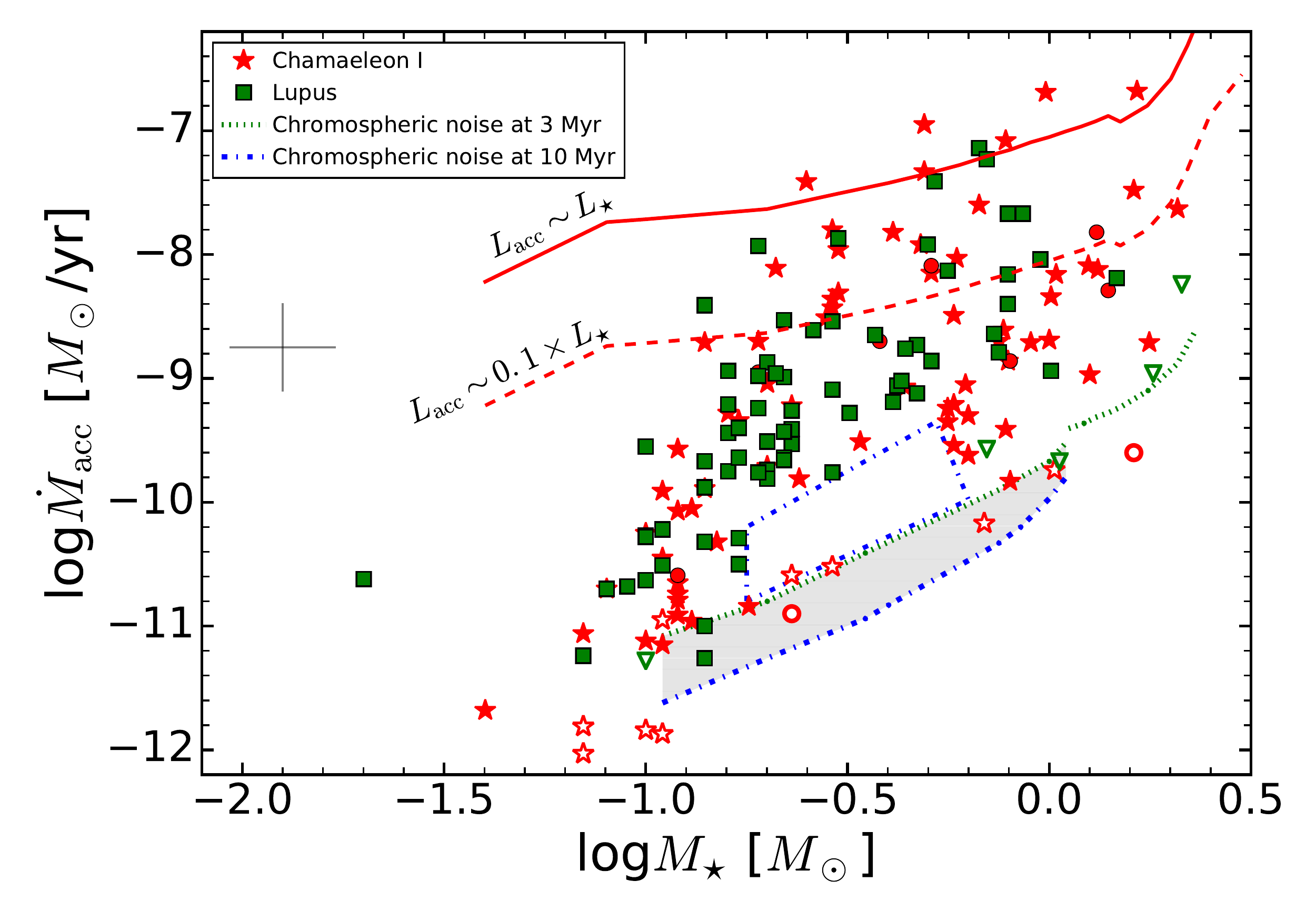}
\caption{Accretion rate vs stellar mass for the objects with disks in the Chamaeleon~I star-forming region and those in the Lupus star-forming regions studied by \citet{Alcala14,Alcala17}. Symbols are the same as in Fig.~\ref{fig::lacc_lstar_all}. The upper boundary expected by theory of \lacc=\lstar \ and the lower limit imposed by chromospheric emission are shown. The blue dot-dashed box highlights the extent of the region that is empty of data, possibly as a result of photoevaporation of the disks.
     \label{fig::macc_mstar_all}}
\end{figure}

\subsection{Spread of the \macc-\mstar \ relation}

It was argued by several authors, including \citet{Clarke06} and \citet{Tilling08}, that the observed \macc \ fill the whole observable range of values. If this is the case, then it is not possible to derive constraints on the viscosity law or the dependence
of disk on stellar mass  from the \lacc-\lstar \ or \macc-\mstar \ relations. 
It is thus relevant to test on this complete sample whether the whole range of observable \macc \ is filled. This observable range extends from the upper boundary \lacc=\lstar \ \citep{Tilling08} to the lower boundary imposed by chromospheric emission. As discussed in Sect.~\ref{sect::noacc} and by \citet{Manara13a}, the emission by active chromospheres in young stars prevents detecting accretion below the typical values of chromospheric emission. This is then an observational limitation. We therefore expect to be able to detect any accretion rate between these two boundaries.

We show in Fig.~\ref{fig::macc_mstar_all} the \macc-\mstar \ relation together with the upper and lower boundaries. We include in this plot all the objects with disks in the Chamaeleon~I region that have been studied with X-Shooter and the similarly complete sample of stars with disks in the Lupus region that was studied by \citet{Alcala14,Alcala17}. The Lupus region has a similar age to Chamaeleon~I and a similar distribution of stellar masses. The data have been analyzed in the same way, therefore the inclusion of this sample allows us to examine a more statistically robust sample. The boundary \lacc=\lstar \ is calculated using the 2 Myr isochrone of the \citet{Baraffe15} models for \mstar$\le$ 1.4 \msun, and the 3 Myr isochrone by \citet{Siess00} for higher \mstar. The points slightly above this line correspond to the same targets above the \lacc=\lstar \ line in Fig.~\ref{fig::lacc_lstar_all}. In general, this line is a good upper boundary of the distribution of points for \mstar$\gtrsim$0.2-0.3 \msun. However, fewer targets are present in the lower-mass range, where stars seem to have much lower values of \lacc/\lstar, as already observed in Fig.~\ref{fig::lacc_lstar_all}. Again, this could be an evolutionary effect, in the sense that these objects already have a much lower accretion rate than the initial one, which could have been \lacc=\lstar.

The lower boundary is taken from the typical value of the chromospheric emission (\laccnoise) derived by \citet{Manara13a} for $\sim$3 Myr old objects, completed with the similar result for higher-mass stars by \citet{Manara16d}. As expected, the dubious accretors nicely follow this lower boundary. Moreover, stars with \mstar $\gtrsim$ 0.5 \msun \ have values of \macc \ that are spread throughout the observable range. Objects with \mstar $\lesssim$ 0.2 \msun \ also\ show values of \macc \ as low as this observational limit. This is not the case for stars with 0.2 \msun $\lesssim$ \mstar $ \lesssim$ 0.5 \msun: an empty region just above the chromospheric noise limit is present in this stellar mass range, at \macc$\sim 10^{-10}$ \msun/yr. This region is highlighted in Fig.~\ref{fig::macc_mstar_all} with a blue dot-dashed box. This empty region is also present in the \macc-\mstar \ relation derived in the Lupus star-forming region \citep{Alcala17}. This region of the plot that is empty of data is of particular interest because the lack of objects with \macc \ and \mstar \ values within the highlighted  box in Fig.~\ref{fig::macc_mstar_all} could imply that disks around stars with this mass evolve very fast once their \macc \ decreases below a certain threshold. This is what photoevaporation predicts: a fast dispersal of the inner disk, thus a fast drop of \macc, once \macc$\lesssim 10^{-9}$ \msun/yr \citep[e.g.,][]{AlexanderPPVI,Gorti16}. Therefore, this empty region of the plot could be explainable by these models. Further work must be carried out to confirm this hypothesis.

%

\section{Conclusions}\label{sect::conclusions}

We have presented a study of a sample of 94 young stars with disks in the Chamaeleon~I star-forming regions. This sample represents 97\% of the stars with disks and with \mstar$\gtrsim$0.1 \msun \ in this region. 

We have analyzed the spectra of these objects obtained with the ESO VLT/X-Shooter to derive the stellar and accretion parameters self-consistently. The objects are distributed on the HRD with a wide spread between the 1 Myr and 10 Myr isochrones of the \citet{Baraffe15} models, with only three objects located well below the 30 Myr isochrone. Two of these three underluminous targets are known to have an edge-on disk, and we suggest that the third also has a similar disk-viewing geometry. By comparing the measured \lacc \ with typical chromospheric noise for young stars with similar spectral type as our targets, we found that the measured excess emission for 13 targets is compatible with being largely due to chromospheric emission. 

We analyzed the logarithmic dependence of \lacc \ on \lstar \ and of \macc \ on \mstar \ and found a positive correlation between these quantities. Moreover, stars surrounded by transition disks are well mixed with those harboring full disks in these plots. We have further investigated these logarithmic relations with different statistical tests. Non-parametric analyses suggest that the relation could be described with a model more complex than a single line, in particular, that a break might be present and lead to a steeper relation at lower \lstar \ or \mstar, and a linear relation at higher \lstar \ or \mstar. 
We then fit both relations with either a single line or a segmented line. In the former case, we obtain a slope of 1.9$\pm$0.1 and 2.3$\pm$0.3 for the \lacc-\lstar \ and the \macc-\mstar \ relations, respectively. These values are consistent with previous results and with theoretical expectations from various studies. However, the segmented line fit with a steeper relation for \lstar$\lesssim 0.45$ \lsun, or \mstar$\lesssim$ 0.3 \msun, is a statistically
slightly preferred description of the \macc-\mstar \ relation according to different statistical estimators, in line with the findings in a similar survey with VLT/X-Shooter in the Lupus star-forming region \citep{Alcala17}, although the single-line fit is not statistically excluded. We suggest that the steeper relation for lower-mass stars is due to a faster evolution of the accretion process around these objects, as was found with previous photometric studies of complete samples of young stars with disks in different star-forming regions \citep{Rigliaco11a,Manara12}. This result is in agreement with theoretical predictions of faster evolution of disks around very low-mass stars because these disks are smaller, highly ionized, and MRI active \citep{Mohanty05,Hartmann06}. Another possibility is that two different accretion regimes are present, with gravitational instability governing accretion in disks around solar-mass stars, and viscosity at lower masses, as predicted by \citet{Vorobyov09}. 

Finally, by exploring the distribution of measured accretion rates in comparison with the values that might be observed as
a result of physical and observational boundaries, we find two main features. First, a lack of very low-mass stars with high accretion rates, possibly due to the same evolutionary effect that causes the steepening of the \macc-\mstar \ relation. Second, a lack of targets with \mstar$\sim$0.3-0.4 \msun \ just above the observational limits imposed by chromospheric emission, as if the disks around these stars were rapidly dissipated once \macc \ is below a certain threshold. This is what photoevaporation theory predicts, and the distribution of data in our sample may be a sign of ongoing photoevaporation.

Future theoretical work is needed to constrain the hypotheses on this steeper relation of accretion with stellar parameters, and to use this information to better describe the evolution
of protoplanetary disks.

\begin{acknowledgements}
We thank the anonymous referee for the useful comments that helped improved the presentation of the results.
      CFM gratefully acknowledges an ESA Research Fellowship and support from the ESO Scientific Visitor Programme. This work was partly supported by the Gothenburg Centre for Advanced Studies in Science and Technology as part of the GoCAS program Origins of Habitable Planets and by the Italian Ministero dell'Istruzione, Universit\`a e Ricerca through the grant Progetti Premiali 2012-iALMA (CUP C52I13000140001). I.P. acknowledges support from an NSF Astronomy \& Astrophysics Research Grant (ID: 1515392). AN acknowledges funding from Science Foundation Ireland (Grant 13/ERC/I2907). DF acknowledges support from the Italian Ministry of Education, Universities and Research project SIR (RBSI14ZRHR). We thank E. Whelan for sharing additional information on ISO-ChaI 127. We acknowledge particularly insightful discussions with C. Clarke. This research made use of Astropy, a community-developed core Python package for Astronomy at http://www.astropy.org, and of the SIMBAD database, operated at CDS, Strasbourg, France.
\end{acknowledgements}


\appendix

\section{Additional data}

Additional information from the literature on the new targets discussed here is reported in Table~\ref{tab::lit}.

\begin{table*}
\begin{center}
\footnotesize
\caption{\label{tab::lit} Sample and data available in the literature for the targets in Chamaeleon~I included in this work }
\begin{tabular}{l|l|cc|ccc|c|c}  
\hline \hline
 2MASS & Object/other name &  RA(2000)  & DEC(2000) &  SpT & A$_J$ &  Type & Notes & References \\ 
   &   &  h \, :m \, :s & $^\circ$ \, ' \, ''   &  \hbox{} & [mag] & \hbox{} & \hbox{} & \hbox{} \\         
\hline

\multicolumn{9}{c}{Sample from Pr.Id. 095.C-0378 (PI Testi)}\\
\hline
J10533978-7712338& \nodata & 10:53:39 & $-$77:12:33 & M2.75 & & II & \nodata & 13, 14 \\
J10561638-7630530 & ESO H$\alpha$ 553 & 10:56:16 & $-$76:30:53 & M5.6 & & Disk from SED & \nodata & 1\\
J10574219-7659356 & T5 / Sz4    &10:57:42 & $-$76:59:35 & M3.25 & & II & Binary (0.16\arcsec) & 1,6,9\\
J10580597-7711501       & \nodata & 10:58:05 & $-$77:11:50 & M5.25 & & II &\nodata  & 1,2\\
J11004022-7619280       & T10 / Sz8 & 11:00:40& $-$76:19:28& M3.75 & & Disk from SED & \nodata & 1 \\
J11023265-7729129       & CHXR71 & 11:02:32 & $-$77:29:12 & M3 & & II & Binary (0.56\arcsec) & 1,2,5 \\
J11040425-7639328        & CHSM1715 & 11:04:04 & $-$76:39:32& M4.25 & & II &\nodata  & 1,2\\
J11045701-7715569        & T16 / GU Cha & 11:04:57& $-$77:15:56& M3 & & II &\nodata  & 1,2\\
J11062554-7633418       & ESO H$\alpha$ 559 & 11:06:25& $-$76:33:41 & M5.25 & & II & \nodata & 1,2 \\
J11063276-7625210        & CHSM 7869 & 11:06:32 & $-$76:25:21 & M6 & & II &\nodata  & 1,2\\
J11063945-7736052 & ISO-ChaI 79 & 11:06:39 & $-$77:36:05 & M5.25 & & II &\nodata  & 1,2\\
J11064510-7727023        & CHXR20 & 11:06:45 & $-$77:27:02 & K6 & & II & Binary (28.5\arcsec) & 1,2,7 \\
J11065939-7530559        & \nodata & 11:06:59 & $-$75:30:55 & M5.25 & & II & \nodata & 1,2 \\
J11071181-7625501       & CHSM 9484 & 11:07:11&  $-$76:25:50 & M5.25 &  & Disk from SED & \nodata & 1\\
J11072825-7652118 & T27 / VV Cha &11:07:28& $-$76:52:11 & M3 & & II & Binary (0.78\arcsec) & 1,2,5 \\
J11074245-7733593        & Cha-H$\alpha$-2 & 11:07:42 & $-$77:33:59 & M5.25 & & II & Binary (0.17\arcsec) & 1,2,12\\
J11074366-7739411       & T28 / FI Cha & 11:07:43 & $-$77:39:41 & M0 & & II & Binary (28.8\arcsec) & 1,2,7 \\
J11074656-7615174        & CHSM 10862 & 11:07:46 &  $-$76:15:17 & M5.75 && II &\nodata  & 1,2 \\
J11075809-7742413        & T30 & 11:07:58 & $-$77:42:41 & M2.5 && II &\nodata  & 1,2 \\
J11080002-7717304        & CHXR30A &11:08:00 & $-$77:17:30 & K8 & & II & Binary (0.5\arcsec) & 1,2,12\\
J11081850-7730408 & ISO-ChaI 138 &11:08:18&  $-$77:30:40 & M6.5 & & II & Binary (18.2\arcsec)  & 1,2,7\\
J11082650-7715550        & ISO-ChaI 147 & 11:08:26&  $-$77:15:55 & M5.75 && II &\nodata  & 1,2 \\
J11085090-7625135 & T37 / Sz28& 11:08:50 & $-$76:25:13 & M5.25 && II &\nodata  & 1,2 \\
J11085367-7521359        & \nodata & 11:08:53 & $-$75:21:35 & M1.5 & & Disk from SED & \nodata & 13 \\
J11085497-7632410       & ISO-ChaI 165 / HS Cha & 11:08:54 & $-$76:32:41 & M5.5 & & II & \nodata & 1,2\\
J11092266-7634320       & C 1-6 / HV Cha & 11:09:22 & $-$76:34:32 &  M1.25 &  &  II & \nodata & 1,2 \\
J11095336-7728365        & ISO-ChaI 220 & 11:09:53 & $-$77:28:36 & M5.75 && II &\nodata  & 1,2 \\
J11095873-7737088        & T45 / WX Cha & 11:09:58 & $-$77:37:08 & M1.25 & & II & Binary (0.74\arcsec) & 1,2,4,5 \\
J11100369-7633291       & Hn11& 11:10:03 & $-$76:33:29&  K8 &  &  II & \nodata & 1,2 \\
J11100704-7629376        & T46 / WY Cha & 11:10:07& $-$76:29:37& M0 & & II & Binary (0.12\arcsec) & 1,6,9\\
J11100785-7727480        & ISO-ChaI 235 & 11:10:07 & $-$77:27:48  & M5.5 && II &\nodata  & 1,2 \\
J11103801-7732399 & CHXR 47 &   11:10:38 & $-$77:32:39 & K3 & &II & Binary (0.17\arcsec) & 1,2,5\\
J11104141-7720480        & ISO-ChaI 252 & 11:10:41& $-$77:20:48&  M6 &  &  II &\nodata  & 1,2 \\
J11105333-7634319       & T48 / WZ Cha & 11:10:53 & $-$76:34:31&  M3.75 &  &  II &\nodata  & 1,2 \\
J11105359-7725004 & ISO-ChaI 256 &      11:10:53 & $-$77:25:00 &  M4.5 &  &  II &\nodata  & 1,2 \\
J11105597-7645325       & Hn13& 11:10:55& $-$76:45:32&  M5.75 &  &  II & Binary (0.13\arcsec) & 1,2,12 \\
J11111083-7641574 & ESO H$\alpha$ 569 & 11:11:10 & $-$76:41:57 &  M2.5 &  &  II &\nodata  & 1,2 \\
J11120351-7726009 & ISO-ChaI 282 &      11:12:03 & $-$77:26:00 &  M4.75 &  &  II &\nodata  & 1,2 \\
J11120984-7634366 & T50 / IN Cha &      11:12:09 & $-$76:34:36 &  M5 &  &  II &\nodata  & 1,2 \\
J11175211-7629392       & \nodata &11:17:52 & $-$76:29:39 & M4.5 & & Disk from SED & New binary & 1 \\
J11183572-7935548  &\nodata  & 11:18:36 & $-$79:35:55 &  M4.75 &  & II/TD & $\eta$ Cha member? &1,6 \\
J11241186-7630425        & \nodata & 11:24:11& $-$76:30:42 &  M5 &  &  II &\nodata  & 1,2 \\
J11432669-7804454       & \nodata & 11:43:26 &$-$78:04:45       & M5 & & II &$\eta$ Cha member? &1,6 \\
\hline
\multicolumn{9}{c}{Sample from Pr.Id. 090.C-0253 (PI Antoniucci)}\\
\hline

J11072074-7738073        & T26 / Sz19 & 11:07:20 & $-$77:38:07 & G2    &  &  II & Binary (4.6\arcsec) & 1,2,5,15,16,17 \\
J11091812-7630292        & CHXR79 & 11:09:18 & $-$76:30:29 & M1.25   &  &  II & Binary (0.88\arcsec)  & 1,2,5,15,16 \\
J11094621-7634463 & Hn 10E & 11:09:46 & $-$76:34:46 & M3.25     &  &  II &\nodata  & 1,2,15,16 \\
J11094742-7726290        & ISO-ChaI 207 / B43 & 11:09:47 & $-$77:26:29 & M3.25 &  &  II &\nodata  & 1,2,15,16 \\
J11095340-7634255        & T42 / Sz32 & 11:09:53& $-$76:34:25 & K5    &  &  II & SB2  & 1,2,9,15,16 \\
J11095407-7629253       & T43 / Sz33 & 11:09:54 & $-$76:29:25 & M2    &  &  II & Binary (0.78\arcsec) & 1,2,5,15,16 \\
J11104959-7717517       & T47 / Sz37 & 11:10:49 & $-$77:17:51 & M2    &  &  II &\nodata  & 1,2,15,16 \\
J11123092-7644241       & T53 / CW Cha & 11:12:30& $-$76:44:24 &       M1      &  &  II &\nodata  & 1,2,15,16 \\

\hline

\end{tabular}
\tablefoot{ Spectral types, extinction, disk classification, accretion indication, and binarity are adopted from the following studies: 1. \citet{Luhman07}; 2. \citet{Luhman08a}; 3. \citet{Luhman04};  4. \citet{Costigan12}; 5. \citet{Daemgen13}; 6. \citet{Manoj11}; 7. \citet{Kraus07}; 8. \citet{Ghez97}; 9. \citet{Nguyen12};
10. \citet{Winston12}; 11. \citet{Schmidt13}; 12. \citet{Lafreniere08}; 13. \citet{LuhmanCha}; 14. \citet{Robberto12}  . Complete name on SIMBAD for T\# is Ass Cha T 2$-$\#. 
Flux calibration of sources observed by Antoniucci is based on photometry from 15. \citet[NOMAD catalogue][$BVRJHK$ bands]{Zacharias04}; 16. \citet[DENIS catalogue][$I$ band]{Epchtein99}; 17. \citet[UCAC4 catalogue][$BVri$ bands]{Zacharias12}
}

\end{center}
\end{table*}

\subsection{2MASS J11175211$-$7629392: a newly identified binary}\label{sect::bin}
The target 2MASS J11175211$-$7629392 was identified as a member of the Chamaeleon~I region by \citet{Luhman07} and the membership was also confirmed by \citet{Lopez-Marti13}  based on proper motion analysis. However, the latter report the proper motion to be dubious. 

In the acquisition image of the VLT and in the raw data we clearly see two components for this system with a separation of $\sim$2\arcsec, equivalent to $\sim$320 au at the distance of Chamaeleon~I. Since the target was observed at a seeing of $\sim$0.9\arcsec, the two components are resolved and we are able to extract the spectra of the two components separately. 
The fact that the system is a binary can explain the dubious proper motions obtained by \citet{Lopez-Marti13}.

The analysis of the spectra of the two components shows that both objects have the same spectral type, which is consistent with the type reported by \citet{Luhman07}, that is, M4.5. None of the objects is accreting, and there is no sign of excess emission with respect to the photosphere in the near-infrared part of the spectrum. The only sign of excess emission is from \textit{Spitzer} 12 $\mu$m photometry, but at the resolution of this telescope it is not possible to separate the emission from one or the other component. Moreover, the millimeter ALMA observation of this system did not detect any continuum emission \citep{Pascucci16}. Therefore, the presence of a disk around these stars is not confirmed, and we thus do not include these objects in the analysis of the dependence of accretion properties with stellar properties.

\clearpage

\section{Additional statistical tests}\label{sect::add_stats}

Here we present some additional tests that we carried out to study the dependence of the accretion on stellar parameters, as discussed in Sect.~\ref{sect::results}.

\subsection{Non-parametric statistics}
Non-parametric statistics allows one to explore the data without assuming an underlying model that describes the data. Therefore, we use this technique to understand what types of models we should use to fit the distribution of data in the \lacc-\lstar \ and \macc-\mstar \ planes. In the following, we describe two different types of tests we carried out.

\subsubsection{Medians}
\begin{figure}[!t]
\centering
\includegraphics[width=0.5\textwidth]{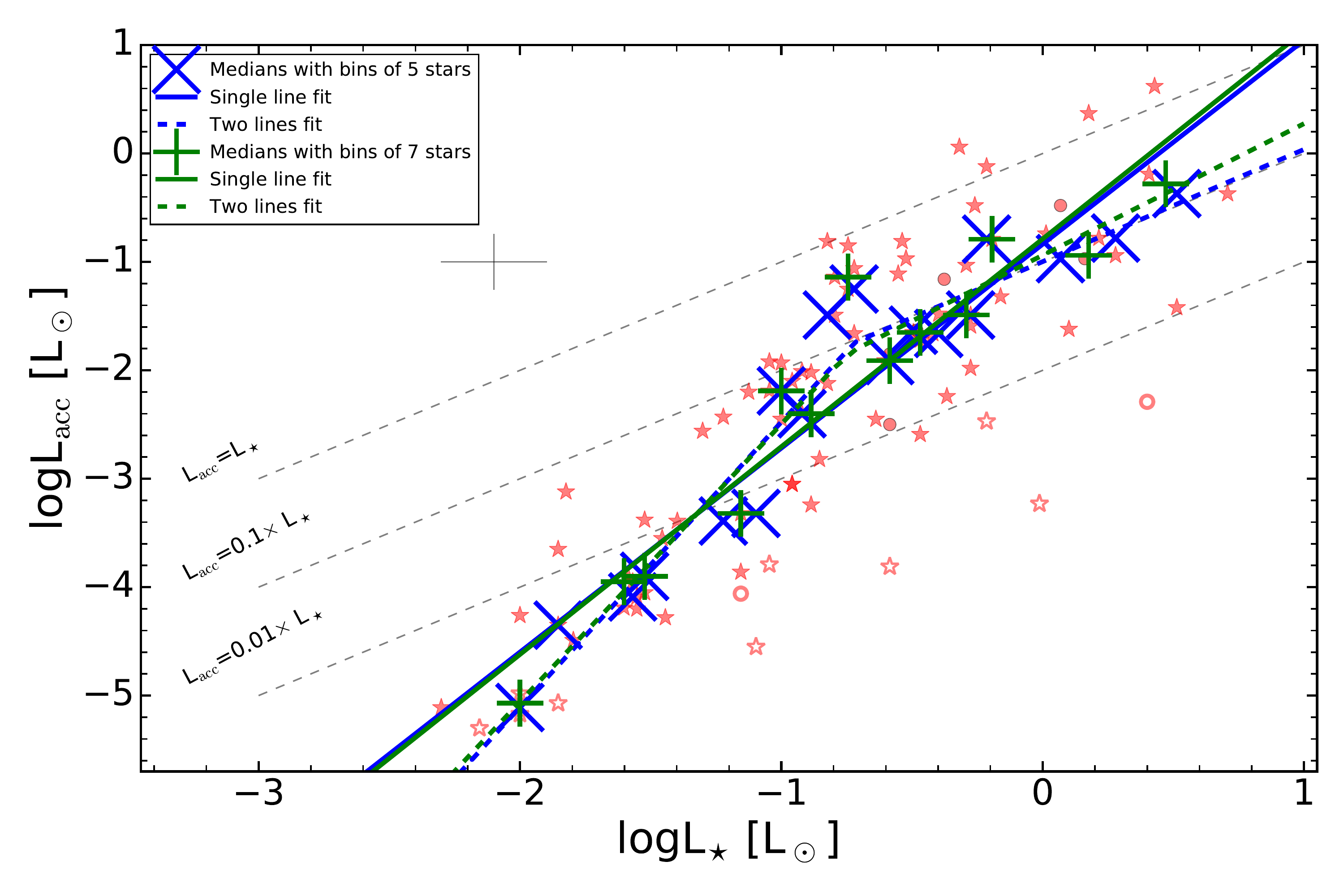}
\caption{Accretion luminosity vs stellar luminosity for the objects with a disk in the Chamaeleon~I star-forming region. Symbols are the same as in Fig.~\ref{fig::lacc_lstar_all}. The median values in bins with the same number of objects are shown with symbols as reported in the legend. Best fits of these median values are also shown.
     \label{fig::lacc_lstar_medians}}
\end{figure}

\begin{figure}[!t]
\centering
\includegraphics[width=0.5\textwidth]{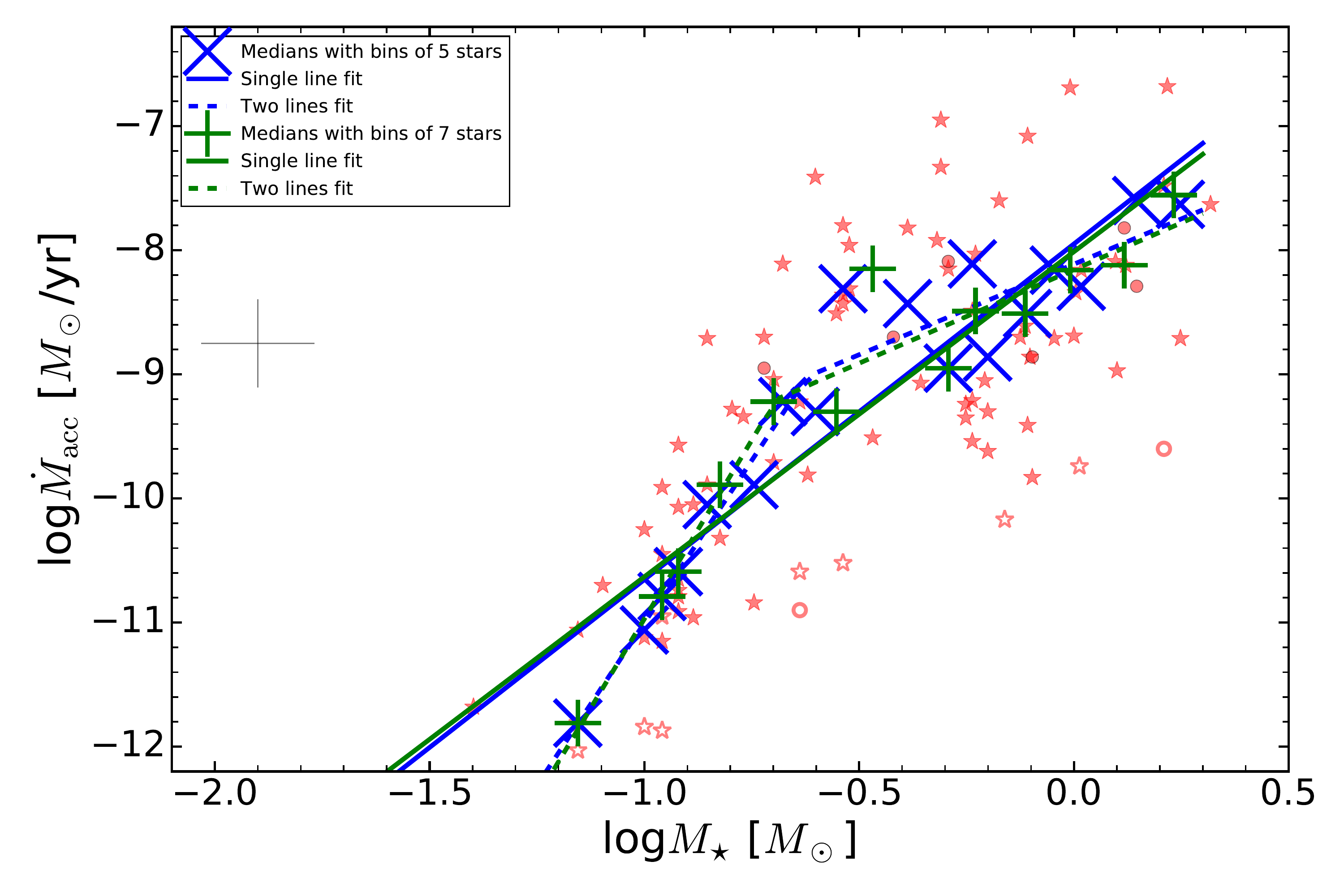}
\caption{Mass accretion rate vs stellar mass for the objects with a disk in the Chamaeleon~I star-forming region. Symbols are the same as in Fig.~\ref{fig::lacc_lstar_all}. The median values in different bins with the same number of objects are shown with symbols as reported in the legend. Best fits of these median values are also shown.
     \label{fig::macc_mstar_medians}}
\end{figure}

We divided the sample into bins comprising an equal number of objects. We carried out the analysis with bins comprising $\text{about
}$fve or seven objects. We then computed the median value of both log\lacc \ and log\lstar \ in one case, and log\macc \ and log\mstar \ in the other. As always, we excluded the edge-on targets from the analysis, while we included the dubious accretors in the sample. The results are shown in Fig.~\ref{fig::lacc_lstar_medians} for log\lacc-log\lstar, and Fig.~\ref{fig::macc_mstar_medians} for log\macc-log\mstar. In both cases the median values show a steeper slope at low \lstar, or \mstar, than at higher values, where the slope is almost linear. The break is located at log(\lstar/\lsun)$\sim -0.7$ and log(\mstar/\msun)$\sim -0.6$, respectively, as derived from a fit of these median values with a segmented line with the same functional shape as Eq.~\ref{eqfit}. The best fit of these median values for a single or broken power-law  is also shown in Fig.~\ref{fig::lacc_lstar_medians} for \lacc-\lstar, and in Fig.~\ref{fig::macc_mstar_medians} for \macc-\mstar, using different colors for the results using five or seen objects per bin. The fit with a broken power-law matches the median values
better.  

\subsubsection{Non-parametric fit}\label{sect::nnpar_fit}
We used two different non-parametric fit methods available with the Python package PyQt-fit, namely the Nadaraya-Watson, a spatial averaging technique, and the local-polynomial fit on the \lacc-\lstar \ and \macc-\mstar \ planes. The local-polynomial fit is carried out using a quadratic and a cubic polynomial. The results are as follows.

\begin{figure}[!t]
\centering
\includegraphics[width=0.5\textwidth]{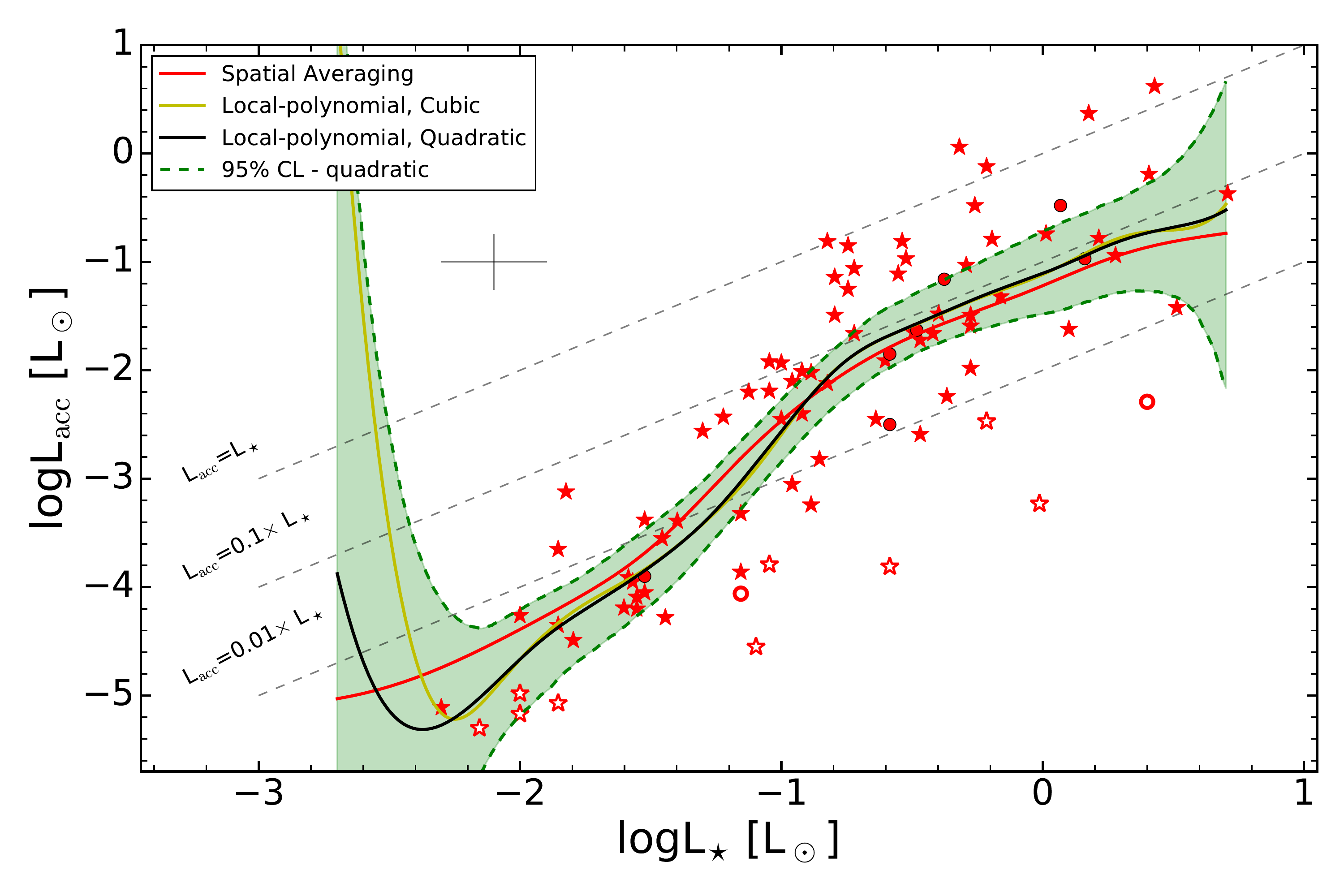}
\caption{Accretion luminosity vs stellar luminosity for the objects with a disk in the Chamaeleon~I star-forming region. Symbols are as in Fig.~\ref{fig::lacc_lstar_all}. The results from different non-parametric fitting procedures are shown with different colors. 
     \label{fig::lacc_lstar_nnpar}}
\end{figure}

For \lacc-\lstar \ there is a $\sim$1 slope at log\lstar$\ge$-0.6, and a steeper slope for lower \lstar. In the plot (Fig.~\ref{fig::lacc_lstar_nnpar}) the best fit is shown with a solid line, where different colors correspond to different methods, as reported in the legend. The green filled region represents the 95\% confidence level interval on the local-polynomial fit using a quadratic polynomial, derived with a bootstrap technique.

\begin{figure}[!t]
\centering
\includegraphics[width=0.5\textwidth]{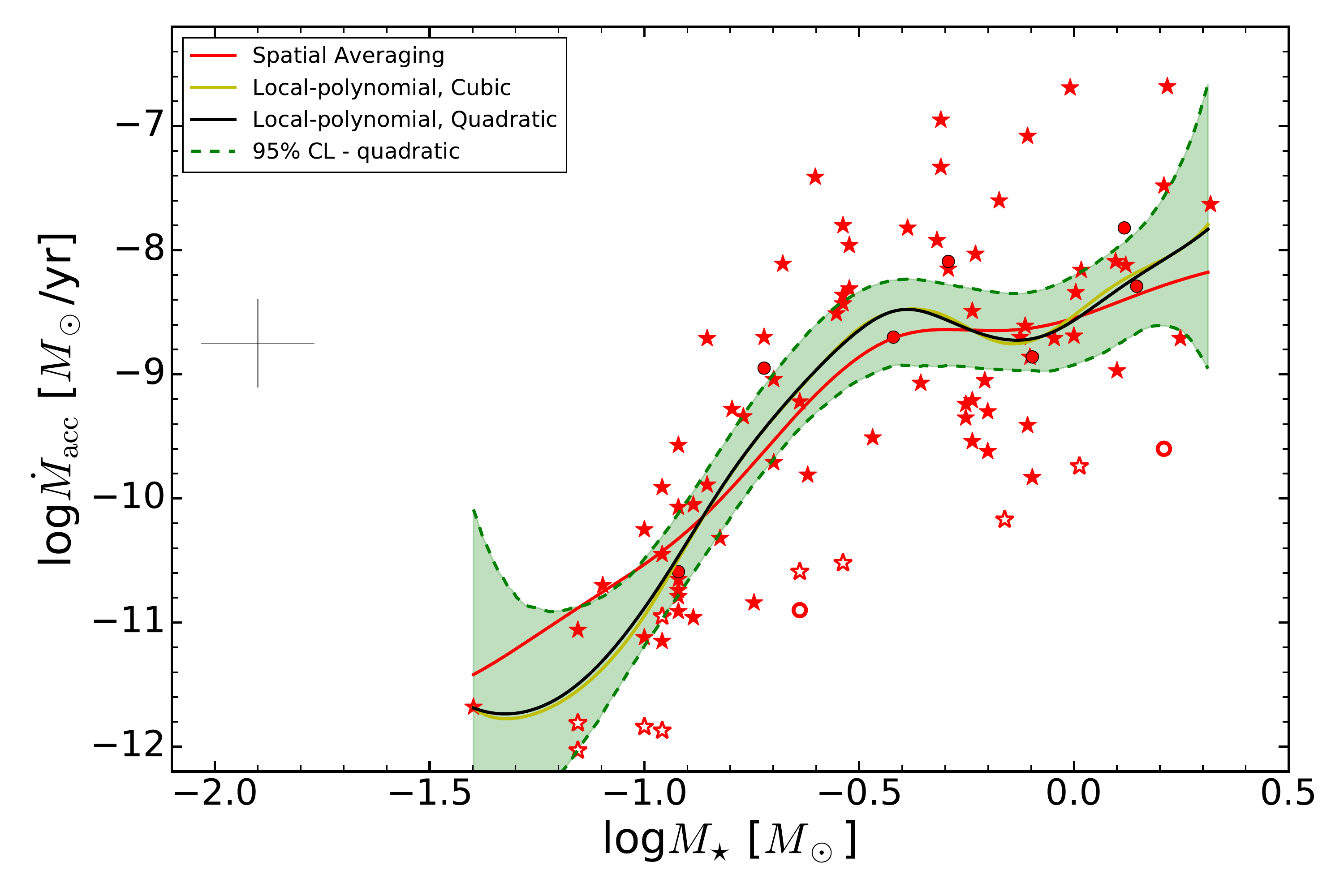}
\caption{Mass accretion rate vs stellar mass for the objects with a disk in the Chamaeleon~I star-forming region. Symbols are the same as in Fig.~\ref{fig::lacc_lstar_all}. The results from different non-parametric fitting procedures are shown with different colors. 
     \label{fig::macc_mstar_nnpar}}
\end{figure}

The fits of the \macc-\mstar \ are more uncertain (see Fig. B.4), probably due to the larger scatter of points. However, a flatter slope is present for log(\mstar/\msun)$\gtrsim$-0.5, and steeper below this value.

\subsection{Treating dubious-accretors in the fit}\label{sect::lowacc_stat}

As discussed in Sect.~\ref{sect::noacc}, 13 objects in our sample have an accretion luminosity below the typical chromospheric contribution to the excess emission with respect to the photospheric one, thus they are considered as dubious accretors. Here we show that considering their measured value of \lacc \ as an upper limit or as a detection does not affect the fit of the \lacc-\lstar \ and \macc-\mstar \ relations.

\begin{figure}[!t]
\centering
\includegraphics[width=0.5\textwidth]{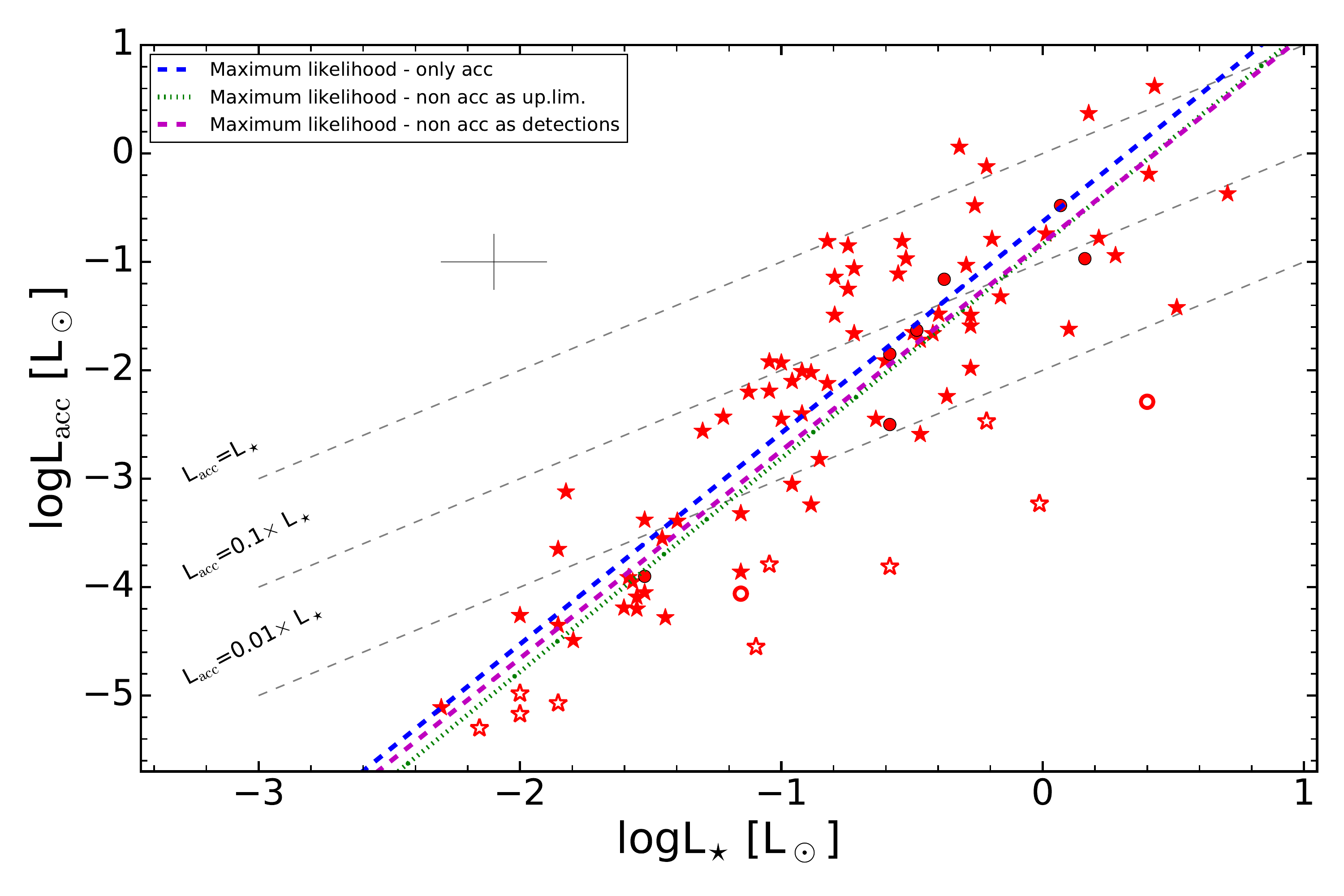}
\caption{Accretion vs stellar luminosity for the objects with a disk in the Chamaeleon~I star-forming region. Symbols are the
same as in Fig.~\ref{fig::lacc_lstar_all}. Lines of equal \lacc/\lstar \ from 1, to 0.1, to 0.01 are labeled. The results of the fit performed with \textit{linmix} are shown using different colors depending on how the dubious accretors are treated in the analysis.
     \label{fig::lacc_lstar}}
\end{figure}

First, we fit the data with a single power-law relation using the maximum-likelihood Bayesian tool \textit{linmix} by \citet{Kelly07}, which derives the linear dependence between log\lacc \ and log\lstar \ considering measurement uncertainties on both axes. We ran the fit with three different assumptions on the dubious accretors. First, we only included clearly accreting targets, second, we included the dubious accretors considering the \lacc \ measured for these objects as an upper limit, and finally, we included the dubious accretors considering their measured \lacc \ as detection. 
The three cases consistently led to the same slope  of 1.9$\pm$0.1 of the linear relation, while the intercept varies from $-0.6\pm0.1$ when only accreting objects are included, to $-0.8\pm0.2$ when non-accreting targets are included either as detection or upper limits, as shown in Fig.~\ref{fig::lacc_lstar}. Therefore, the choice of including dubious accretors in the analysis is relevant only when determining the intercept of the relation, but the choice of considering them as upper limit or detection does not change the results. 

\begin{figure}[!t]
\centering
\includegraphics[width=0.5\textwidth]{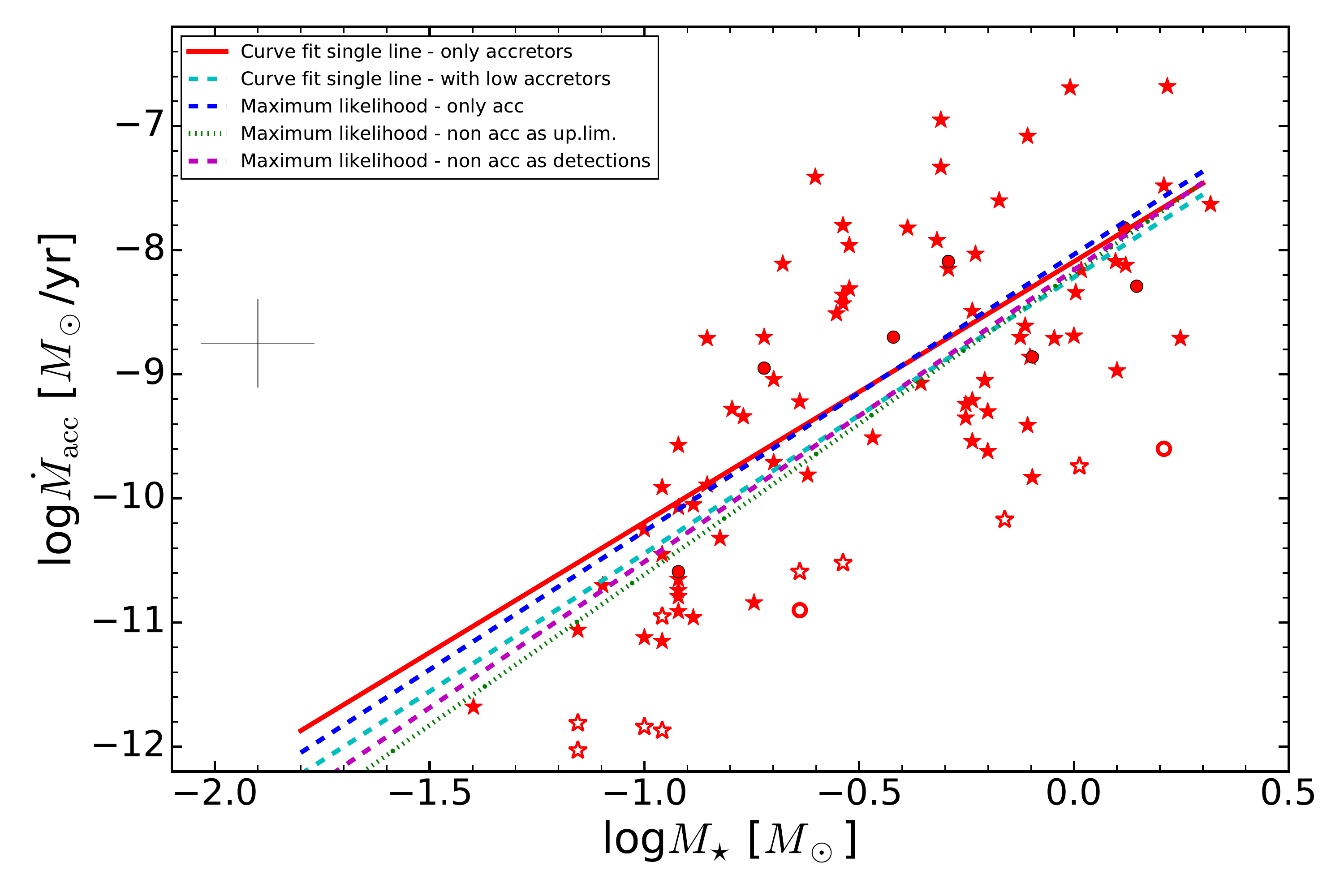}
\caption{Accretion rate vs stellar mass for the objects with a disk in the Chamaeleon~I star-forming region. Symbols are the
same as in Fig.~\ref{fig::lacc_lstar_all}. The results of the fit with a single line performed with \textit{linmix} (maximum likelihood) or with \textit{scipy.optimize.curve\_fit} (curve fit) are shown using different colors depending on how the dubious accretors are treated in the analysis.
     \label{fig::macc_mstar}}
\end{figure}

The same result is found when analyzing the \macc-\mstar \ relation. The slope of the best fit is compatible when considering dubious
accretors as detection, or upper limits, or even neglecting them. These results are shown in Fig.~\ref{fig::macc_mstar}. 

\begin{figure}[!t]
\centering
\includegraphics[width=0.5\textwidth]{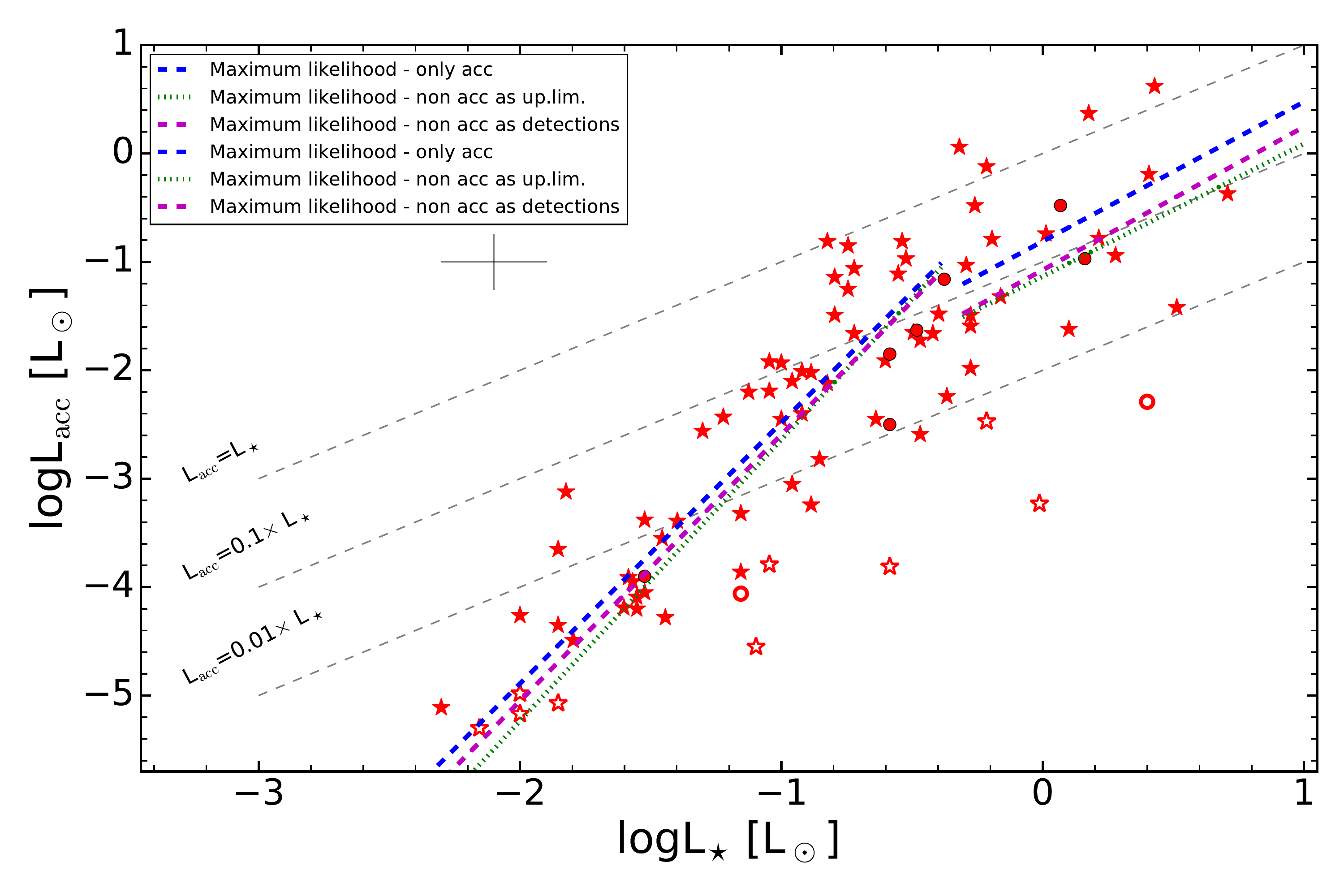}
\caption{Accretion luminosity vs stellar luminosity for the objects with a disk in the Chamaeleon~I star-forming region. Symbols are the same as in Fig.~\ref{fig::lacc_lstar_all}. The results of the fit with a segmented line performed with \textit{linmix} (maximum likelihood) are shown using different colors depending on how the dubious accretors are treated in the analysis.
     \label{fig::lacc_lstar_doublelin_test}}
\end{figure}

\begin{figure}[!t]
\centering
\includegraphics[width=0.5\textwidth]{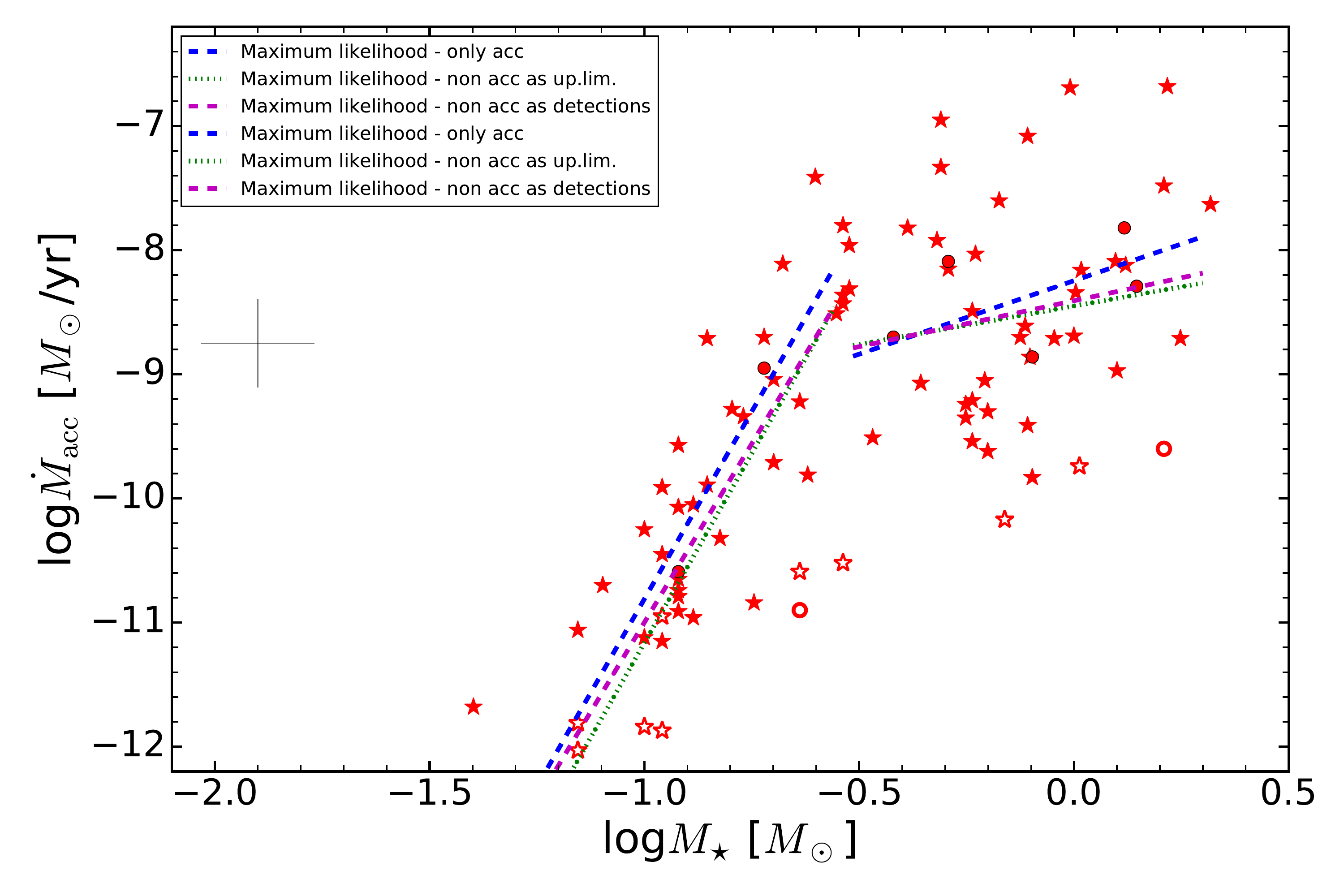}
\caption{Accretion rate vs stellar mass for the objects with a disk in the Chamaeleon~I star-forming region. Symbols are the
same as in Fig.~\ref{fig::lacc_lstar_all}. The results of the fit with a segmented line performed with \textit{linmix} (maximum likelihood) are shown using different colors depending on how the dubious accretors are treated in the analysis.
     \label{fig::macc_mstar_doublelin_test}}
\end{figure}

Similarly, the choice of considering the accretion rate for dubious accretors as detections is solid even when the sample is divided into two to fit the two power-laws (see Sect.~\ref{sect::results}), and the slope of the two parts of the segmented line are independent of how the dubious accretors are treated, as we show in Fig.~\ref{fig::lacc_lstar_doublelin_test} for the \lacc-\lstar \ relation, and in Fig.~\ref{fig::macc_mstar_doublelin_test} for the \macc-\mstar \ relation. 

\begin{figure}[!t]
\centering
\includegraphics[width=0.5\textwidth]{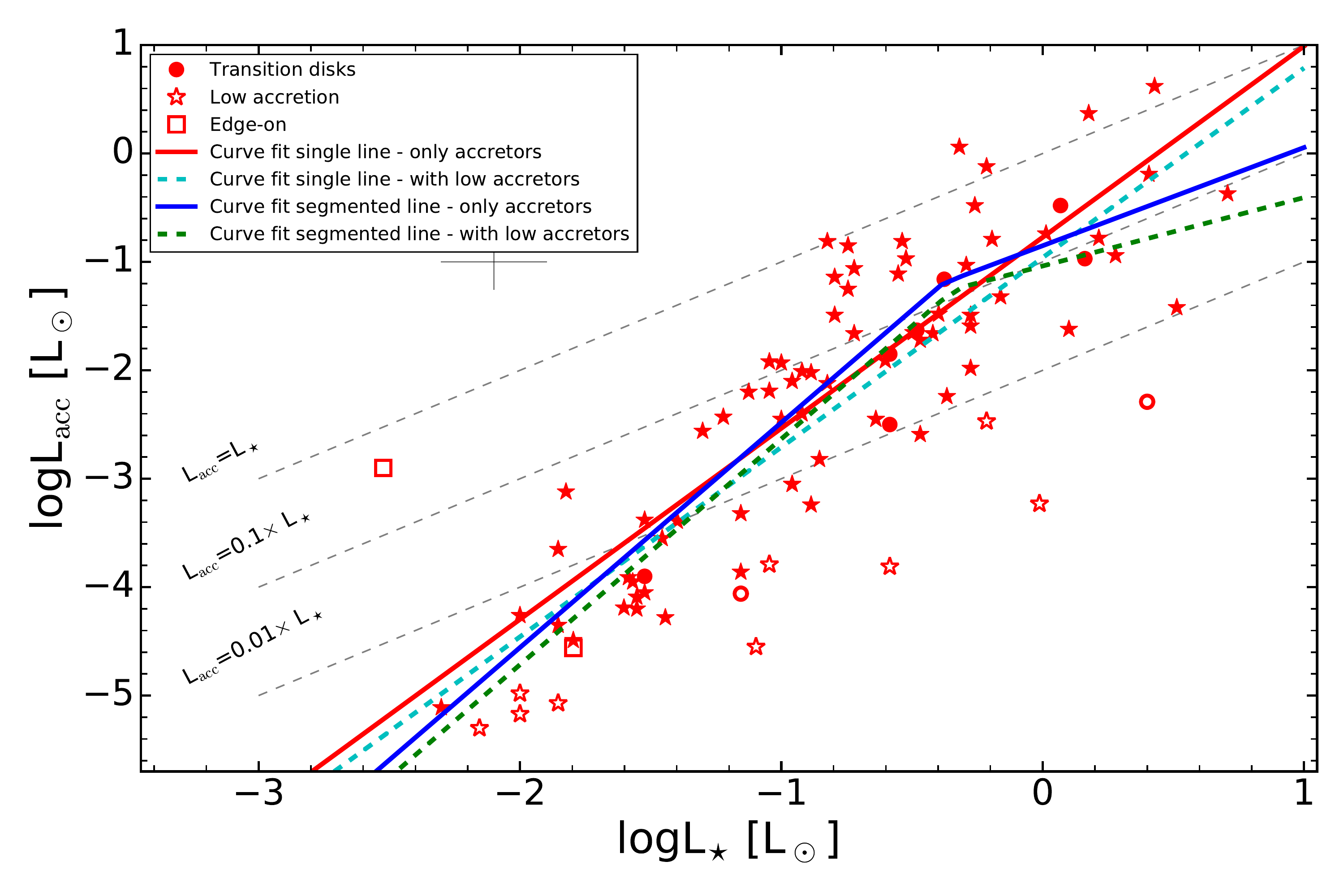}
\caption{Accretion vs stellar luminosity for the objects with a disk in the Chamaeleon~I star-forming region. Symbols are the
same as in Fig.~\ref{fig::lacc_lstar_all}, and the fit shown was performed using \textit{scipy.optimize.curve\_fit} with either a single linear fit or a segmented line.
     \label{fig::lacc_lstar_segmfit}}
\end{figure}

\begin{figure}[!t]
\centering
\includegraphics[width=0.5\textwidth]{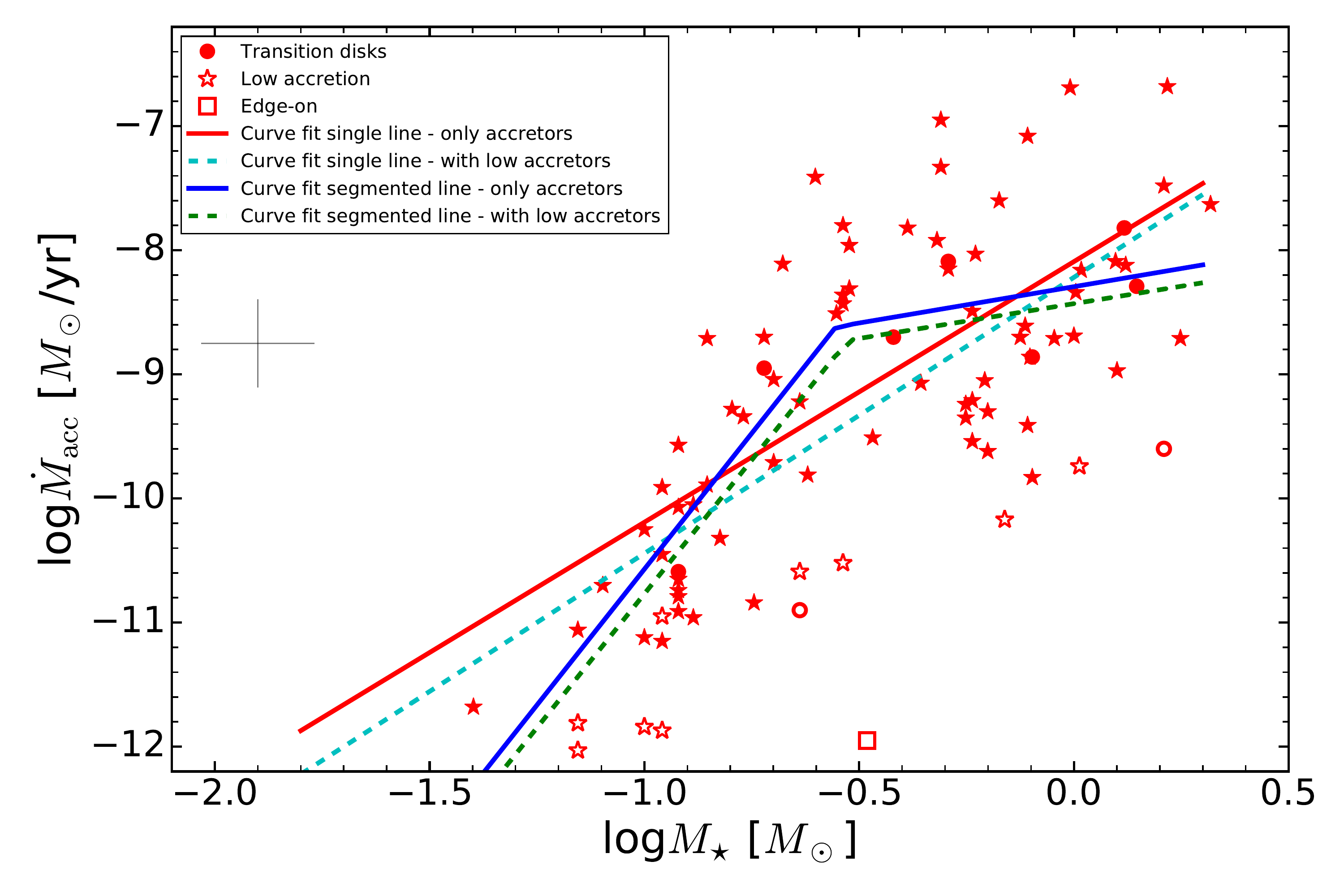}
\caption{Accretion rate vs stellar mass for the objects with a disk in the Chamaeleon~I star-forming region. Symbols are the
same as in Fig.~\ref{fig::lacc_lstar_all}. The results of the fit with a segmented line performed with \textit{scipy.optimize.curve\_fit} are shown using different colors depending on how the dubious accretors are treated in the analysis.
     \label{fig::macc_mstar_segm}}
\end{figure}

We then test how strongly the results obtained with \textit{scipy.optimize.curve\_fit} are affected by including or excluding the dubious accretors. This tool does not allow including upper limits in the test. The result is shown in Fig.~\ref{fig::lacc_lstar_segmfit} for the \lacc-\lstar \ relation, and in Fig.~\ref{fig::macc_mstar} and Fig.~\ref{fig::macc_mstar_segm} for the \macc-\mstar \ relation. The slope of the single power-law fit is the same regardless
of whether the dubious accretors are included, while small differences are present in the intercept for both the single power-law and
the segmented line fit. 

\subsection{Comparison between methods considering or excluding measurement errors}

We tested whether the uncertainties on the individual measurements contribute to the estimate of the best fit parameters more than the scatter of the data itself. We find that the scatter is the most important effect in determining the fit parameters following a similar approach to \citet{Pascucci16}. We fit the observations using \textit{linmix}, where the uncertainties are considered, and two methods that derive the best fit without considering the uncertainty: \textit{scipy.optimize.curve\_fit} and \textit{cenken} in the NADA R package. We discussed in Sect.~\ref{sect::results} that the former leads to best-fit values that are compatible with \textit{linmix}. The same is true for the latter, both when considering dubious accretors as detection or upper limits and when fitting the whole sample or the two subsamples at high and low \lstar \ and \mstar \ separately. Therefore, we conclude that the scatter dominates the fit determination.

\Online

\section{Best fit}

Here we show the best fit of the Balmer continuum region obtained with our method as described in Sect.~\ref{sect::method}.

\begin{figure*}[!t]
        \centering
        \begin{subfigure}[b]{0.45\textwidth}
               \includegraphics[width=\textwidth]{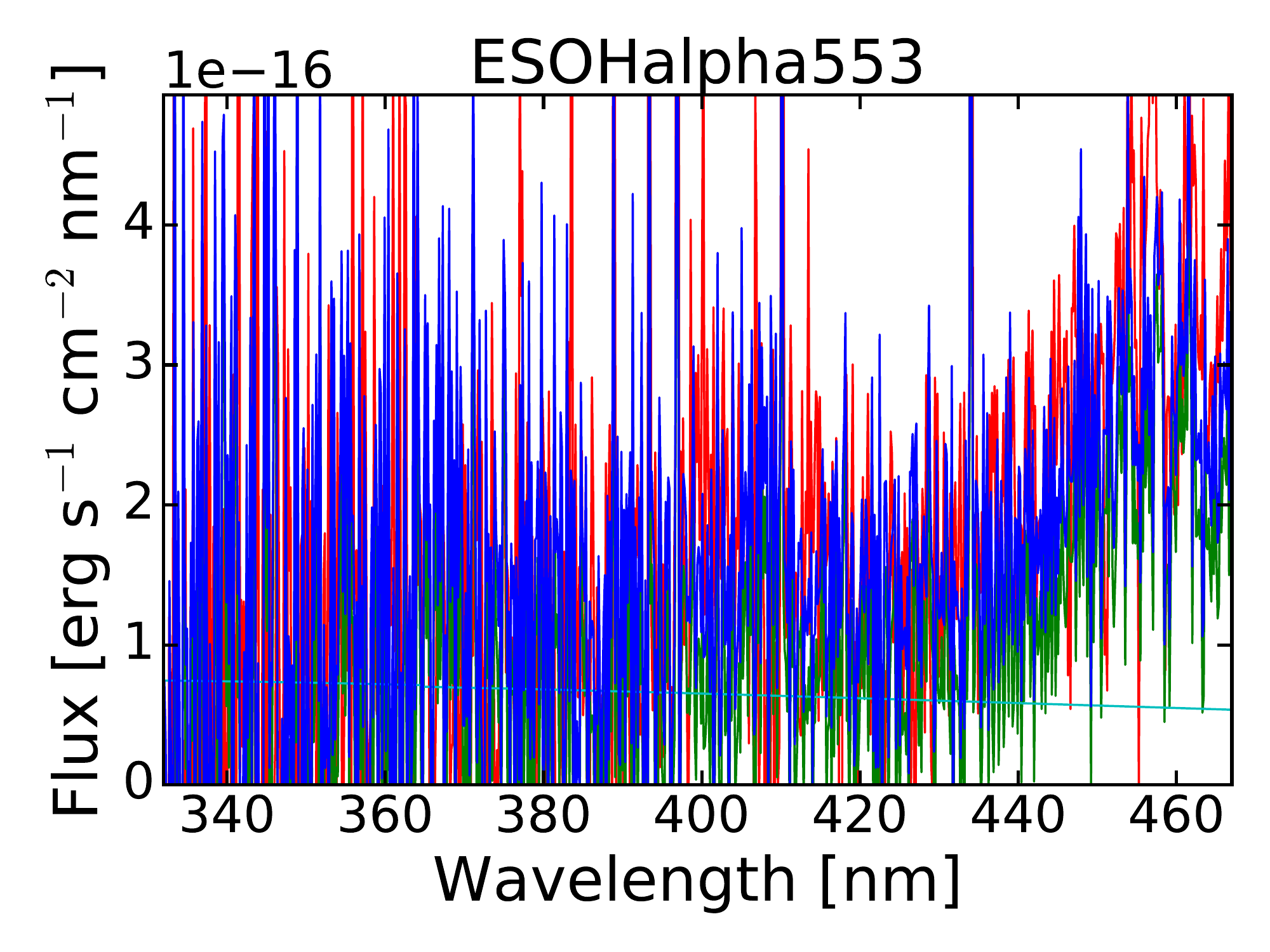}
        \end{subfigure}%
        ~ 
        \begin{subfigure}[b]{0.45\textwidth}
               \includegraphics[width=\textwidth]{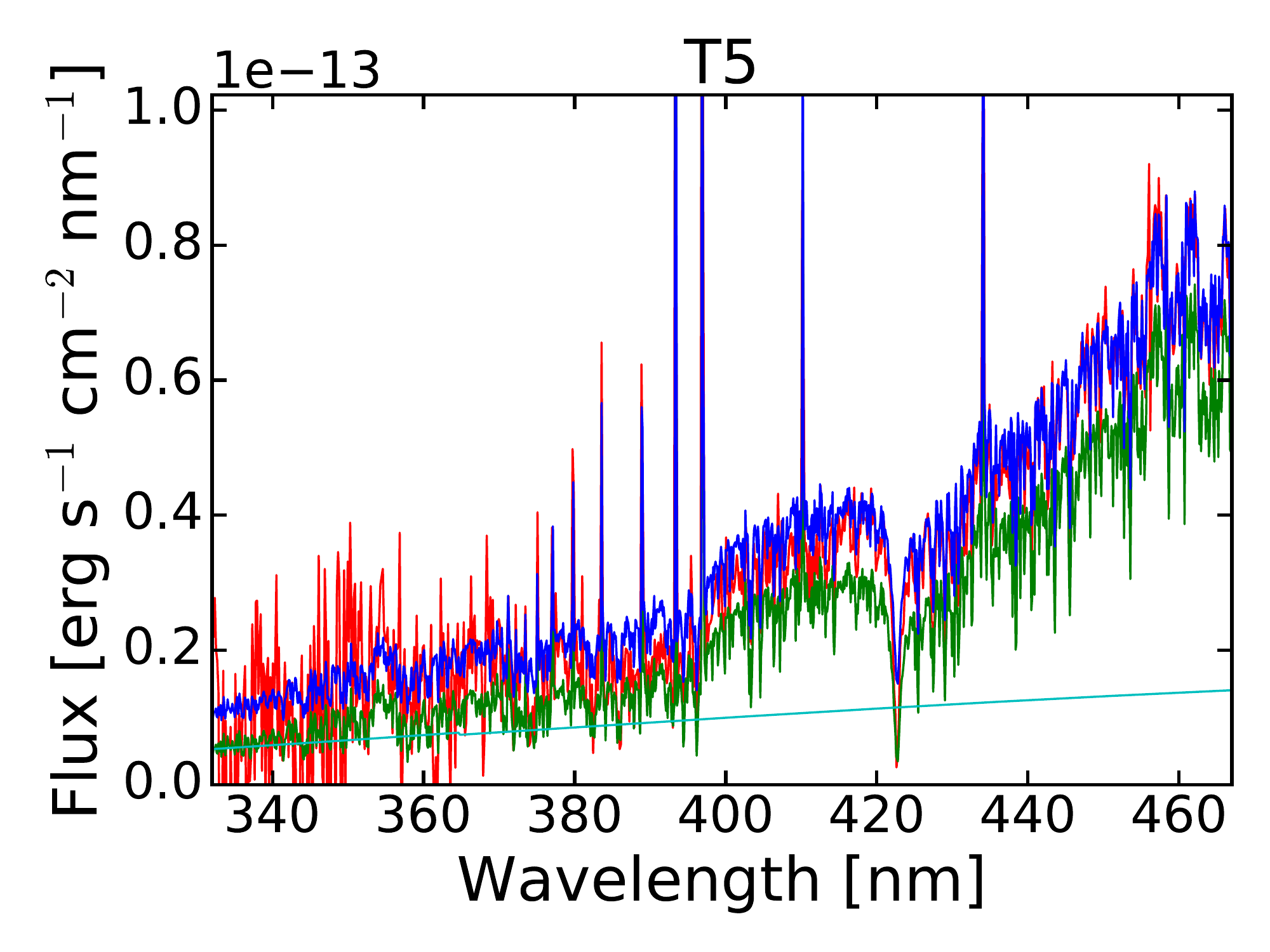}
        \end{subfigure}
        \begin{subfigure}[b]{0.45\textwidth}
               \includegraphics[width=\textwidth]{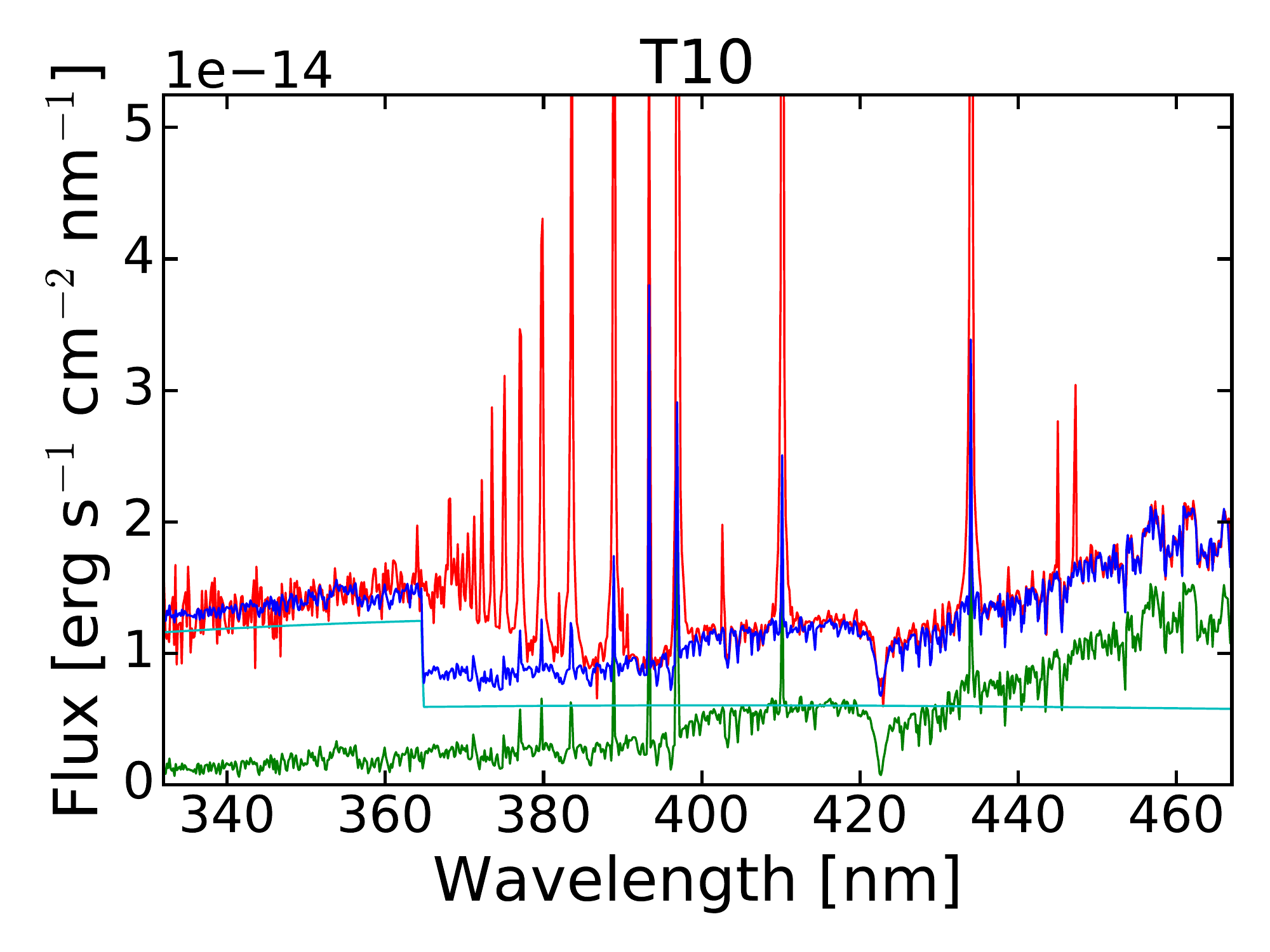}
        \end{subfigure}%
        ~ 
        \begin{subfigure}[b]{0.45\textwidth}
               \includegraphics[width=\textwidth]{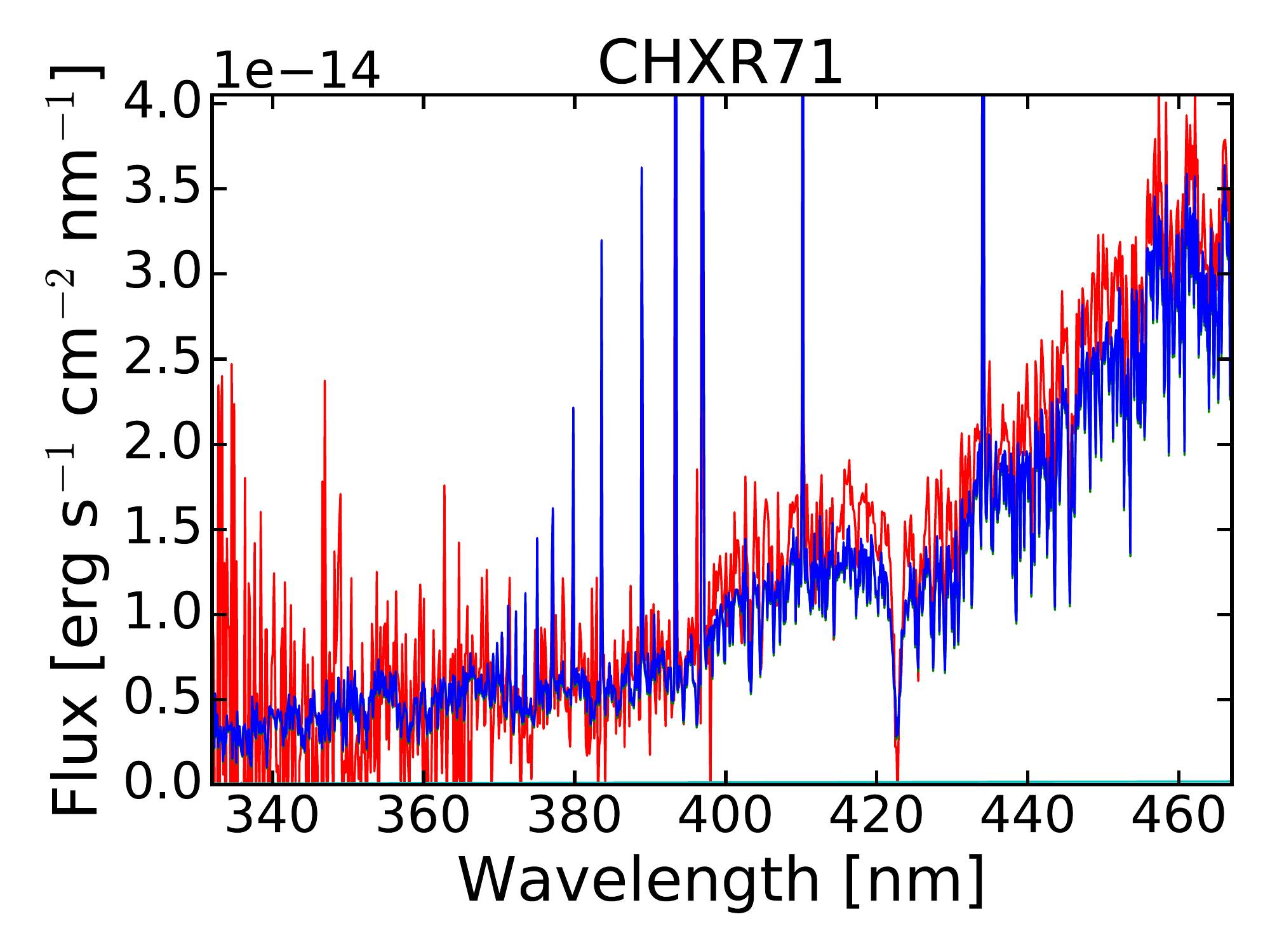}
        \end{subfigure}
        \begin{subfigure}[b]{0.45\textwidth}
               \includegraphics[width=\textwidth]{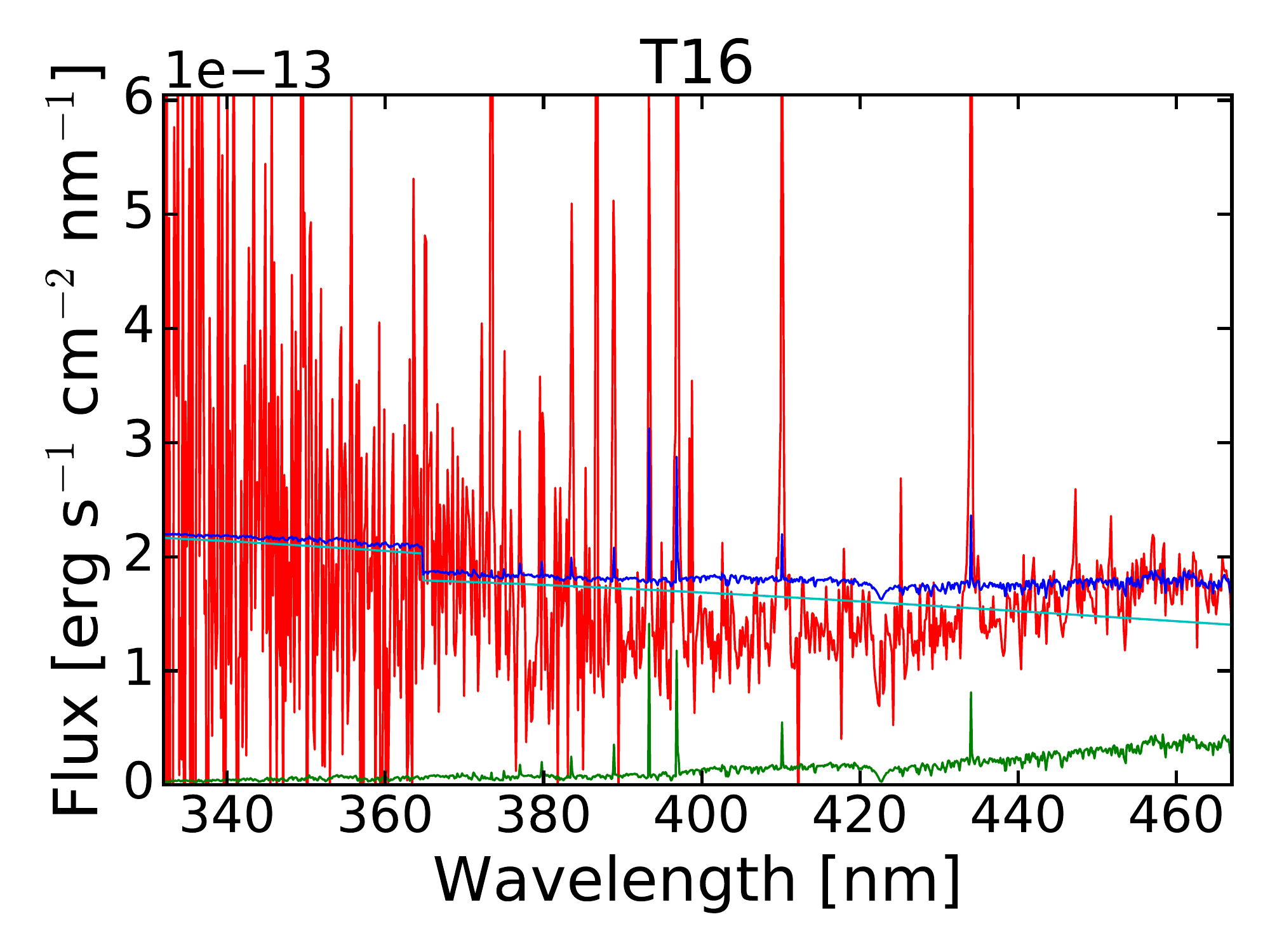}
        \end{subfigure}%
        ~ 
        \begin{subfigure}[b]{0.45\textwidth}
               \includegraphics[width=\textwidth]{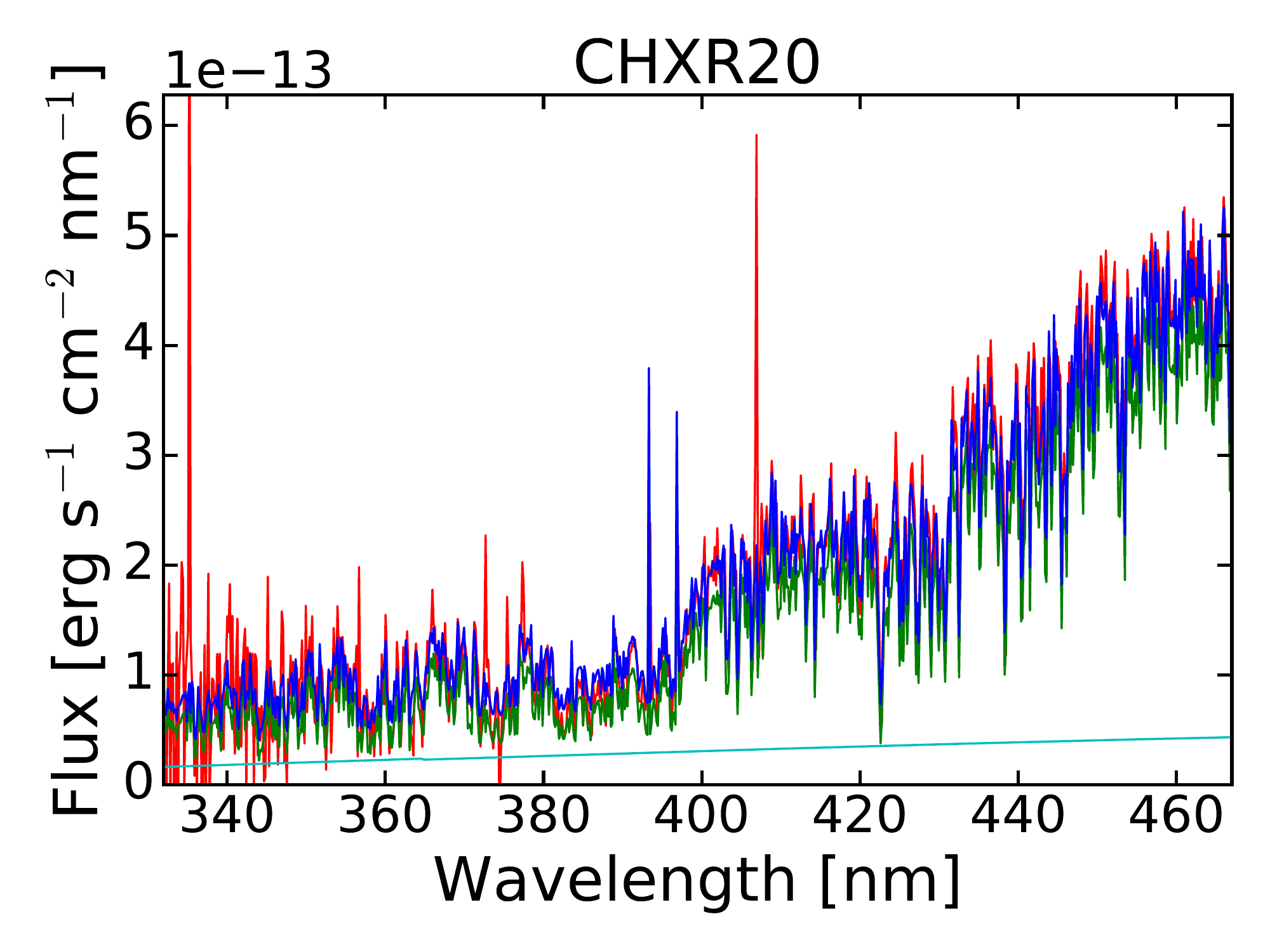}
        \end{subfigure}
         \begin{subfigure}[b]{0.45\textwidth}
               \includegraphics[width=\textwidth]{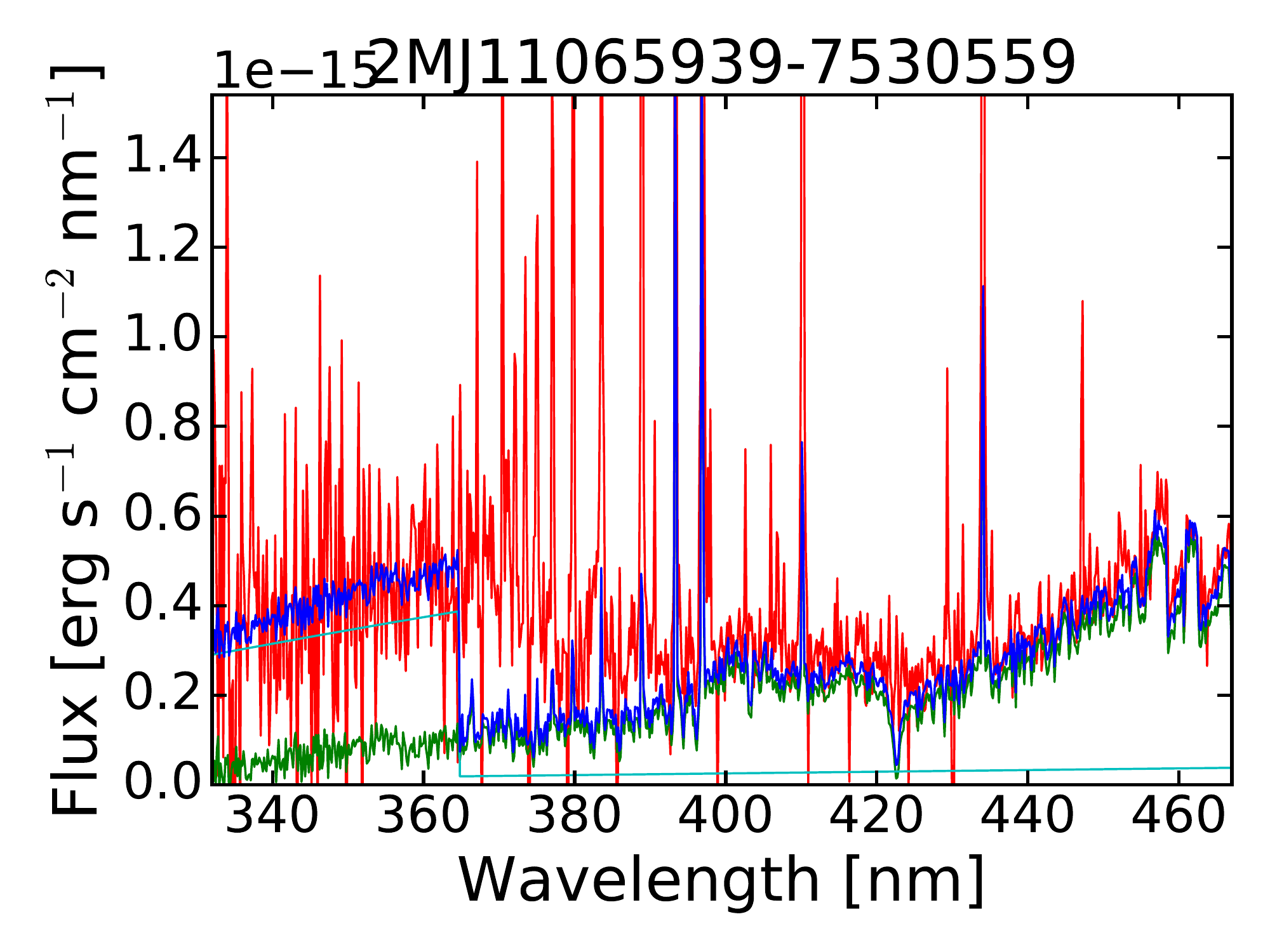}
        \end{subfigure}%
        ~ 
        \begin{subfigure}[b]{0.45\textwidth}
               \includegraphics[width=\textwidth]{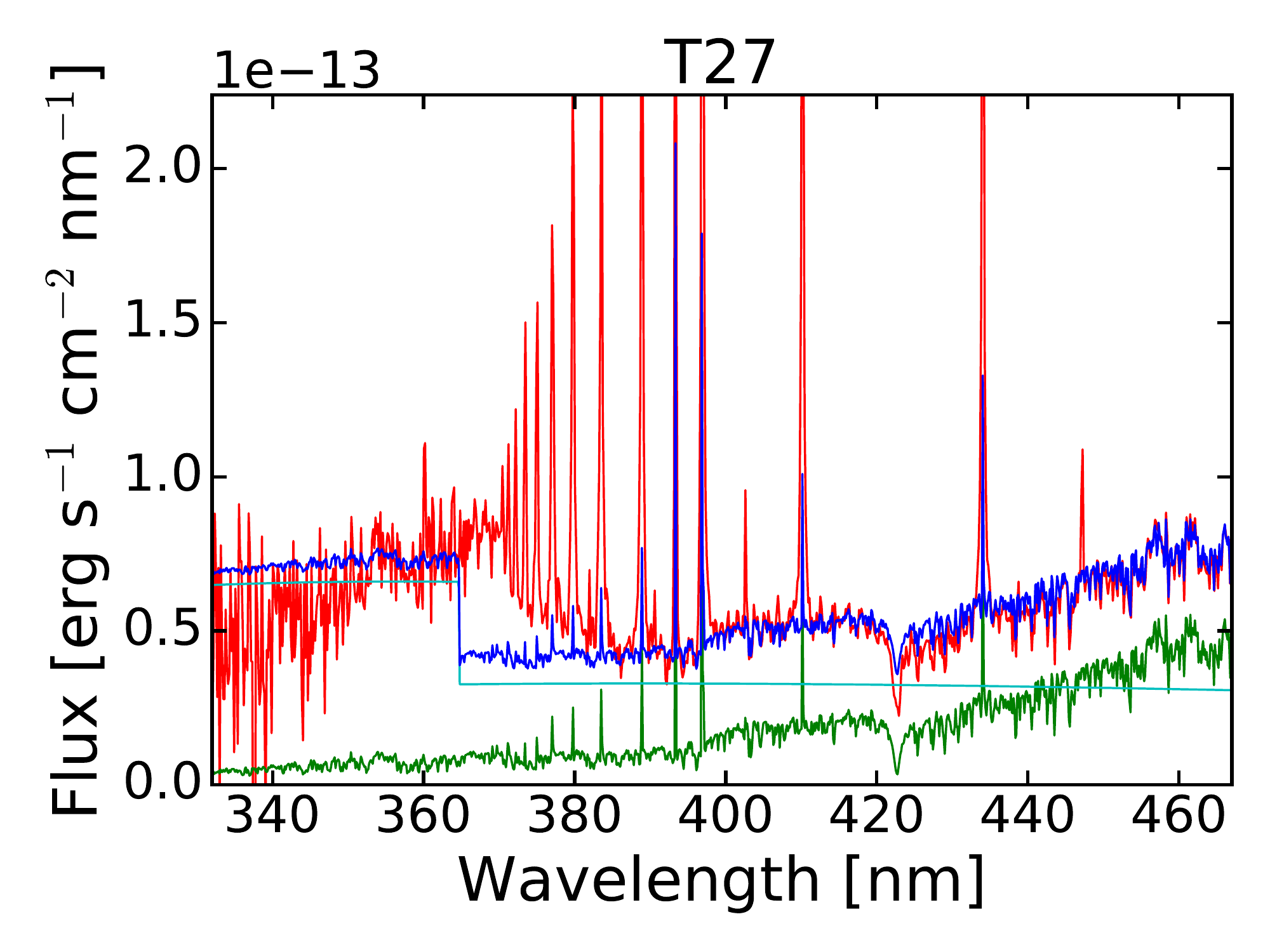}
        \end{subfigure}
       \caption{Best fit for the Chamaeleon~I targets studied here. Names are reported in the title of each subplot. The red line is the reddening-corrected spectrum of the target, the blue line the best fit, which is the sum of the photospheric template (green line) and the slab model (cyan line). The input spectrum and the photospheric templates are smoothed for better visualization.
}\label{fig:best_fits}

\end{figure*}
%
%
\begin{figure*}[!t]
        \centering
        \begin{subfigure}[b]{0.45\textwidth}
               \includegraphics[width=\textwidth]{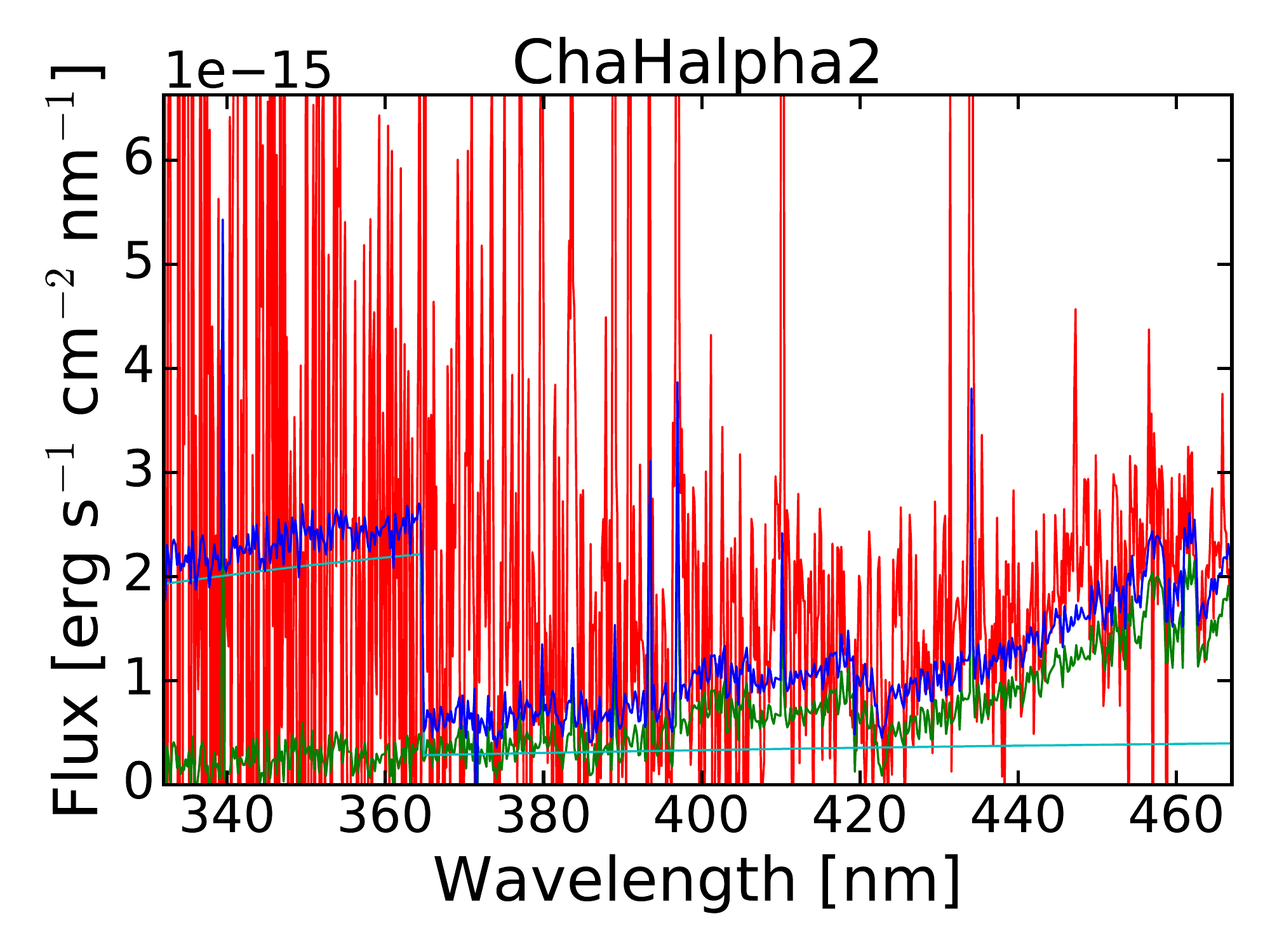}
        \end{subfigure}%
        ~ 
        \begin{subfigure}[b]{0.45\textwidth}
               \includegraphics[width=\textwidth]{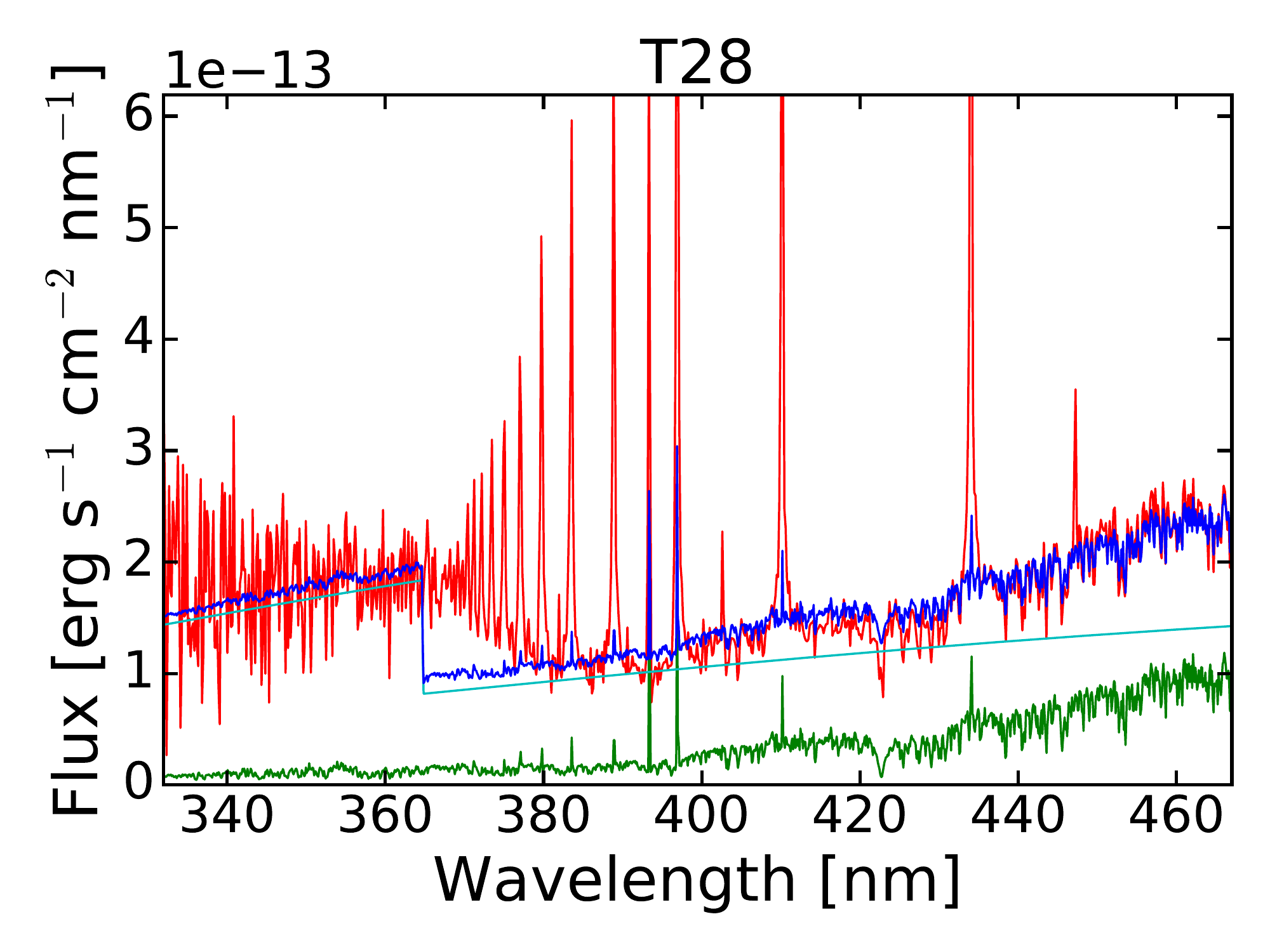}
        \end{subfigure}
        \begin{subfigure}[b]{0.45\textwidth}
               \includegraphics[width=\textwidth]{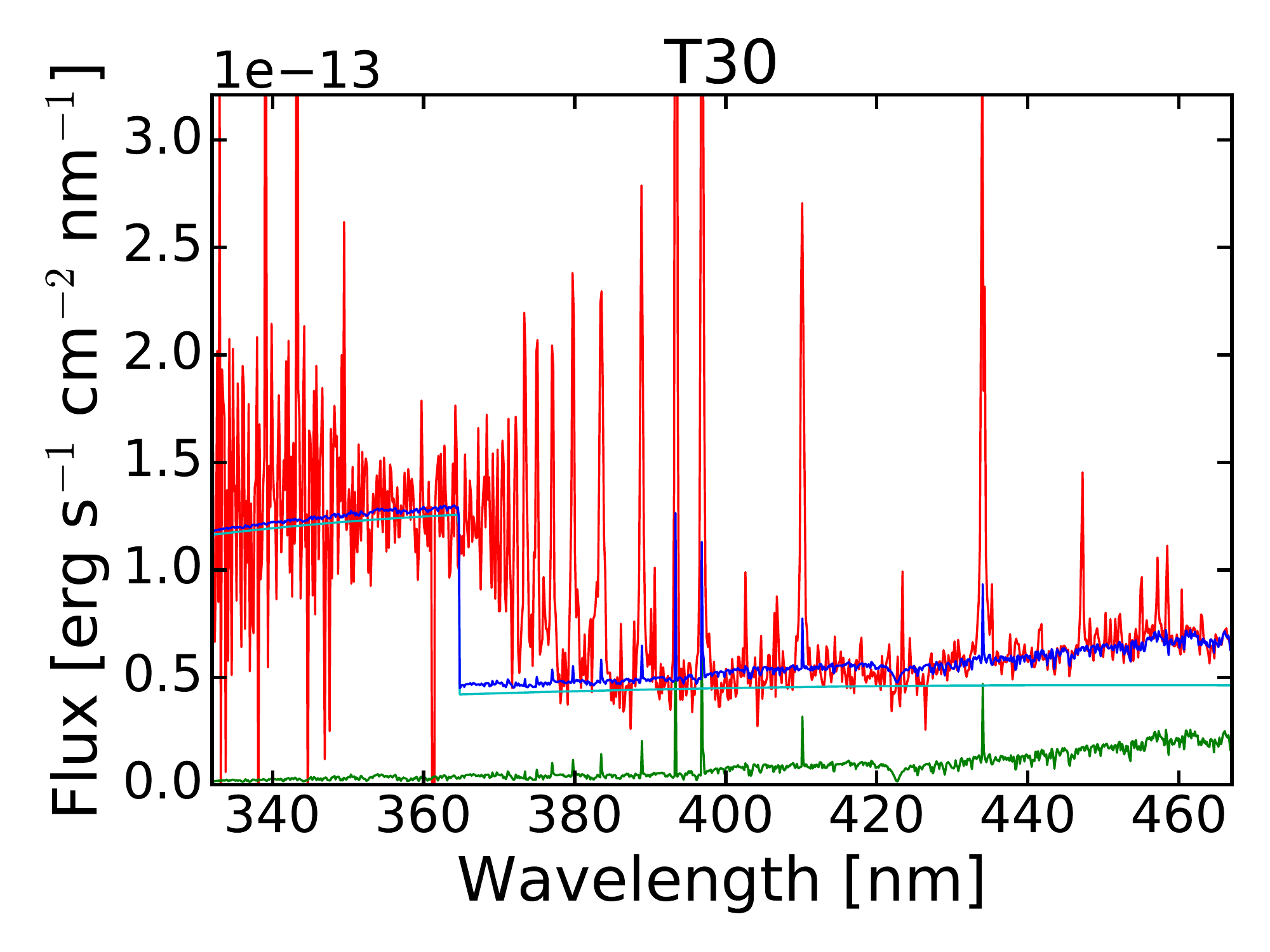}
        \end{subfigure}%
        ~ 
        \begin{subfigure}[b]{0.45\textwidth}
               \includegraphics[width=\textwidth]{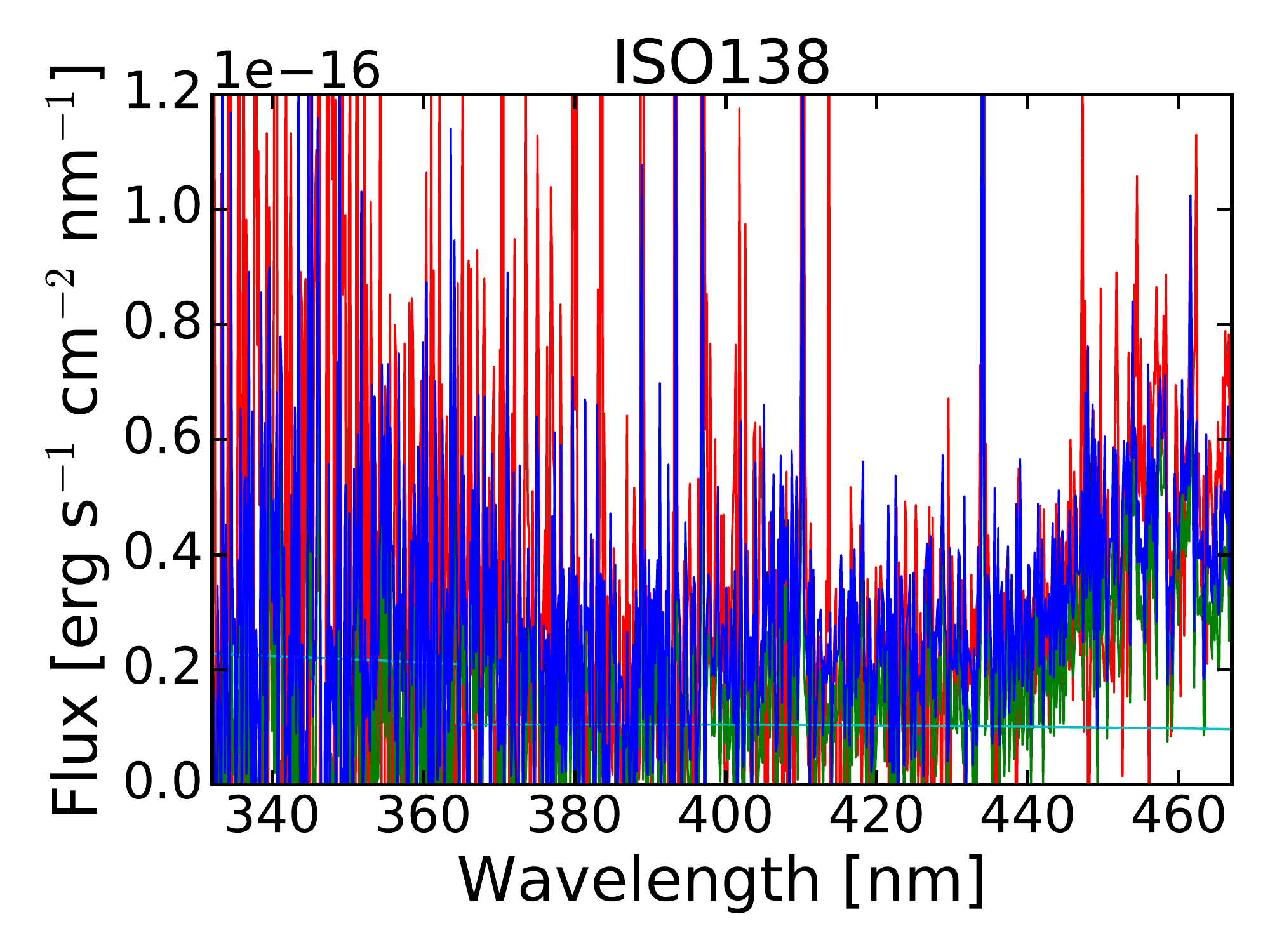}
        \end{subfigure}
         \begin{subfigure}[b]{0.45\textwidth}
               \includegraphics[width=\textwidth]{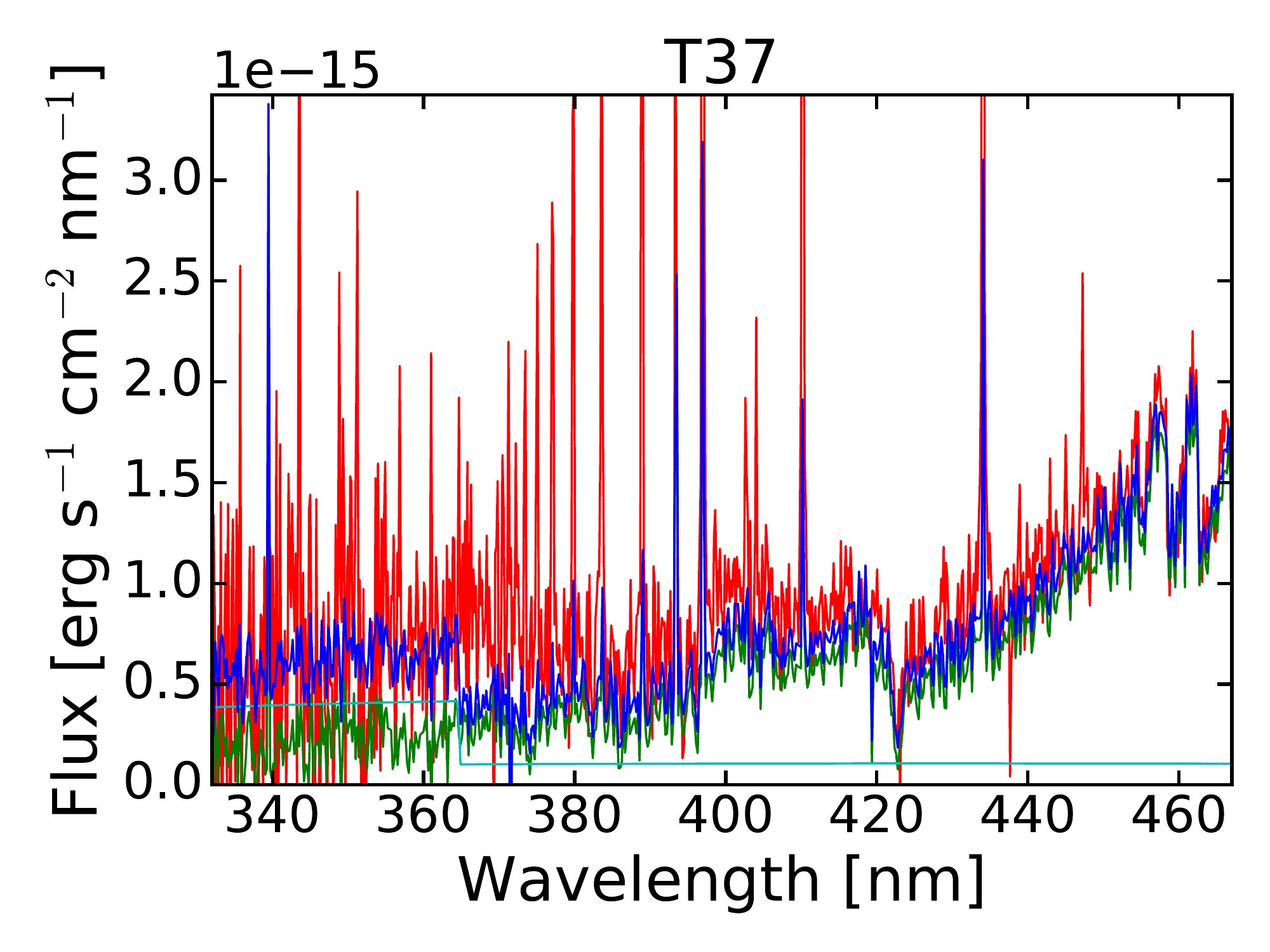}
        \end{subfigure}%
        ~ 
        \begin{subfigure}[b]{0.45\textwidth}
               \includegraphics[width=\textwidth]{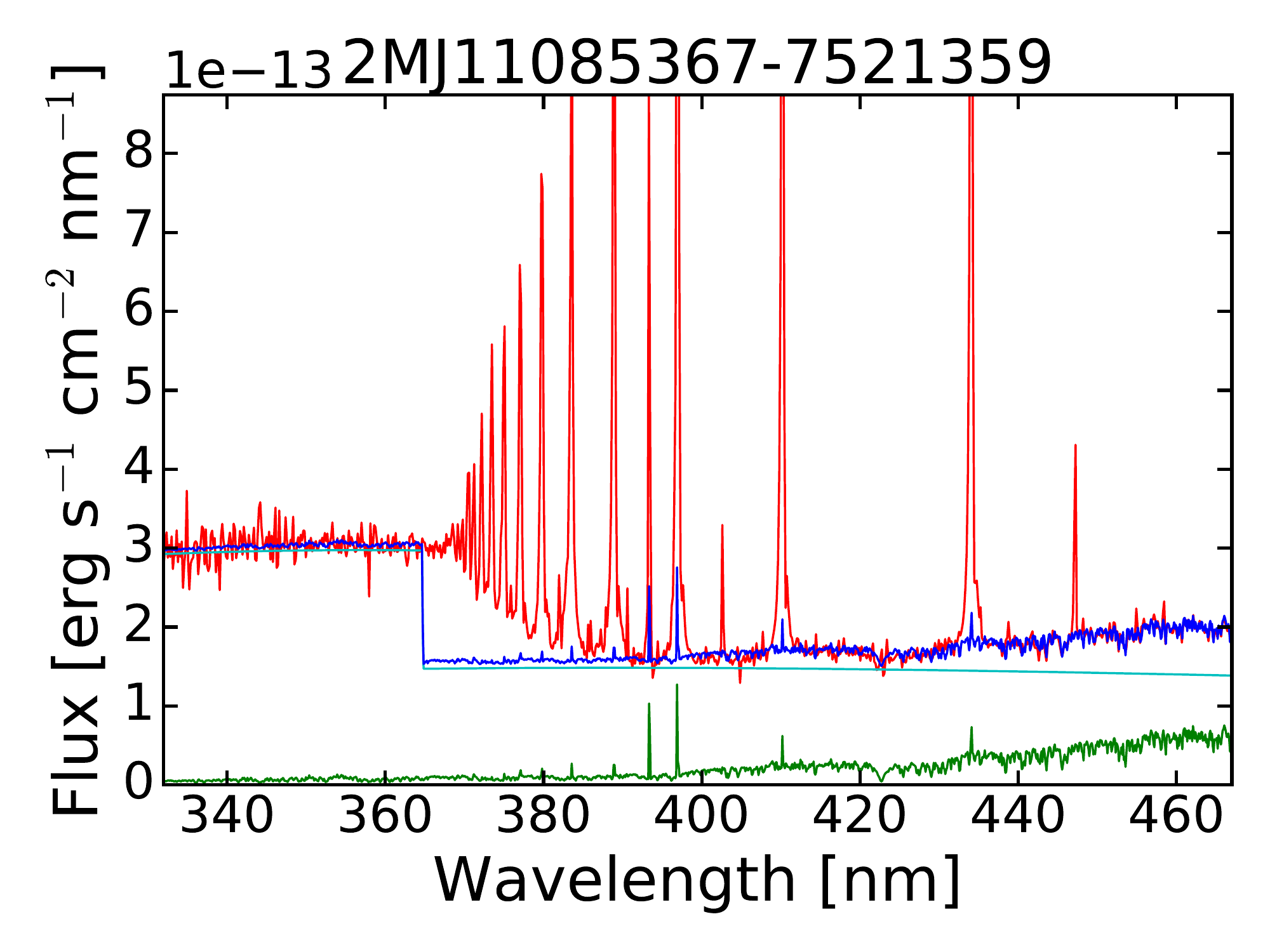}
        \end{subfigure}
        \begin{subfigure}[b]{0.45\textwidth}
               \includegraphics[width=\textwidth]{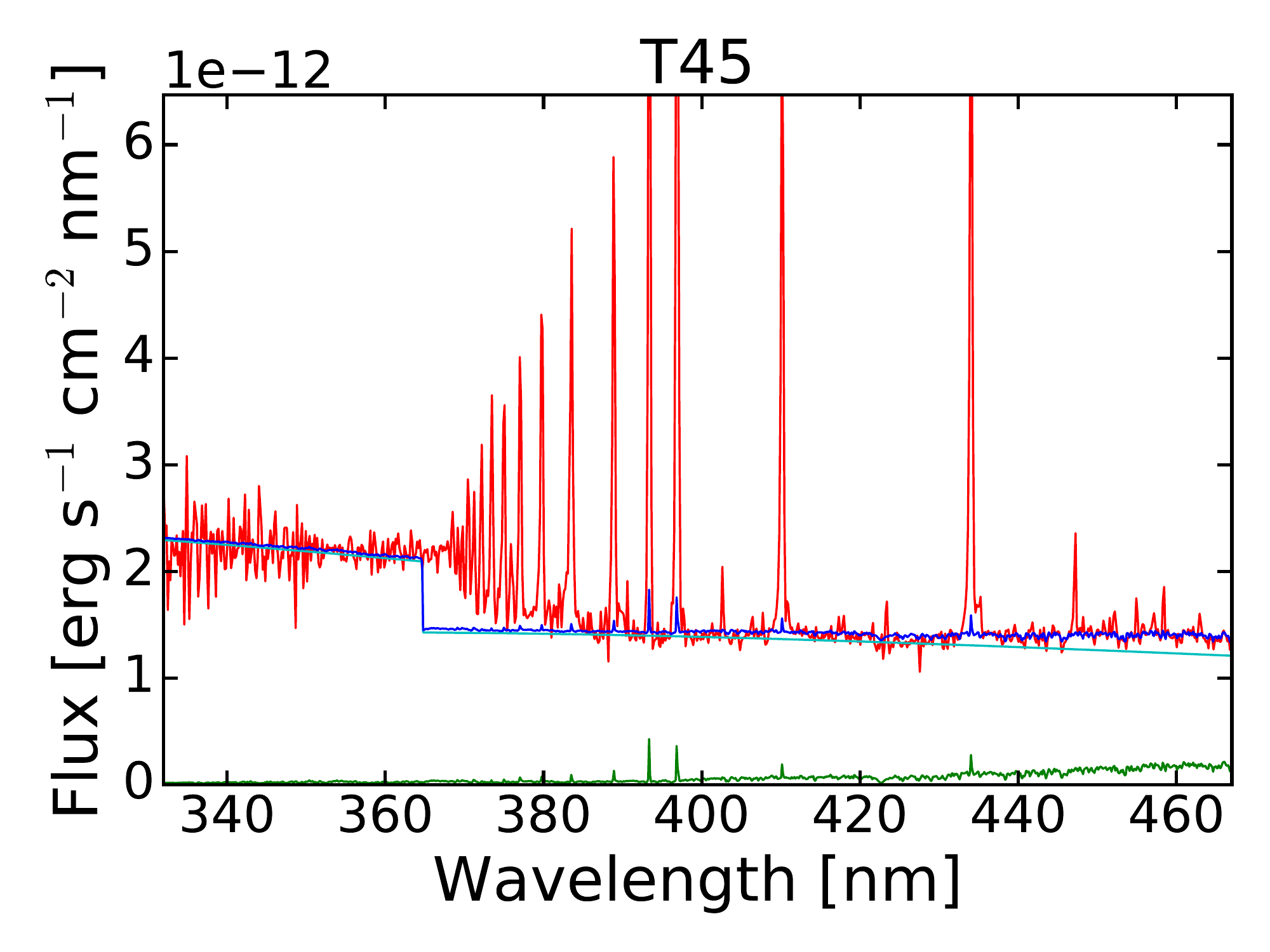}
        \end{subfigure}%
        ~ 
        \begin{subfigure}[b]{0.45\textwidth}
               \includegraphics[width=\textwidth]{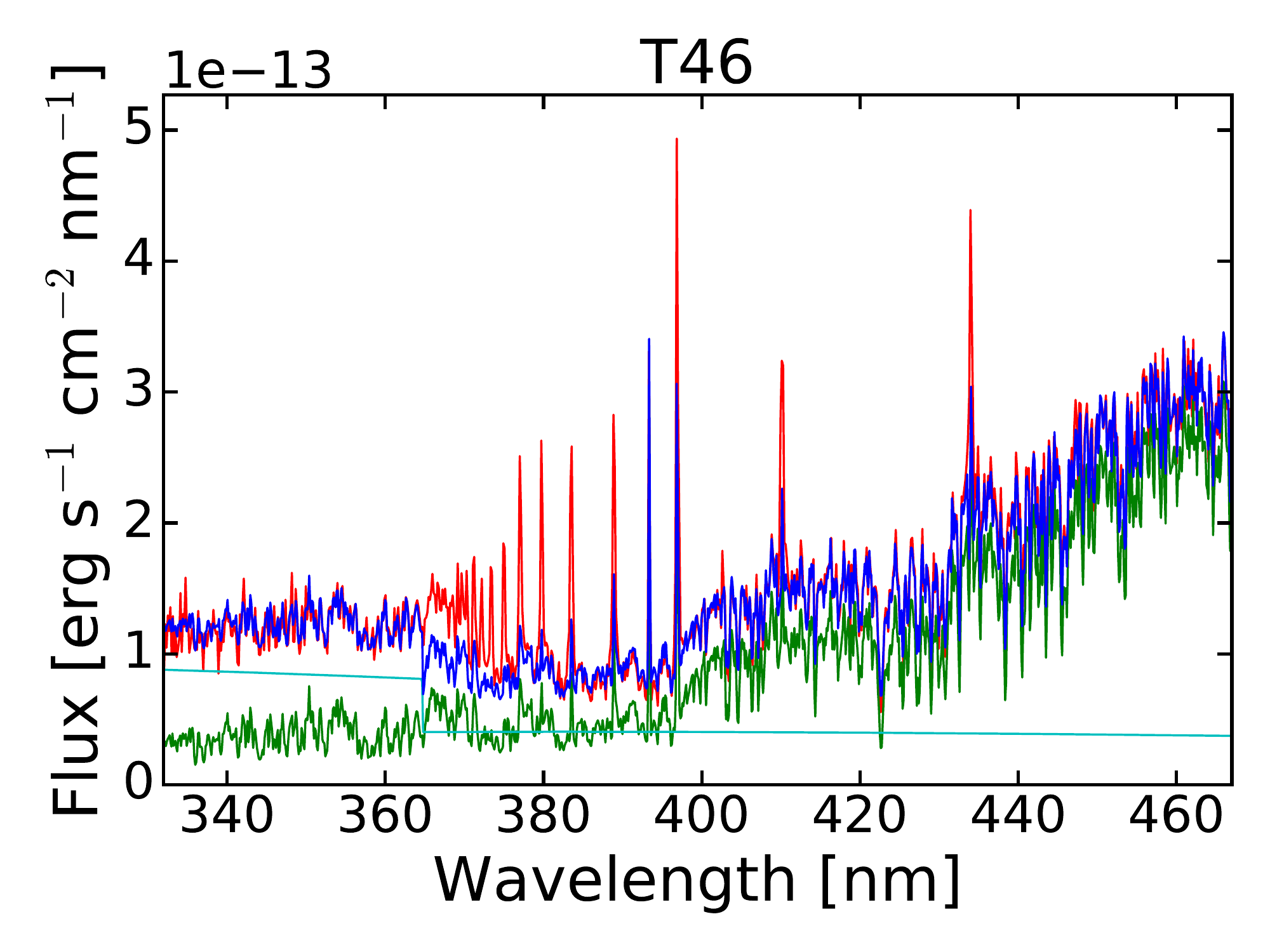}
        \end{subfigure}
       \caption{Same as \ref{fig:best_fits}.
}\label{fig:best_fits2}

\end{figure*}
%
%
\begin{figure*}[!t]
        \centering
        \begin{subfigure}[b]{0.45\textwidth}
               \includegraphics[width=\textwidth]{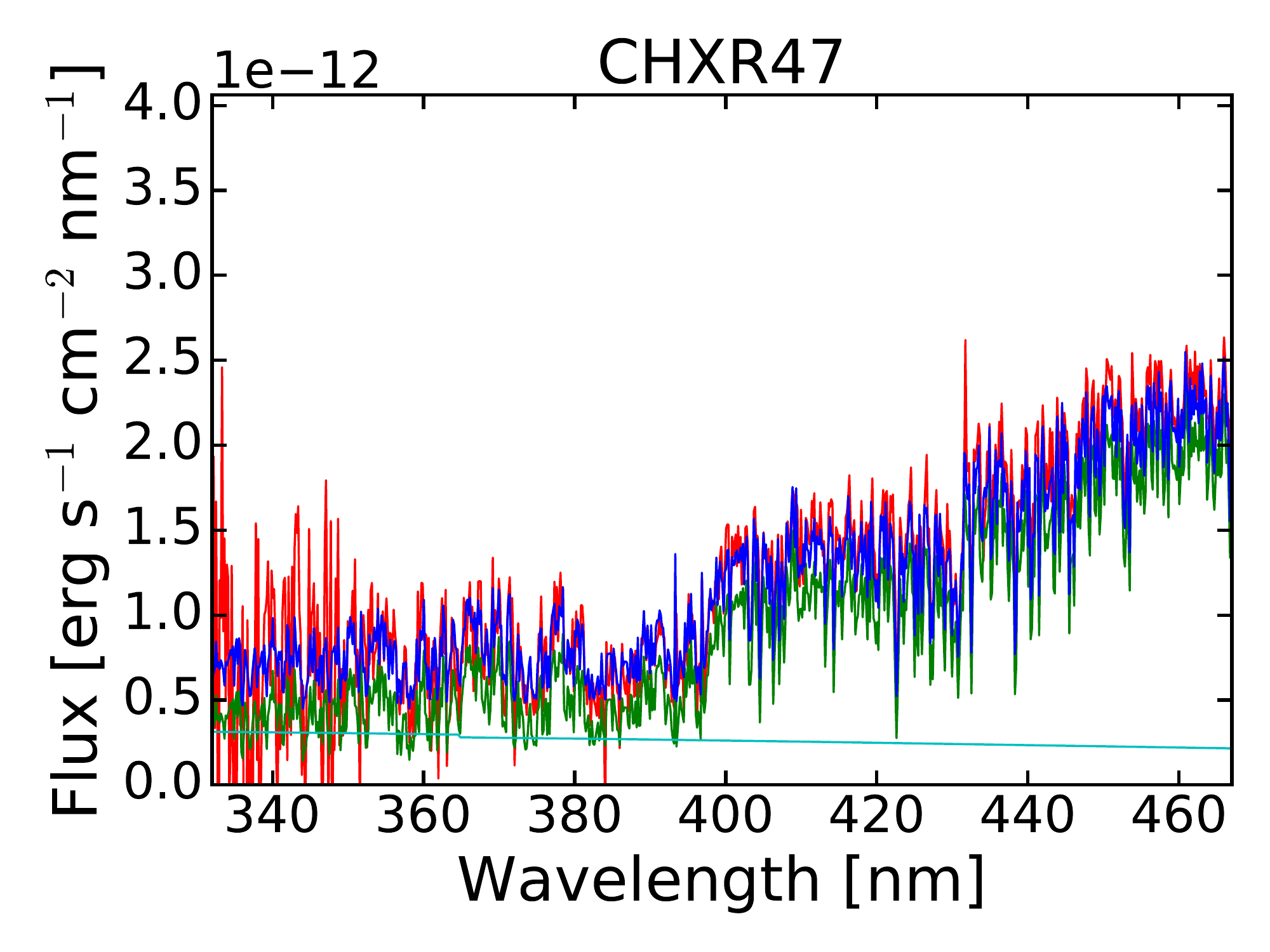}
        \end{subfigure}%
        ~ 
        \begin{subfigure}[b]{0.45\textwidth}
               \includegraphics[width=\textwidth]{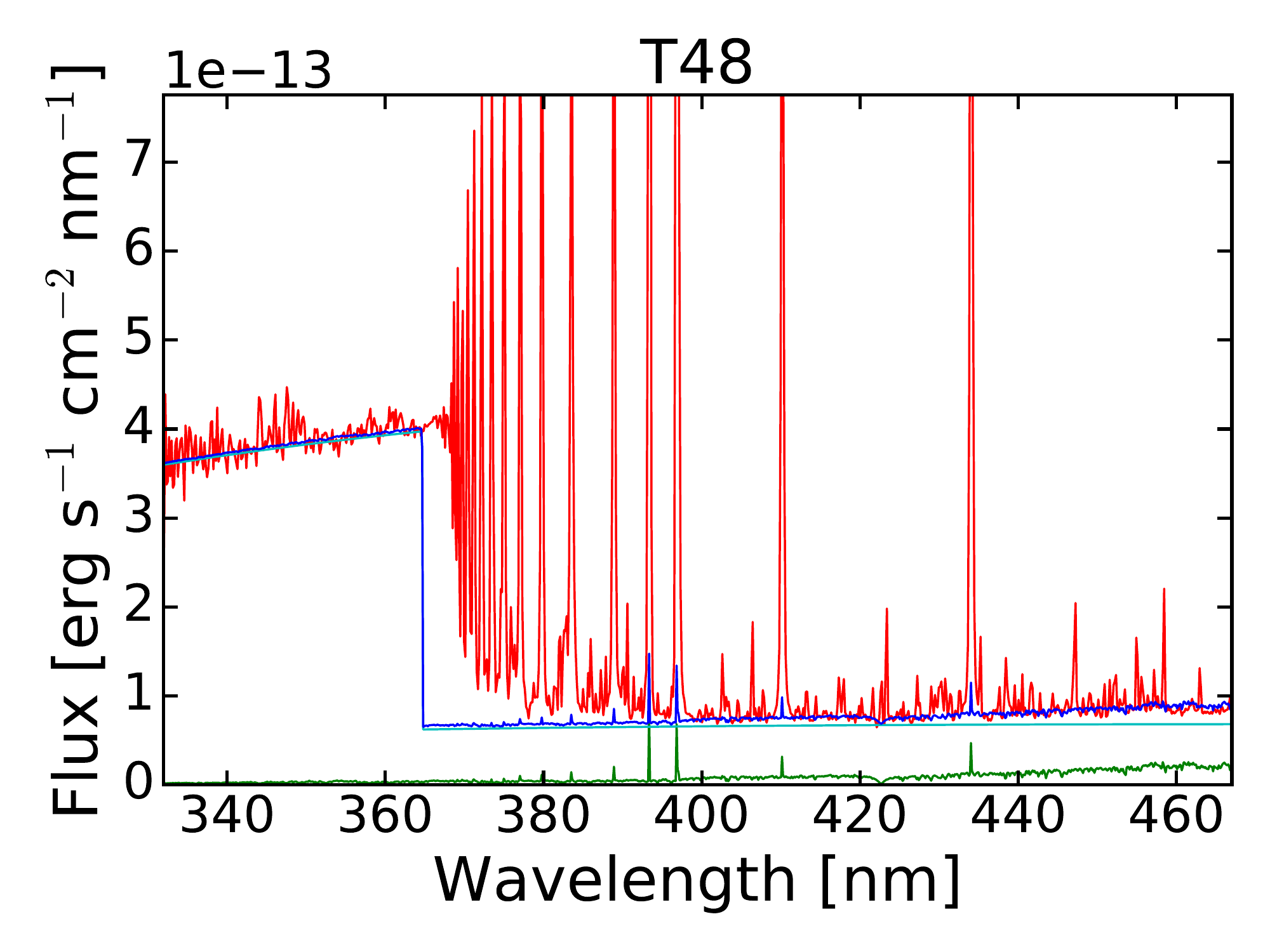}
        \end{subfigure}
        \begin{subfigure}[b]{0.45\textwidth}
               \includegraphics[width=\textwidth]{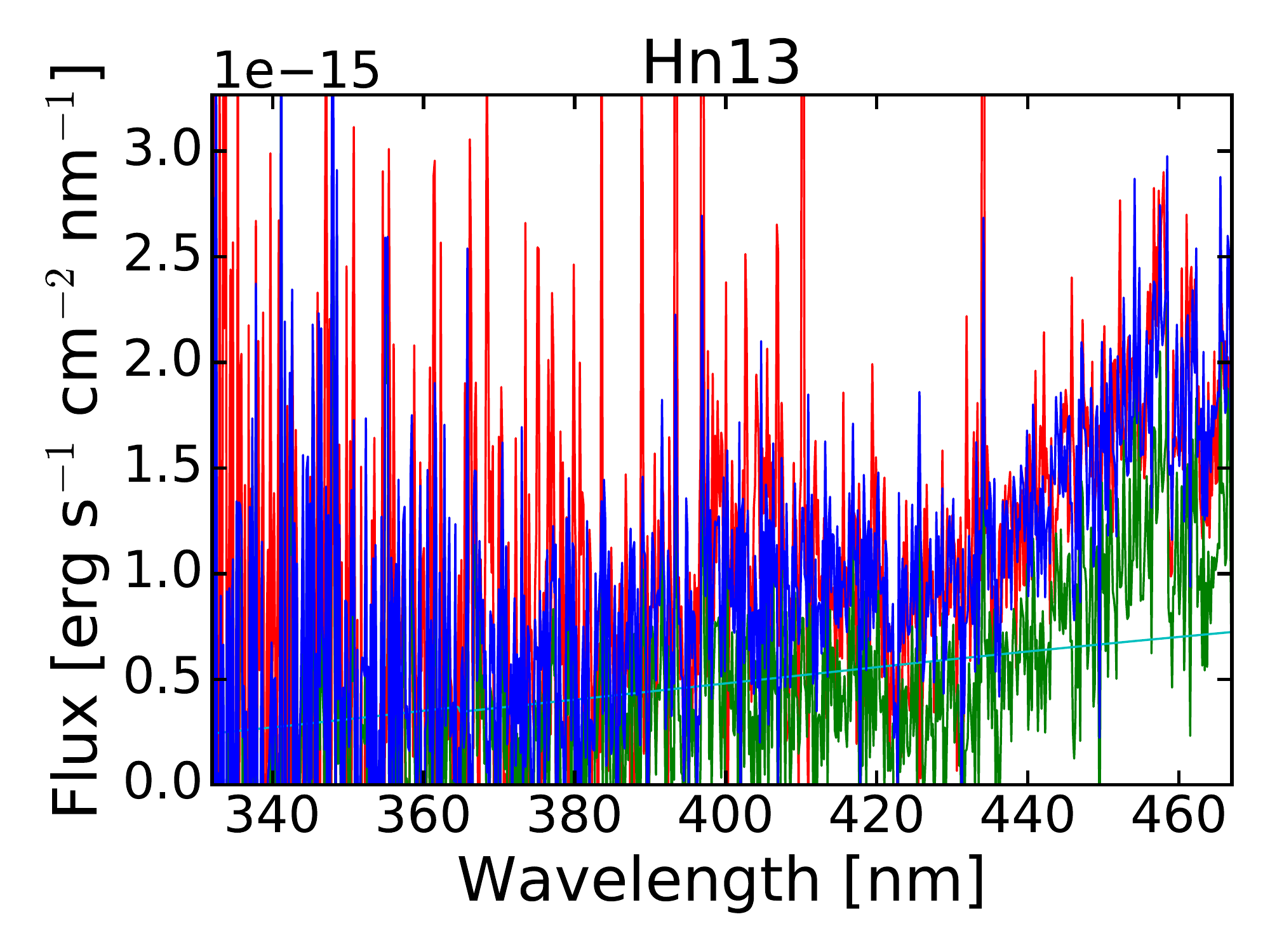}
        \end{subfigure}%
        ~ 
        \begin{subfigure}[b]{0.45\textwidth}
               \includegraphics[width=\textwidth]{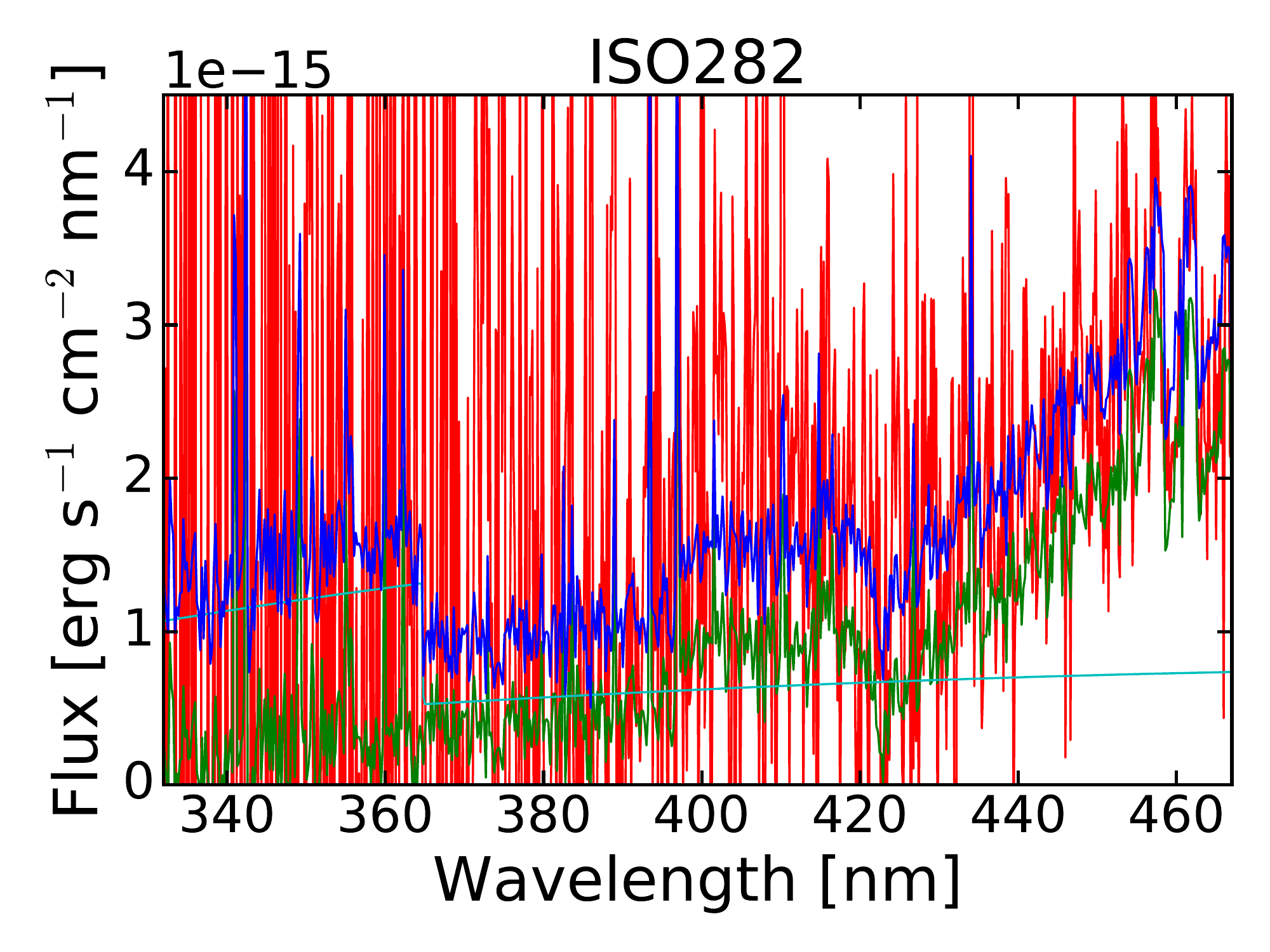}
        \end{subfigure}
         \begin{subfigure}[b]{0.45\textwidth}
               \includegraphics[width=\textwidth]{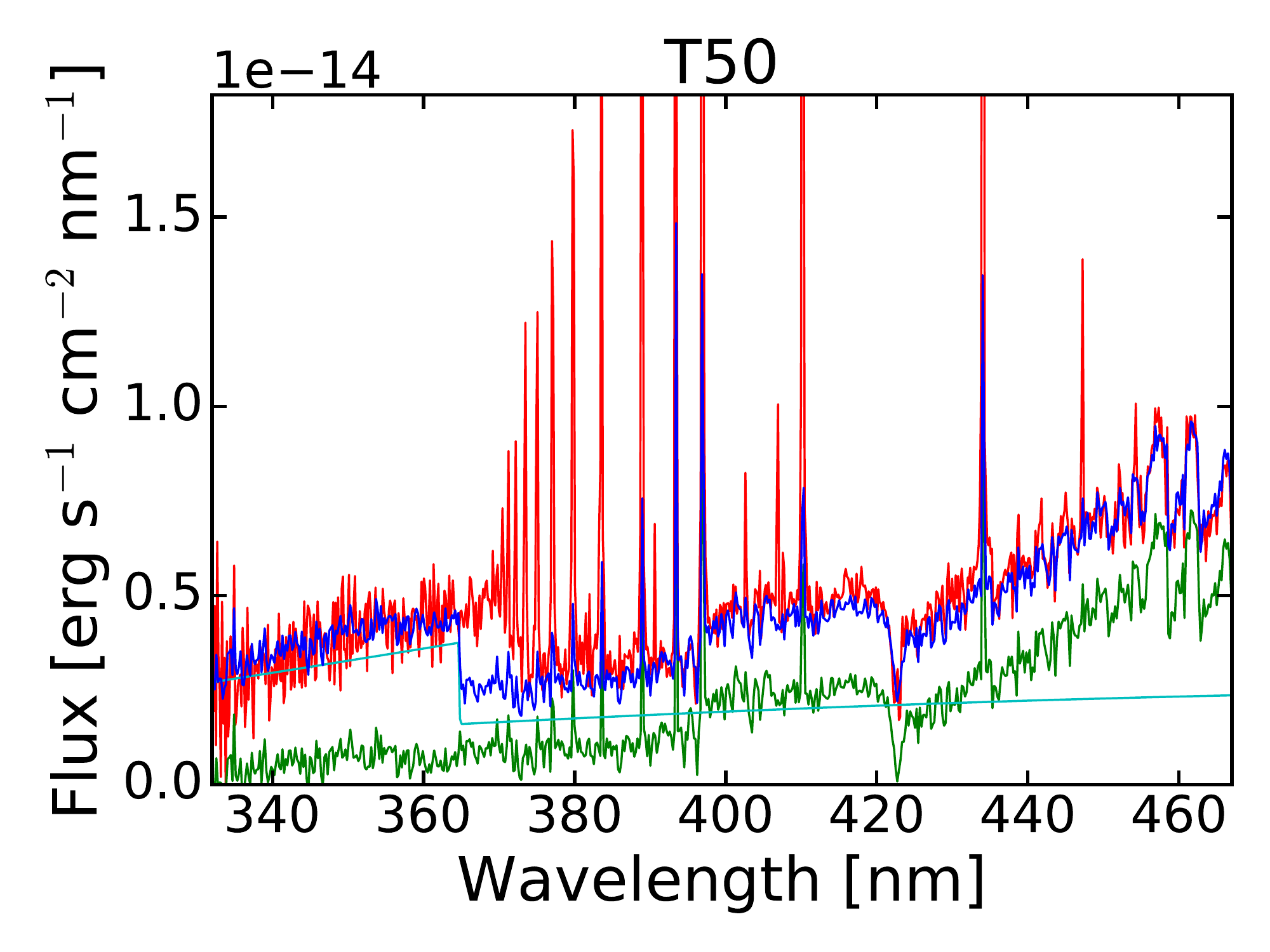}
        \end{subfigure}%
        ~ 
        \begin{subfigure}[b]{0.45\textwidth}
               \includegraphics[width=\textwidth]{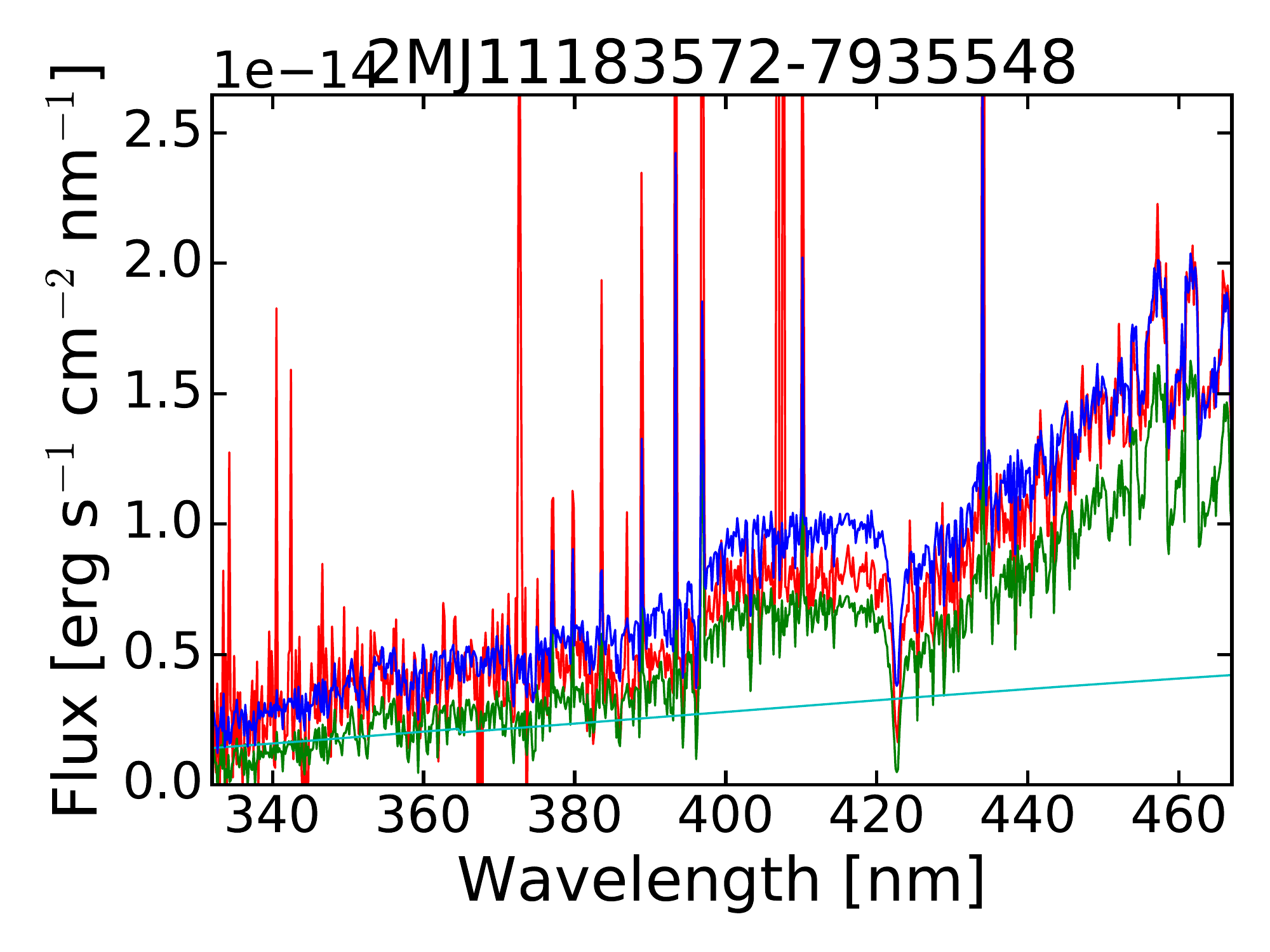}
        \end{subfigure}
         \begin{subfigure}[b]{0.45\textwidth}
               \includegraphics[width=\textwidth]{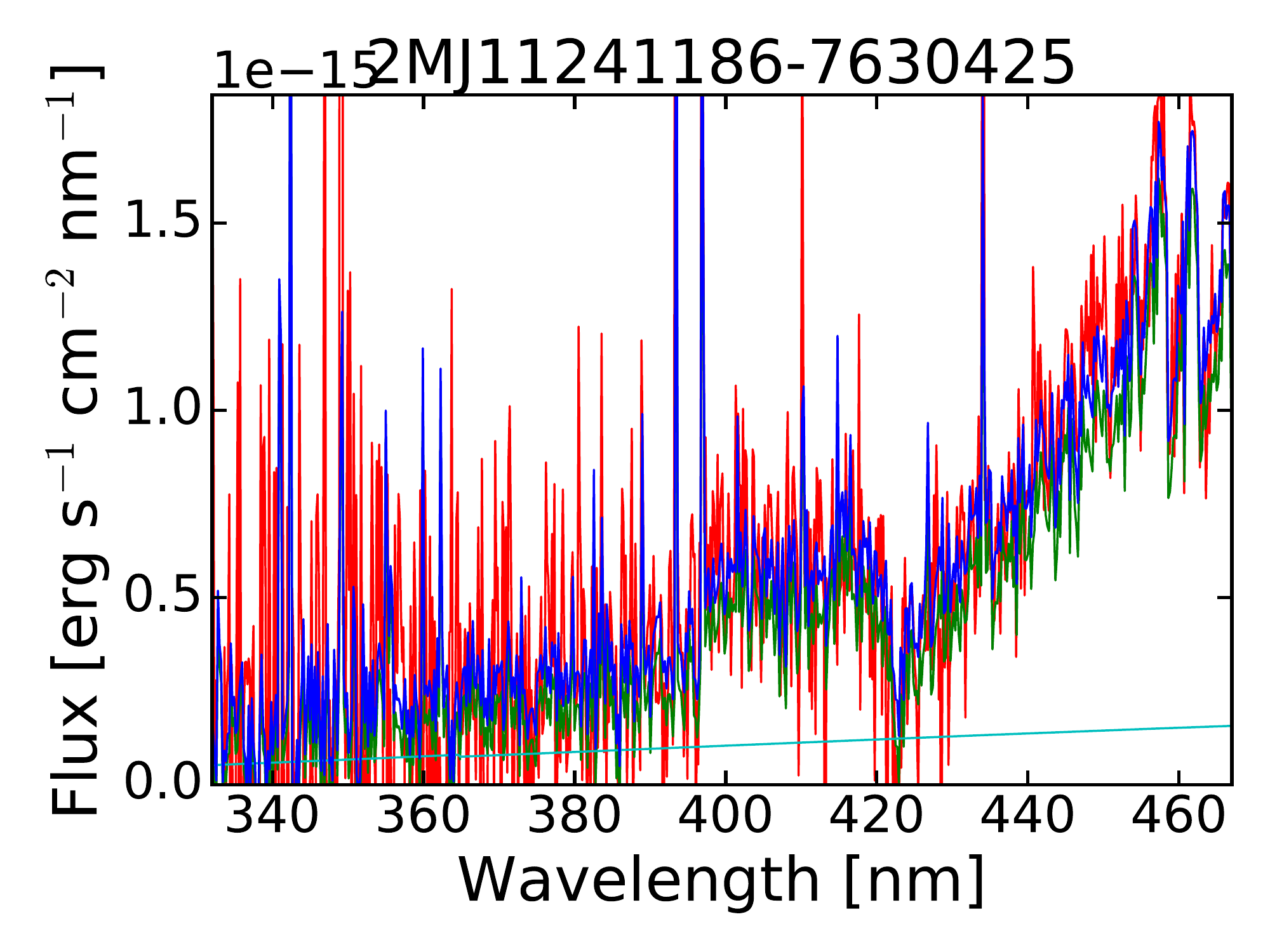}
        \end{subfigure}%
        ~ 
        \begin{subfigure}[b]{0.45\textwidth}
               \includegraphics[width=\textwidth]{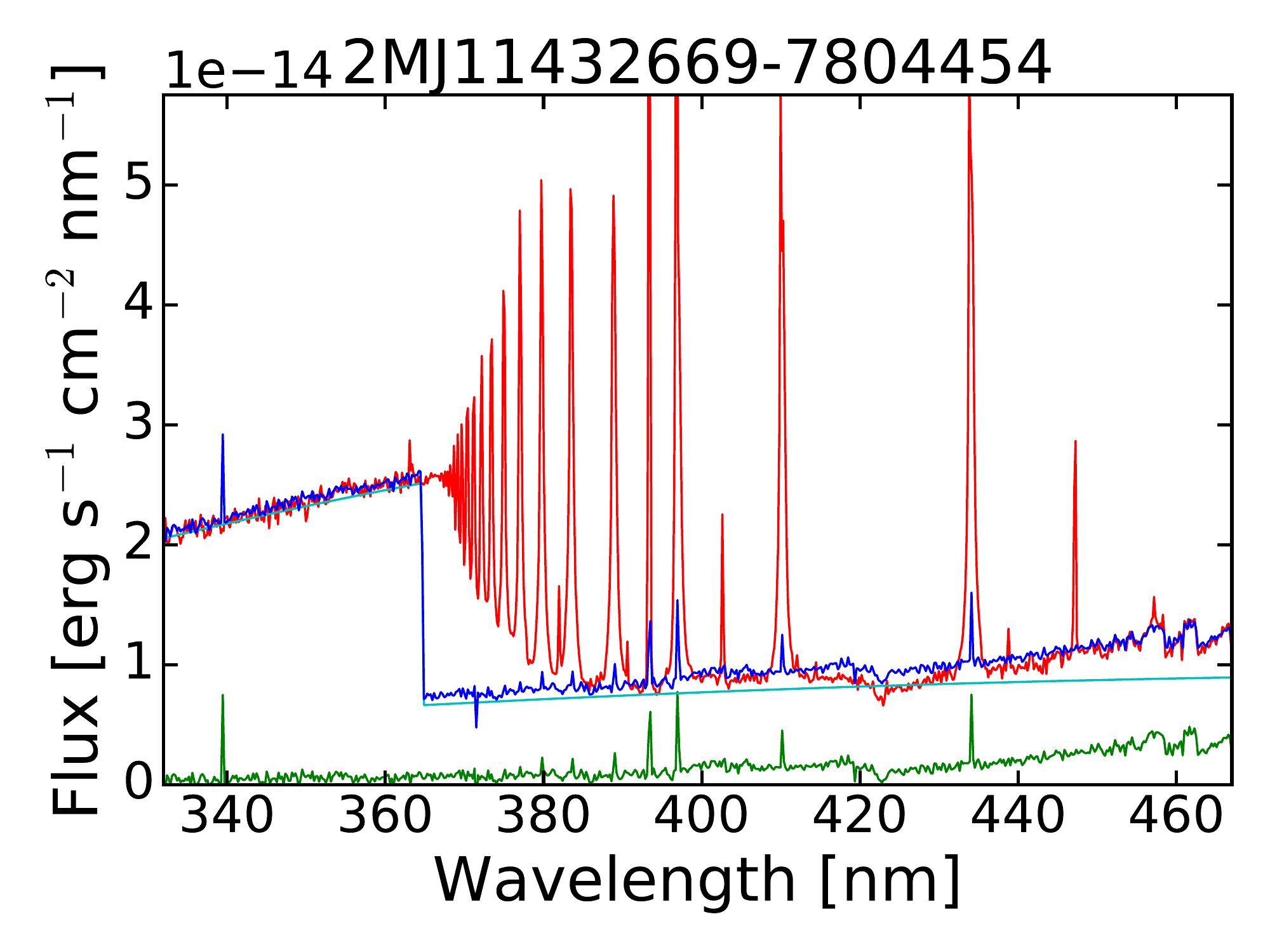}
        \end{subfigure}
      \caption{Same as \ref{fig:best_fits}.
}\label{fig:best_fits3}

\end{figure*}
%
%
%
\begin{figure*}[!t]
        \centering
        \begin{subfigure}[b]{0.45\textwidth}
               \includegraphics[width=\textwidth]{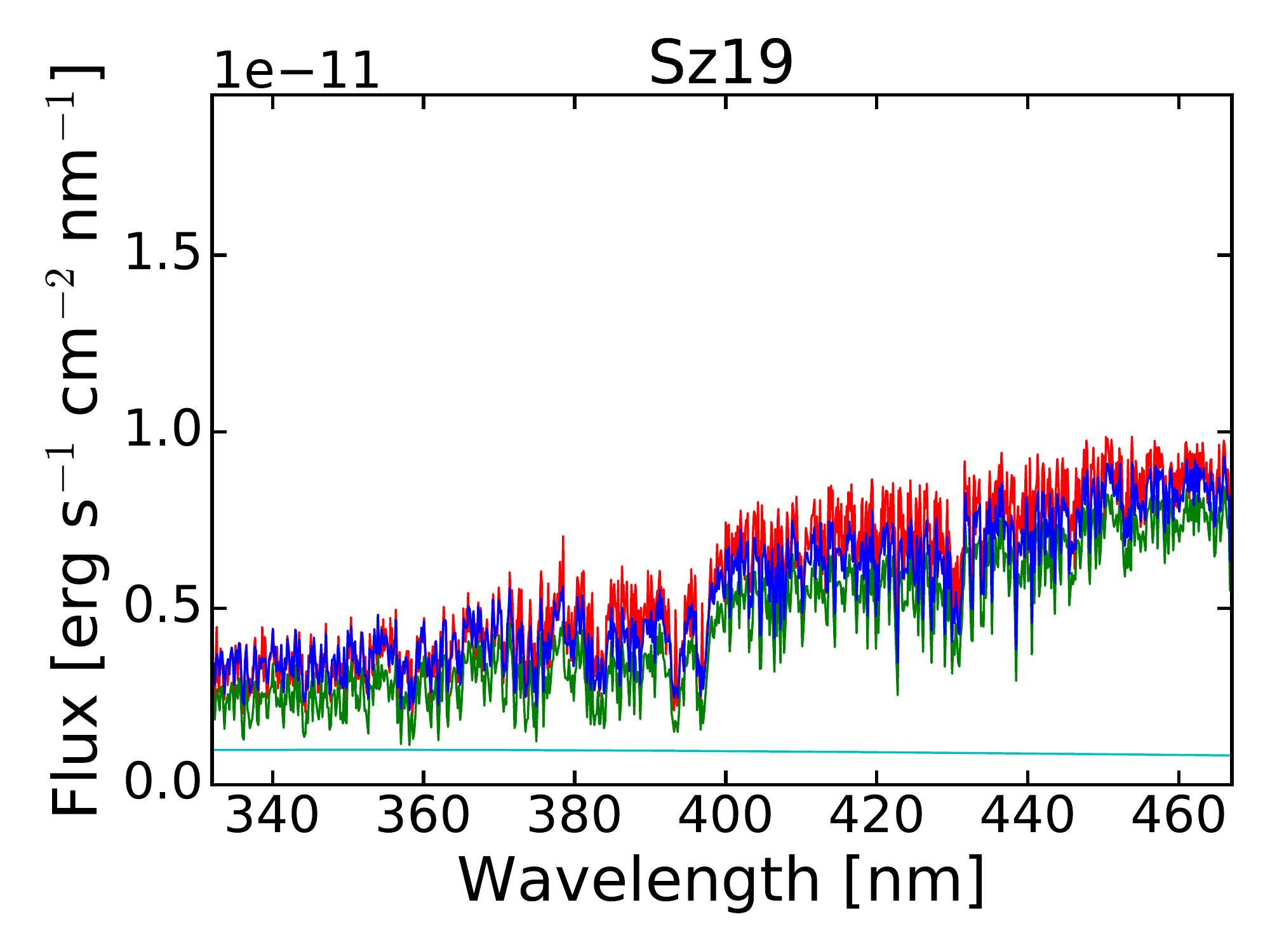}
        \end{subfigure}%
        ~ 
        \begin{subfigure}[b]{0.45\textwidth}
               \includegraphics[width=\textwidth]{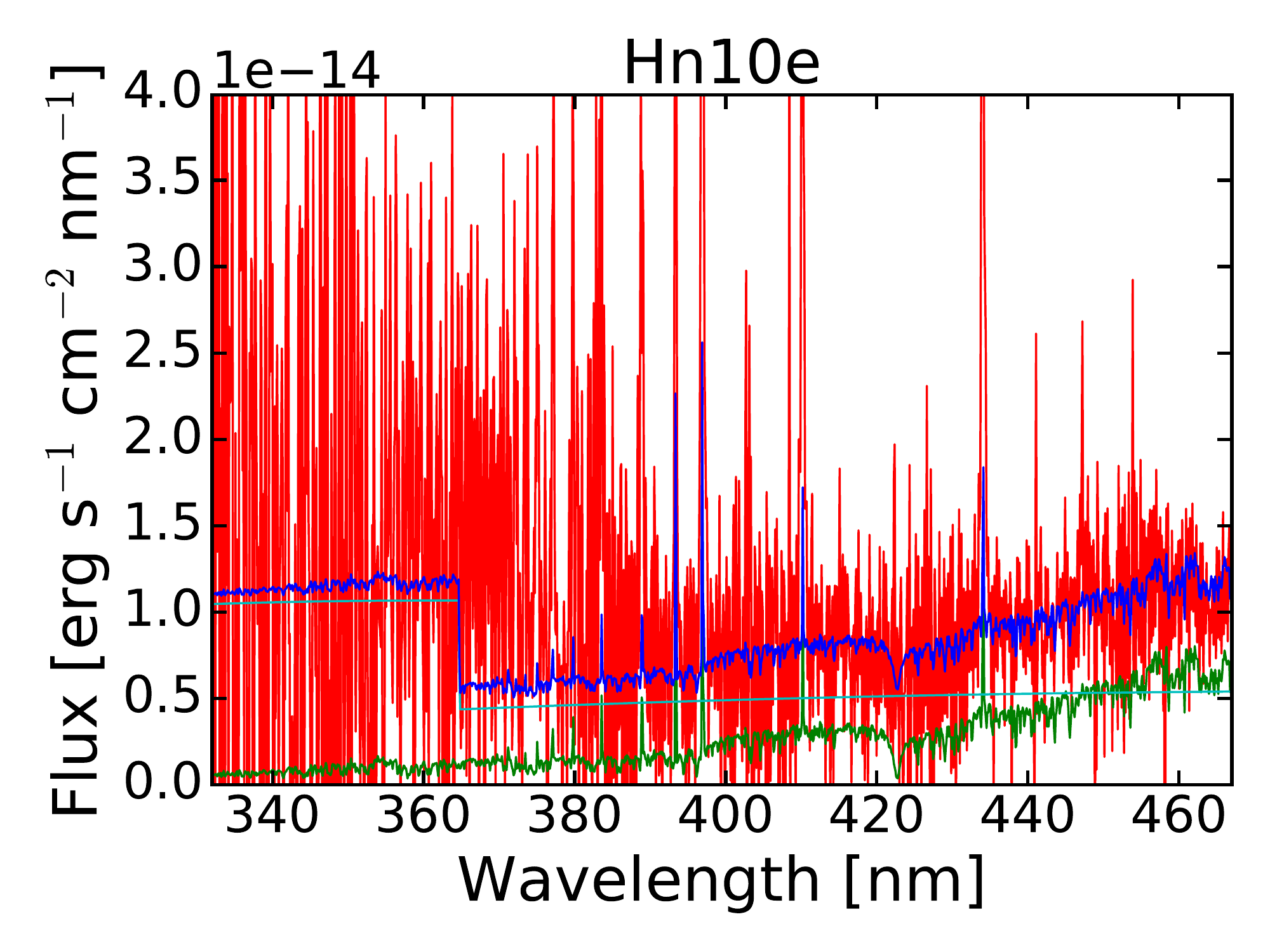}
        \end{subfigure}
        \begin{subfigure}[b]{0.45\textwidth}
               \includegraphics[width=\textwidth]{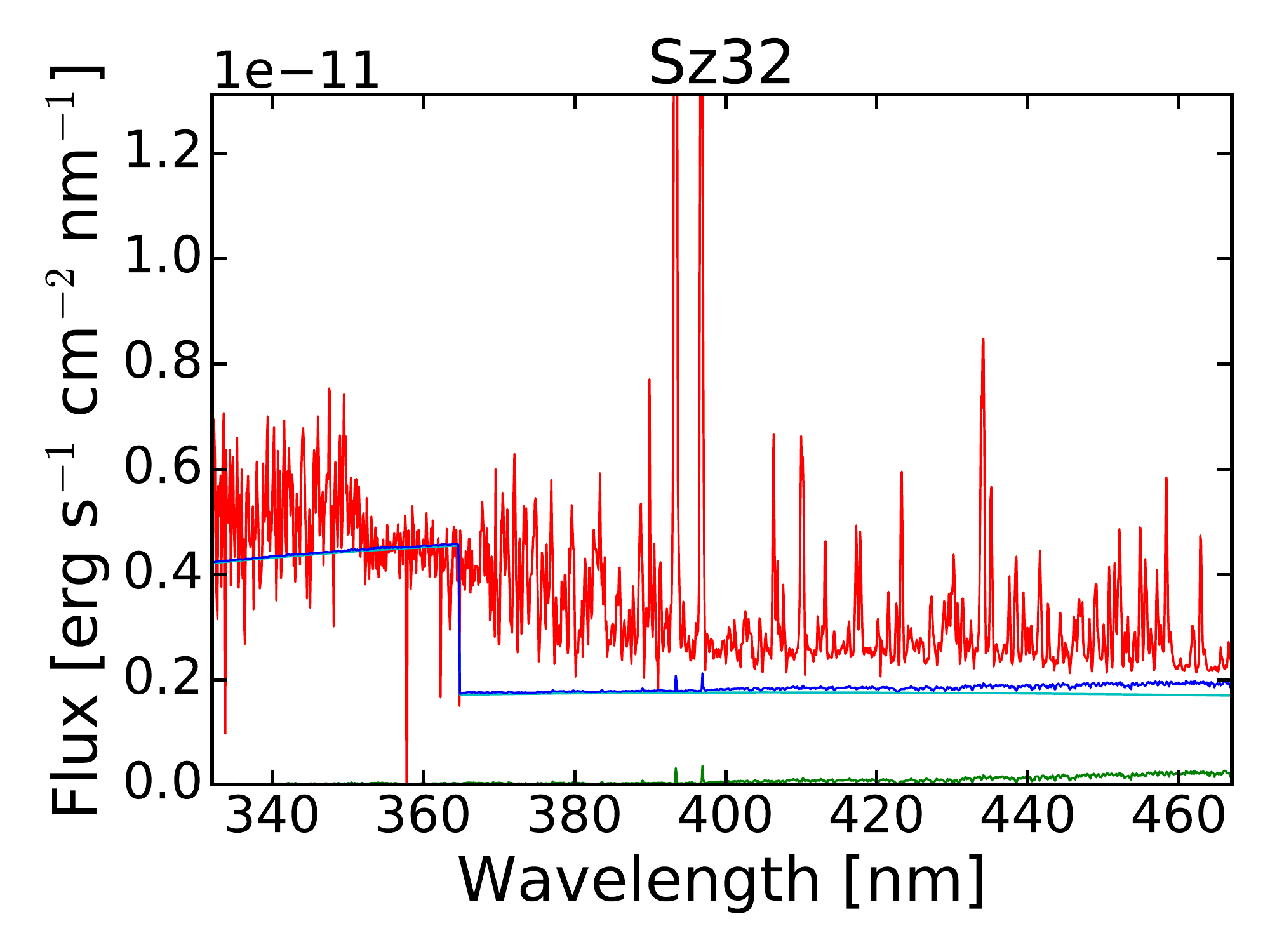}
        \end{subfigure}%
        ~ 
        \begin{subfigure}[b]{0.45\textwidth}
               \includegraphics[width=\textwidth]{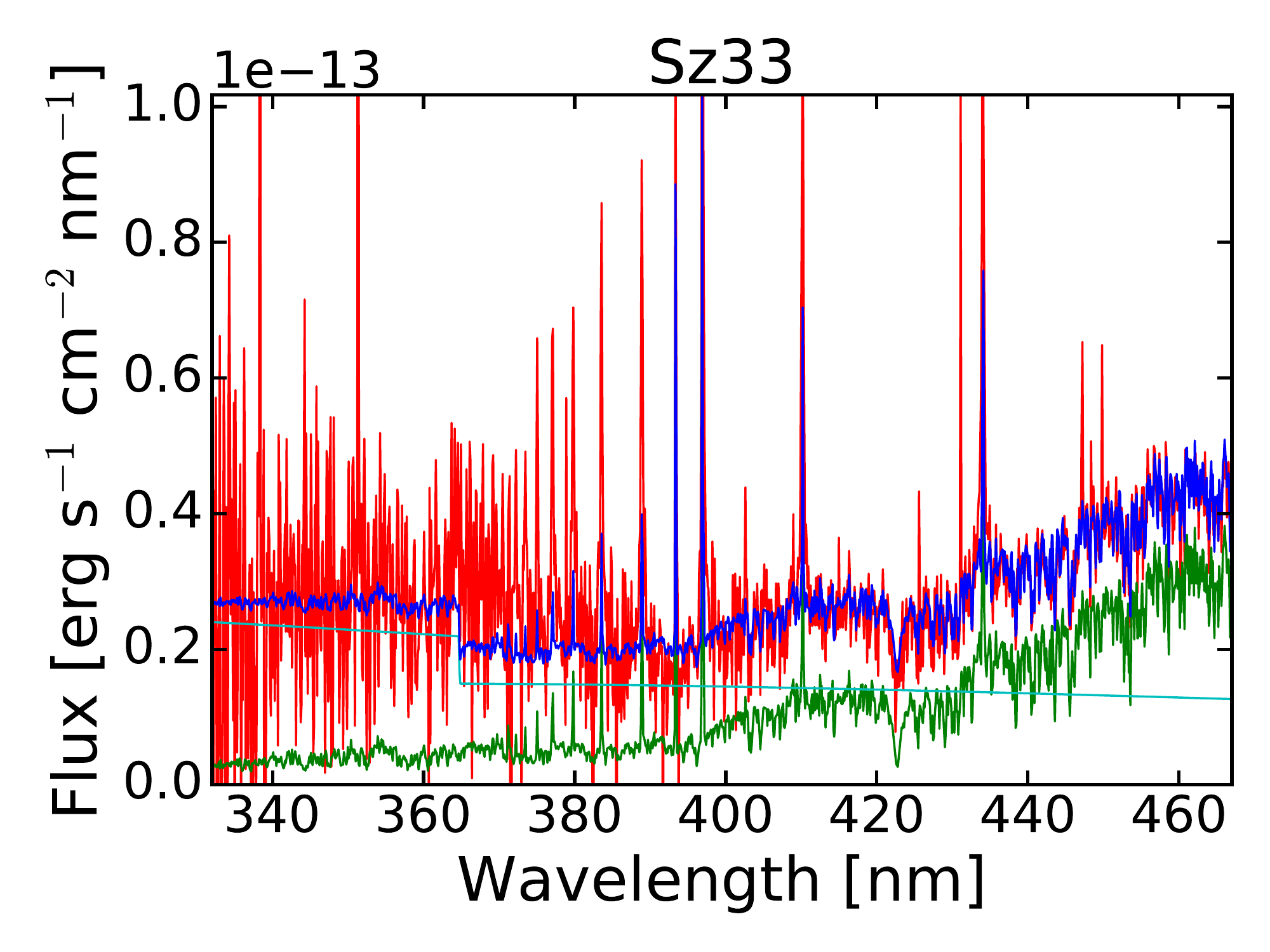}
        \end{subfigure}
         \begin{subfigure}[b]{0.45\textwidth}
               \includegraphics[width=\textwidth]{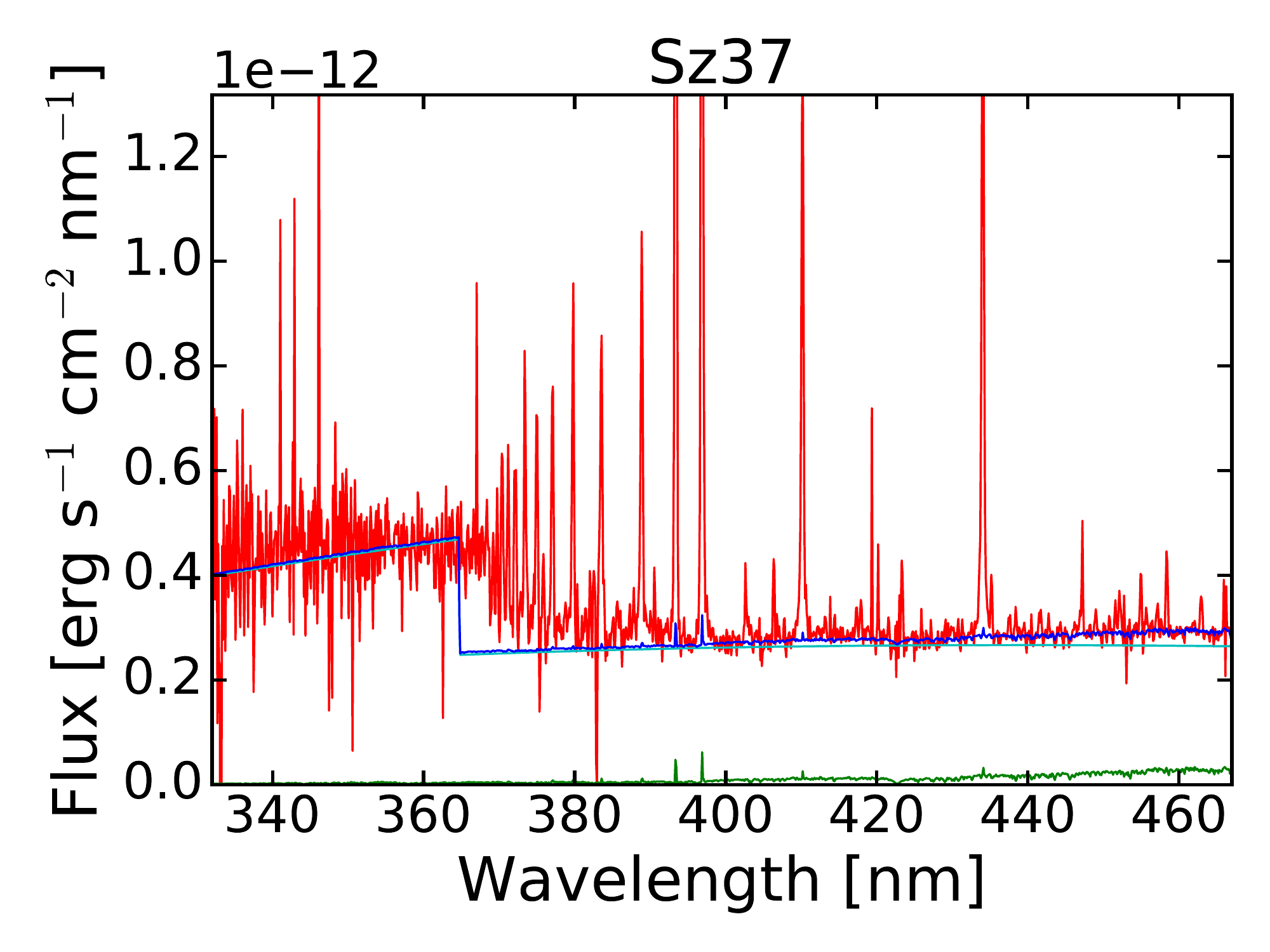}
        \end{subfigure}%
        ~ 
        \begin{subfigure}[b]{0.45\textwidth}
               \includegraphics[width=\textwidth]{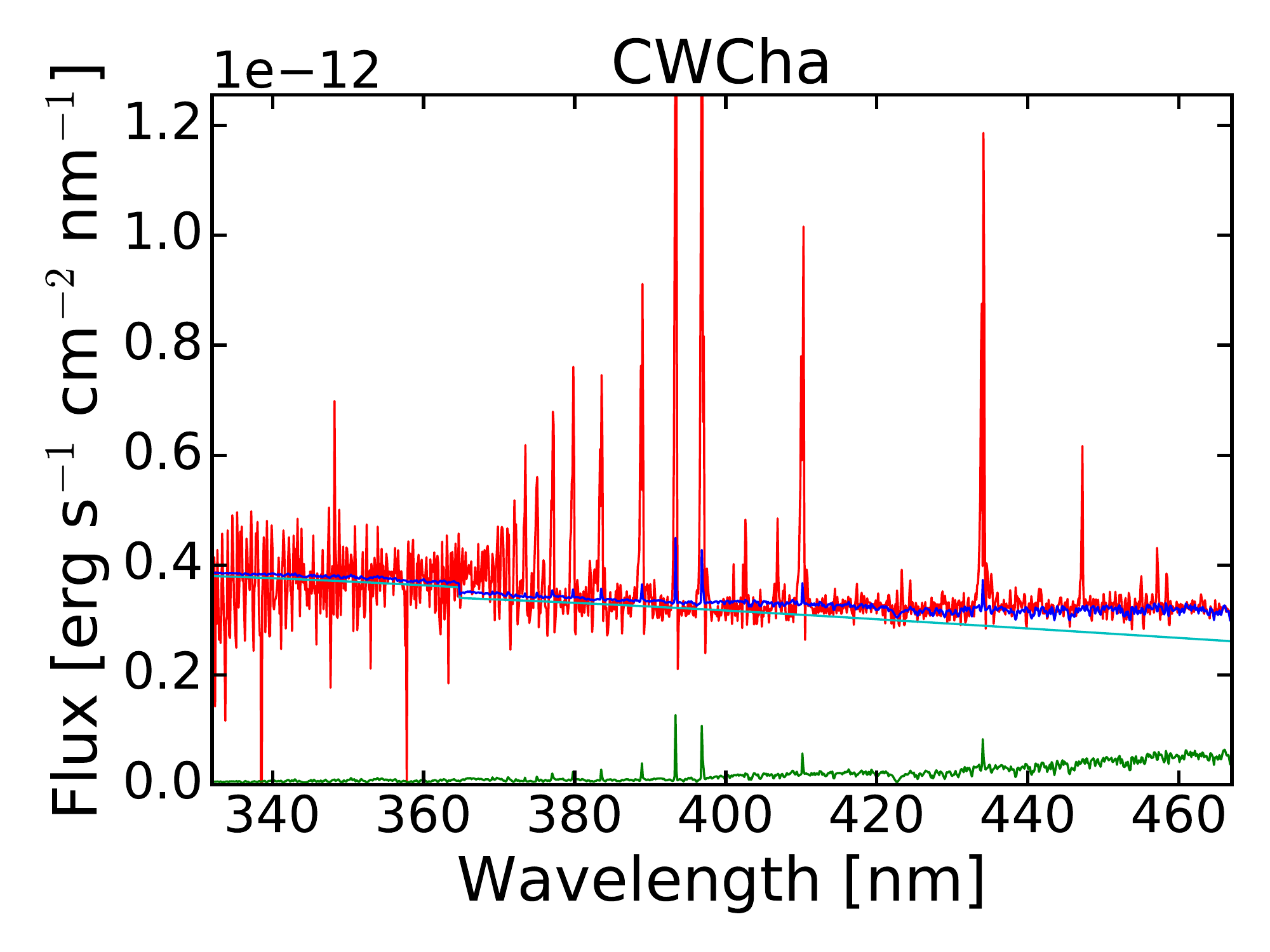}
        \end{subfigure}
       \caption{Same as \ref{fig:best_fits}.
}\label{fig::best_fits4}

\end{figure*}

\clearpage


\section{Observation log}

\begin{table*}  
\begin{center}  
\footnotesize  
\caption{\label{tab::log} Night log and basic information on the spectra }  
\begin{tabular}{l|c|cc| cc | cc | cc }    
\hline \hline  
 2MASS  &  Date of observation [UT] &  \multicolumn{2}{c}{Exp. Time [Nexp x (s)]} |& \multicolumn{2}{c}{Slit Width [\arcsec]} &  \multicolumn{2}{c}{S/N}  & H$\alpha$ & Li \\    
   &   &  UVB & VIS & UVB & VIS & $\lambda$ 450 nm & $\lambda$ 700 nm&  &  \\    
\hline  

\multicolumn{10}{c}{Sample from Pr.Id. 095.C-0378 (PI Testi)}\\   
\hline  
J10533978-7712338 & 2016-01-27T03:27:47.881 & 4 x 675 & 4 x 585 & 1.0 & 0.9 & 1 & 18 & Y& Y \\ 
J10561638-7630530  & 2015-04-05T02:07:11.735 & 4 x 675 & 4 x 585 & 1.0 & 0.4 & 1 & 10 & Y& Y \\ 
J10574219-7659356  & 2015-04-03T07:38:12.483 & 4 x 150 & 4 x 90 & 0.5 & 0.4 & 10 & 20 & Y& Y \\ 
J10580597-7711501        & 2015-06-04T02:01:09.058 & 4 x 735 & 4 x 645 & 1.0 & 0.9 & 1 & 2 & Y& N \\ 
J11004022-7619280        & 2015-05-01T02:50:08.240 & 4 x 450 & 4 x 355 & 1.0 & 0.4 & 14 & 10 & Y& Y \\ 
J11023265-7729129        & 2015-04-07T06:09:08.627 & 4 x 225 & 4 x 145 & 0.5 & 0.4 & 6 & 20 & Y& Y \\ 
J11040425-7639328         & 2016-02-16T02:35:03.092 & 4 x 675 & 4 x 585 & 1.0 & 0.4 & 1 & 13 & Y& N \\ 
J11045701-7715569         & 2016-01-29T06:52:31.844 & 4 x 675 & 4 x 585 & 1.0 & 0.9 & 4 & 27 & Y& Y \\ 
J11062554-7633418        & 2015-04-05T03:17:45.617 & 4 x 675 & 4 x 585 & 1.3 & 0.9 & 0 & 9 & Y& N \\ 
J11063276-7625210         & 2015-04-03T05:03:18.113 & 4 x 675 & 4 x 585 & 1.3 & 0.9 & 0 & 7 & Y& N \\ 
J11063945-7736052  & 2015-05-30T00:35:03.030 & 4 x 675 & 4 x 585 & 1.3 & 0.9 & 0 & 1 & Y& N \\ 
J11064510-7727023         & 2015-06-17T01:02:07.452 & 4 x 450 & 4 x 355 & 1.0 & 0.4 & 28 & 52 & Y& Y \\ 
J11065939-7530559         & 2015-04-20T01:52:49.561 & 4 x 735 & 4 x 645 & 1.0 & 0.9 & 3 & 15 & Y& Y \\ 
J11071181-7625501        & 2015-04-04T04:54:53.369 & 4 x 735 & 4 x 645 & 1.0 & 0.9 & 1 & 10 & Y& N \\ 
J11072825-7652118  & 2015-04-03T06:33:02.140 & 4 x 150 & 4 x 90 & 0.5 & 0.4 & 12 & 14 & Y& Y \\ 
J11074245-7733593         & 2016-02-16T03:42:49.652 & 4 x 675 & 4 x 585 & 1.0 & 0.9 & 1 & 18 & Y& Y \\ 
J11074366-7739411        & 2015-04-14T02:45:28.404 & 4 x 225 & 4 x 145 & 0.5 & 0.4 & 9 & 30 & Y& Y \\ 
J11074656-7615174         & 2015-04-03T04:02:27.543 & 4 x 675 & 4 x 585 & 1.3 & 0.9 & 0 & 6 & Y& N \\ 
J11075809-7742413         & 2016-01-29T08:04:03.596 & 4 x 675 & 4 x 585 & 1.0 & 0.9 & 6 & 34 & Y& Y \\ 
J11080002-7717304         & 2016-01-28T05:00:15.639 & 4 x 675 & 4 x 585 & 1.3 & 0.9 & 1 & 25 & Y& Y \\ 
J11081850-7730408  & 2015-04-23T03:15:29.168 & 4 x 735 & 4 x 645 & 1.0 & 0.9 & 1 & 9 & Y& N \\ 
J11082650-7715550         & 2015-04-28T01:52:15.781 & 4 x 675 & 4 x 585 & 1.3 & 0.9 & 0 & 7 & Y& N \\ 
J11085090-7625135  & 2015-05-03T03:28:19.945 & 4 x 675 & 4 x 585 & 1.0 & 0.4 & 3 & 15 & Y& Y \\ 
J11085367-7521359         & 2015-04-14T02:09:15.906 & 4 x 150 & 4 x 90 & 0.5 & 0.4 & 26 & 22 & Y& Y \\ 
J11085497-7632410        & 2015-04-06T03:59:56.930 & 4 x 735 & 4 x 645 & 1.0 & 0.9 & 1 & 15 & Y& Y \\ 
J11092266-7634320        & 2015-04-03T02:52:56.195 & 4 x 675 & 4 x 585 & 1.3 & 0.9 & 0 & 2 & Y& N \\ 
J11095336-7728365         & 2015-04-28T03:08:09.993 & 4 x 675 & 4 x 585 & 1.3 & 0.9 & 0 & 2 & Y& N \\ 
J11095873-7737088         & 2015-04-03T07:12:09.712 & 4 x 150 & 4 x 90 & 0.5 & 0.4 & 19 & 38 & Y& Y \\ 
J11100369-7633291        & 2015-04-05T04:25:14.823 & 4 x 675 & 4 x 585 & 1.0 & 0.9 & 1 & 29 & Y& Y \\ 
J11100704-7629376         & 2015-04-05T06:19:37.863 & 4 x 150 & 4 x 90 & 0.5 & 0.4 & 10 & 33 & Y& Y \\ 
J11100785-7727480         & 2015-04-30T01:42:57.583 & 4 x 675 & 4 x 585 & 1.3 & 0.9 & 0 & 2 & Y& N \\ 
J11103801-7732399  & 2015-04-03T06:06:52.758 & 4 x 150 & 4 x 90 & 0.5 & 0.4 & 19 & 84 & Y& Y \\ 
J11104141-7720480         & 2015-05-03T02:13:51.777 & 4 x 675 & 4 x 585 & 1.3 & 0.9 & 0 & 3 & Y& N \\ 
J11105333-7634319        & 2015-04-14T03:26:22.548 & 4 x 225 & 4 x 145 & 0.5 & 0.4 & 18 & 26 & Y& Y \\ 
J11105359-7725004  & 2015-05-17T00:59:04.809 & 4 x 675 & 4 x 585 & 1.3 & 0.9 & 0 & 3 & Y& N \\ 
J11105597-7645325        & 2015-04-05T05:26:50.337 & 4 x 450 & 4 x 355 & 1.0 & 0.4 & 2 & 11 & Y& Y \\ 
J11111083-7641574  & 2016-02-16T04:56:59.416 & 4 x 675 & 4 x 585 & 1.3 & 0.9 & 2 & 7 & Y& N \\ 
J11120351-7726009  & 2015-05-28T01:28:57.685 & 4 x 675 & 4 x 585 & 1.3 & 0.9 & 1 & 16 & Y& Y \\ 
J11120984-7634366  & 2015-04-14T03:59:22.529 & 4 x 225 & 4 x 145 & 0.5 & 0.4 & 9 & 15 & Y& Y \\ 
J11175211-7629392\_one   & 2015-04-18T01:13:13.804 & 4 x 225 & 4 x 145 & 0.5 & 0.4 & 4 & 10 & Y& Y \\ 
J11175211-7629392\_two   & 2015-04-18T01:13:13.804 & 4 x 225 & 4 x 145 & 0.5 & 0.4 & 5 & 7 & Y& Y \\ 
J11183572-7935548   & 2015-04-03T01:56:53.970 & 4 x 150 & 4 x 90 & 0.5 & 0.4 & 12 & 27 & Y& Y \\ 
J11241186-7630425         & 2015-04-06T02:48:05.668 & 4 x 675 & 4 x 585 & 1.0 & 0.9 & 3 & 17 & Y& Y \\ 
J11432669-7804454        & 2015-04-20T03:09:12.788 & 4 x 450 & 4 x 355 & 1.0 & 0.4 & 36 & 17 & Y& Y \\ 
\hline  
\multicolumn{10}{c}{Sample from Pr.Id. 090.C-0253 (PI Antoniucci)}\\   
\hline  
J11072074-7738073         & 2013-03-14T00:23:08.247 & 2 x 50 & 2 x 90 & 0.5 & 0.4 & 52 & 112 & Y& Y \\ 
J11091812-7630292         & 2013-03-14T02:58:36.662 & 6 x 300 & 6 x 360 & 0.5 & 0.4 & 1 & 14 & Y& Y \\ 
J11094621-7634463  & 2013-03-15T04:55:37.960 & 6 x 300 & 6 x 360 & 0.5 & 0.4 & 1 & 17 & Y& Y \\ 
J11094742-7726290         & 2013-03-14T03:51:55.384 & 8 x 300 & 8 x 360 & 0.5 & 0.4 & 0 & 5 & Y& N \\ 
J11095340-7634255         & 2013-03-15T05:58:40.596 & 4 x 300 & 4 x 360 & 0.5 & 0.4 & 21 & 82 & Y& Y \\ 
J11095407-7629253        & 2013-03-14T00:37:41.949 & 6 x 300 & 6 x 360 & 0.5 & 0.4 & 6 & 33 & Y& Y \\ 
J11104959-7717517        & 2013-03-14T02:08:43.586 & 4 x 300 & 4 x 360 & 0.5 & 0.4 & 13 & 34 & Y& Y \\ 
J11123092-7644241        & 2013-03-14T01:32:05.376 & 4 x 300 & 4 x 360 & 0.5 & 0.4 & 16 & 45 & Y& Y \\ 
\hline 
\end{tabular} 
\end{center} 
\end{table*}  

\section{Stellar masses and mass accretion rates using different evolutionary models}

\begin{table*}  
\begin{center}  
\footnotesize  
\caption{\label{tab::results_all_masses_new} Stellar mass and mass accretion rates for the whole sample }  
\begin{tabular}{l|l| cc | cc | cc | l }    
\hline \hline  
 2MASS & Object &  \mstar(B15) & log\macc(B15) &  \mstar(B98) & log\macc(B98) &  \mstar(S00) & log\macc(S00) & Notes \\    
   &   &  [\msun] & [\msun/yr]  &  \\     
\hline  

\multicolumn{9}{c}{Sample from Pr.Id. 095.C-0378 (PI Testi)}\\   
\hline  
J10533978-7712338  &   \nodata   & \nodata & \nodata & \nodata & \nodata & 0.33 & -11.95 & UL,$^m$ \\ 
J10561638-7630530   &   ESO H$\alpha$ 553  & 0.11 & -10.95 & 0.11 & -10.95 & 0.10 & -10.91 & $^\dag$ \\ 
J10574219-7659356   &   T5        & 0.28 & -8.51 & 0.51 & -8.77 & 0.33 & -8.57 & \nodata \\ 
J10580597-7711501         &   \nodata   & 0.11 & -11.87 & 0.09 & -11.80 & 0.10 & -11.85 & $^\dag$,$^m$ \\ 
J11004022-7619280         &   T10    & 0.23 & -9.22 & 0.26 & -9.27 & 0.23 & -9.21 & \nodata \\ 
J11023265-7729129         &   CHXR71   & 0.29 & -10.52 & 0.40 & -10.65 & 0.32 & -10.55 & $^\dag$ \\ 
J11040425-7639328          &   CHSM1715   & 0.18 & -10.84 & 0.16 & -10.79 & 0.16 & -10.80 & $^m$ \\ 
J11045701-7715569          &   T16    & 0.29 & -7.80 & 0.41 & -7.94 & 0.32 & -7.83 & \nodata \\ 
J11062554-7633418         &   ESO H$\alpha$ 559  & 0.12 & -10.79 & 0.11 & -10.74 & 0.11 & -10.77 & $^m$ \\ 
J11063276-7625210          &   CHSM 7869   & 0.07 & -11.06 & 0.07 & -11.04 & \nodata & \nodata & $^m$ \\ 
J11063945-7736052   &   ISO-ChaI 79   & \nodata & \nodata & \nodata & \nodata & \nodata & \nodata & UL,$^m$ \\ 
J11064510-7727023          &   CHXR20   & 0.90 & -8.71 & 1.12 & -8.80 & 0.97 & -8.74 & \nodata \\ 
J11065939-7530559          &   \nodata   & 0.10 & -11.12 & 0.09 & -11.06 & 0.10 & -11.12 & \nodata \\ 
J11071181-7625501         &   CHSM 9484   & 0.10 & -11.84 & 0.09 & -11.78 & 0.10 & -11.84 & $^\dag$,$^m$ \\ 
J11072825-7652118   &   T27    & 0.29 & -8.36 & 0.43 & -8.53 & 0.32 & -8.41 & \nodata \\ 
J11074245-7733593          &   Cha-H$\alpha$-2   & 0.13 & -10.05 & 0.12 & -10.02 & 0.12 & -10.01 & \nodata \\ 
J11074366-7739411         &   T28    & 0.48 & -7.92 & 0.72 & -8.10 & 0.46 & -7.91 & \nodata \\ 
J11074656-7615174          &   CHSM 10862   & 0.07 & -12.03 & 0.07 & -12.01 & \nodata & \nodata & $^\dag$,$^m$ \\ 
J11075809-7742413          &   T30   & 0.30 & -8.31 & 0.37 & -8.39 & 0.31 & -8.32 & \nodata \\ 
J11080002-7717304          &   CHXR30A   & 0.69 & -10.17 & 1.17 & -10.40 & 0.76 & -10.21 & $^\dag$,$^m$ \\ 
J11081850-7730408   &   ISO-ChaI 138   & 0.07 & -11.81 & 0.07 & -11.80 & \nodata & \nodata & $^\dag$ \\ 
J11082650-7715550          &   ISO-ChaI 147   & 0.11 & -11.15 & 0.09 & -11.08 & 0.10 & -11.13 & $^m$ \\ 
J11085090-7625135   &   T37   & 0.12 & -10.74 & 0.11 & -10.69 & 0.10 & -10.66 & \nodata \\ 
J11085367-7521359          &   \nodata   & 0.51 & -8.15 & 0.68 & -8.27 & 0.45 & -8.09 & \nodata \\ 
J11085497-7632410         &   ISO-ChaI 165    & 0.12 & -10.65 & 0.10 & -10.60 & 0.11 & -10.61 & $^m$ \\ 
J11092266-7634320         &   C 1-6    & 0.58 & -9.54 & 0.61 & -9.56 & 0.44 & -9.42 & $^m$ \\ 
J11095336-7728365          &   ISO-ChaI 220   & 0.11 & -10.45 & 0.09 & -10.38 & 0.10 & -10.43 & $^m$ \\ 
J11095873-7737088          &   T45    & 0.49 & -6.95 & 0.88 & -7.20 & 0.51 & -6.97 & \nodata \\ 
J11100369-7633291         &   Hn11  & 0.63 & -9.62 & 0.83 & -9.74 & 0.58 & -9.58 & $^m$ \\ 
J11100704-7629376          &   T46   & 0.75 & -8.70 & 1.07 & -8.85 & 0.81 & -8.73 & \nodata \\ 
J11100785-7727480          &   ISO-ChaI 235   & 0.13 & -10.96 & 0.12 & -10.91 & 0.11 & -10.87 & $^m$ \\ 
J11103801-7732399   &   CHXR 47   & 1.32 & -8.12 & \nodata & \nodata & 1.51 & -8.18 & \nodata \\ 
J11104141-7720480          &   ISO-ChaI 252   & 0.11 & -9.91 & 0.09 & -9.84 & 0.10 & -9.89 & $^m$ \\ 
J11105333-7634319         &   T48   & 0.30 & -7.96 & 0.37 & -8.04 & 0.31 & -7.97 & \nodata \\ 
J11105359-7725004   &   ISO-ChaI 256   & 0.15 & -10.32 & 0.13 & -10.27 & 0.14 & -10.28 & $^m$ \\ 
J11105597-7645325         &   Hn13  & \nodata & \nodata & \nodata & \nodata & 0.12 & -9.57 & \nodata \\ 
J11111083-7641574   &   ESO H$\alpha$ 569  & \nodata & \nodata & \nodata & \nodata & \nodata & \nodata & UL \\ 
J11120351-7726009   &   ISO-ChaI 282   & 0.14 & -9.89 & 0.14 & -9.89 & 0.14 & -9.89 & \nodata \\ 
J11120984-7634366   &   T50   & 0.17 & -9.34 & 0.16 & -9.31 & 0.19 & -9.39 & \nodata \\ 
J11175211-7629392\_one    &   \nodata   & 0.20 & -11.16 & 0.20 & -11.17 & 0.18 & -11.12 & $^\dag$ \\ 
J11175211-7629392\_two    &   \nodata   & 0.19 & -11.44 & 0.20 & -11.45 & 0.18 & -11.42 & $^\dag$ \\ 
J11183572-7935548    &  \nodata    & 0.19 & -8.95 & \nodata & \nodata & 0.21 & -8.98 & TD \\ 
J11241186-7630425          &   \nodata   & 0.12 & -10.59 & 0.11 & -10.54 & 0.10 & -10.51 & TD \\ 
J11432669-7804454         &   \nodata   & 0.14 & -8.71 & 0.14 & -8.71 & 0.15 & -8.74 & \nodata \\ 
\hline  
\multicolumn{9}{c}{Sample from Pr.Id. 090.C-0253 (PI Antoniucci)}\\   
\hline  
J11072074-7738073          &   Sz19   & \nodata & \nodata & \nodata & \nodata & 2.08 & -7.63 & \nodata \\ 
J11091812-7630292          &   CHXR79   & 0.62 & -9.05 & 0.84 & -9.18 & 0.59 & -9.03 & $^m$ \\ 
J11094621-7634463   &   Hn 10e   & 0.34 & -9.51 & 0.36 & -9.54 & 0.27 & -9.41 & \nodata \\ 
J11094742-7726290          &   ISO-ChaI 207   & 0.58 & -9.21 & 0.64 & -9.25 & 0.45 & -9.10 & $^m$ \\ 
J11095340-7634255          &   Sz32   & 0.78 & -7.08 & 1.05 & -7.22 & 0.82 & -7.11 & \nodata \\ 
J11095407-7629253         &   Sz33   & 0.56 & -9.35 & 0.65 & -9.41 & 0.45 & -9.25 & \nodata \\ 
J11104959-7717517         &   Sz37   & 0.41 & -7.82 & 0.51 & -7.91 & 0.36 & -7.76 & \nodata \\ 
J11123092-7644241         &   CW Cha   & 0.59 & -8.03 & 0.75 & -8.13 & 0.51 & -7.96 & \nodata \\ 
\hline  
\end{tabular} 
\tablefoot{UL = objects located well below the 30 Myr isochrone. $^\dag$ Objects with low accretion, compatible with chromospheric noise. $^m$ Stellar and accretion parameters not derived from UV-excess. TD = transition disks. All stellar parameters have been derived using the \citet{Baraffe15} evolutionary models except for objects with an asterisk, for which the \citet{Siess00} models were used.} 
\end{center} 
\end{table*}  

\begin{table*}  
\begin{center}  
\footnotesize  
\caption{\label{tab::results_all_models} Stellar mass and mass accretion rates for the whole sample }  
\begin{tabular}{l|l| cc | cc | cc | l }    
\hline \hline  
 2MASS & Object &  \mstar(B15) & log\macc(B15) &  \mstar(B98) & log\macc(B98) &  \mstar(S00) & log\macc(S00) & Notes \\    
   &   &  [\msun] & [\msun/yr]  &  \\     
\hline  

\multicolumn{9}{c}{Data from \citet{Manara16} }\\   
\hline  
J10555973-7724399 & T3 & 0.77 & -8.61 & 0.79 & -8.62 & 0.69 & -8.56 & \nodata \\ 
... & T3 B & 0.29 & -8.43 & 0.38 & -8.54 & 0.31 & -8.46 & \nodata \\ 
J10563044-7711393 & T4 & 0.78 & -9.41 & 1.03 & -9.53 & 0.82 & -9.43 & \nodata \\ 
J10590108-7722407 & TW Cha & 0.79 & -8.86 & 1.00 & -8.96 & 0.83 & -8.89 & \nodata \\ 
J10590699-7701404 & CR Cha & \nodata & \nodata & \nodata & \nodata & 1.77 & -8.71 & \nodata,$^*$ \\ 
J11025504-7721508 & T12 & 0.19 & -8.70 & 0.23 & -8.78 & 0.22 & -8.75 & \nodata \\ 
J11040909-7627193 & CT Cha A & 0.98 & -6.69 & 1.40 & -6.85 & 1.09 & -6.74 & \nodata \\ 
J11044258-7741571 & ISO-ChaI 52 & 0.23 & -10.59 & 0.25 & -10.61 & 0.22 & -10.56 & $^\dag$ \\ 
J11064180-7635489 & Hn 5 & 0.16 & -9.28 & 0.16 & -9.27 & 0.15 & -9.25 & \nodata \\ 
J11065906-7718535 & T23 & 0.21 & -8.11 & 0.33 & -8.29 & 0.24 & -8.16 & \nodata \\ 
J11071206-7632232 & T24 & 0.58 & -8.49 & 0.91 & -8.68 & 0.57 & -8.48 & \nodata \\ 
J11071668-7735532 & Cha H$\alpha$1 & 0.04 & -11.68 & 0.05 & -11.69 & \nodata & \nodata & \nodata \\ 
J11071860-7732516 & Cha H$\alpha$ 9 & 0.12 & -10.91 & 0.10 & -10.85 & 0.10 & -10.84 & \nodata \\ 
J11075792-7738449 & Sz 22 & 1.01 & -8.34 & 1.08 & -8.37 & 1.03 & -8.35 & \nodata \\ 
J11080148-7742288 & VW Cha & 0.67 & -7.60 & 1.24 & -7.86 & 0.74 & -7.64 & \nodata \\ 
J11080297-7738425 & ESO H$\alpha$ 562 & 0.56 & -9.24 & 0.66 & -9.31 & 0.45 & -9.14 & \nodata \\ 
J11081509-7733531 & T33 A & 1.26 & -8.97 & 1.15 & -8.93 & 1.23 & -8.96 & \nodata \\ 
... & T33 B & 1.00 & -8.69 & 0.95 & -8.67 & 0.98 & -8.68 & \nodata \\ 
J11082238-7730277 & ISO-ChaI 143 & 0.12 & -10.07 & 0.11 & -10.02 & 0.10 & -9.99 & \nodata \\ 
J11083952-7734166 & Cha H$\alpha$6 & 0.10 & -10.25 & 0.10 & -10.26 & 0.10 & -10.25 & \nodata \\ 
J11085464-7702129 & T38 & 0.63 & -9.30 & 0.71 & -9.35 & 0.52 & -9.22 & \nodata \\ 
J11092379-7623207 & T40 & 0.49 & -7.33 & 0.87 & -7.58 & 0.52 & -7.36 & \nodata \\ 
J11100010-7634578 & T44 & \nodata & \nodata & \nodata & \nodata & 1.65 & -6.68 & \nodata,$^*$ \\ 
J11100469-7635452 & T45a & 0.80 & -9.83 & 0.97 & -9.91 & 0.82 & -9.84 & \nodata \\ 
J11101141-7635292 & ISO-ChaI 237 & 1.03 & -9.74 & 1.15 & -9.79 & 1.07 & -9.76 & $^\dag$ \\ 
J11113965-7620152 & T49 & 0.25 & -7.41 & 0.36 & -7.55 & 0.29 & -7.46 & \nodata \\ 
J11114632-7620092 & CHX18N & 1.25 & -8.09 & 1.17 & -8.06 & 1.22 & -8.08 & \nodata \\ 
J11122441-7637064 & T51 & 1.04 & -8.16 & 0.97 & -8.13 & 1.01 & -8.15 & \nodata \\ 
... & T51 B & 0.44 & -9.07 & 0.51 & -9.13 & 0.35 & -8.97 & \nodata \\ 
J11122772-7644223 & T52 & \nodata & \nodata & \nodata & \nodata & 1.62 & -7.48 & \nodata,$^*$ \\ 
J11124268-7722230 & T54 A & \nodata & \nodata & \nodata & \nodata & 1.62 & -9.60 & $^\dag$,TD,$^*$ \\ 
J11124861-7647066 & Hn17 & 0.20 & -9.71 & 0.22 & -9.76 & 0.20 & -9.73 & \nodata \\ 
J11132446-7629227 & Hn18 & 0.24 & -9.81 & 0.26 & -9.85 & 0.23 & -9.81 & \nodata \\ 
J11142454-7733062 & Hn21W & 0.20 & -9.04 & 0.22 & -9.10 & 0.20 & -9.06 & \nodata \\ 
\hline  
\multicolumn{9}{c}{Data from \citet{Manara14} }\\   
\hline  
J10581677-7717170 & Sz Cha & 1.31 & -7.82 & 1.22 & -7.79 & 1.28 & -7.81 & TD \\ 
J11022491-7733357 & CS Cha & 1.40 & -8.29 & 1.32 & -8.27 & 1.40 & -8.29 & TD \\ 
J11071330-7743498 & CHXR22E & 0.23 & -10.90 & 0.24 & -10.91 & 0.21 & -10.87 & $^\dag$,TD \\ 
J11071915-7603048 & Sz18 & 0.38 & -8.70 & 0.54 & -8.86 & 0.38 & -8.71 & TD \\ 
J11083905-7716042 & Sz27 & 0.80 & -8.86 & 0.96 & -8.94 & 0.81 & -8.87 & TD \\ 
J11173700-7704381 & Sz45 & 0.51 & -8.09 & 0.85 & -8.31 & 0.51 & -8.09 & TD \\ 
\hline  
\multicolumn{9}{c}{Data from \citet{Whelan14} }\\   
\hline  
J11095215-7639128 & ISO-ChaI217 & 0.08 & -10.70 & 0.07 & -10.66 & \nodata & \nodata & \nodata,$^*$ \\ 
\hline 
\end{tabular} 
\tablefoot{UL = objects located well below the 30 Myr isochrone. $^\dag$ Objects with low accretion, compatible with chromospheric noise. $^m$ Stellar and accretion parameters not derived from UV-excess. TD = transition disks. All stellar parameters have been derived using the \citet{Baraffe15} evolutionary models except for objects with an asterisk, for which the \citet{Siess00} models were used.} 
\end{center} 
\end{table*}

\end{document}